\documentclass{ws-procs9x6-notrim}
\usepackage{graphicx,color,wrapfig}
\newcommand{\beq}{\begin{equation*}}
\newcommand{\eeq}{\end{equation*}}
\newcommand{\beqs}{\begin{eqnarray*}}
\newcommand{\eeqs}{\end{eqnarray*}}
\newcommand{\bitem}{\begin{itemize}\item}
\newcommand{\eitem}{\end{itemize}}
\newcommand{\xb}[1]{\textcolor{darkgreen}{x_b(#1)}}
\newcommand{\xa}[1]{\textcolor{red}{x_a(#1)}}
\newcommand{\xt}{\textcolor{magenta}{x(t_2)x(t_1)}}
\newcommand{\phin}{\phi_n(0)}
\newcommand{\phim}{\textcolor{purple}{\phi_m(0)}}
\newcommand{\dd}{\displaystyle}

\newcommand{\ip}[2]{\langle #1\vert #2\rangle}

\newcommand{\Pmat}{\mathbf{P}}
\newcommand{\Wmat}{\mathbf{W}}

\newcommand{\uvec}{\mathbf{u}}
\newcommand{\vvec}{\mathbf{v}}
\newcommand{\wvec}{\mathbf{w}}
\newcommand{\yvec}{\mathbf{y}}

\definecolor{darkgreen}{rgb}{0,0.6,0}
\definecolor{pink}{rgb}{1,0.6,1}
\definecolor{purple}{rgb}{0.9,0,0.9}
\newcommand{\alert}[1]{\textit{#1}}
\newcommand{\bfalert}[1]{\textbf{\textit{#1}}}

\begin{document}

\title{THE MONTE CARLO METHOD\\ IN QUANTUM FIELD THEORY}

\author{COLIN MORNINGSTAR}

\address{Department of Physics,
         Carnegie Mellon University,\\
         Pittsburgh, PA 15213, USA\\
         E-mail: colin\_morningstar@cmu.edu}

\begin{abstract}
This series of six lectures is an introduction to using the Monte Carlo
method to carry out nonperturbative studies in quantum field theories.  
Path integrals in quantum field theory are reviewed, and their evaluation
by the Monte Carlo method with Markov-chain based importance sampling
is presented.  Properties of Markov chains are discussed in detail
and several proofs are presented, culminating in the fundamental limit 
theorem for irreducible Markov chains.
The example of a real scalar field theory is used to illustrate
the Metropolis-Hastings method and to demonstrate the effectiveness
of an action-preserving (microcanonical) local updating algorithm
in reducing autocorrelations.  The goal of these lectures is to provide
the beginner with the basic skills needed to start carrying out Monte
Carlo studies in quantum field theories, as well as to present the 
underlying theoretical foundations of the method.
\end{abstract}


\bodymatter

\section{Introduction}
\label{sec:intro}

Some of the most interesting features of quantum field theories,
such as spontaneous symmetry breaking and bound states of particles,
require computational treatments beyond ordinary perturbation theory.
The Monte Carlo method using Markov-chain based importance sampling 
with a space-time lattice regulator is a powerful tool for carrying
out such studies.  One of the most prominent applications of such
methods is hadron formation and quark confinement in quantum 
chromodynamics (QCD).  This series of six lectures is an introduction to 
using the Monte Carlo method to carry out nonperturbative studies in
quantum field theories.  

In Sec.~\ref{sec:paths}, the path integral method in nonrelativistic
quantum mechanics is briefly reviewed and illustrated using
several simple examples: a free particle in one dimension,
the one-dimensional infinite square well, a free particle
in one dimension with periodic boundary conditions, and the
one-dimensional simple harmonic oscillator.  The extraction
of observables from correlation functions or vacuum expectation
values is discussed, and the evaluation of these correlation
functions using ratios of path integrals is described.  The
crucial trick of Wick rotating to imaginary time is introduced.

The evaluation of path integrals in the imaginary time formalism
using the Monte Carlo method is discussed next in 
Sec.~\ref{sec:montecarlo}.  After a brief
review of probability theory in Sec.~\ref{sec:prob}, simple Monte Carlo
integration is described, and its justification by the law of 
large numbers and the central limit theorem is outlined.
The need for clever importance sampling is then emphasized, leading
to the use of stationary stochastic processes and the modification
of the Monte Carlo method to take autocorrelations into account.
Markov chains, one of the most convenient and useful of stationary
stochastic processes, are introduced and their properties
are discussed in detail in Sec.~\ref{sec:markov}.  This subsection 
is rather technical,
containing a host of definitions, much mathematics, and many proofs
of the properties of Markov chains, culminating in the fundamental 
limit theorem for irreducible Markov chains.
The Metropolis-Hastings method of constructing a Markov chain 
appropriate to the path integral to be evaluated is then described.

Monte Carlo evaluations of the path integrals needed for correlation
functions in a one-dimensional simple harmonic oscillator are presented 
in Sec.~\ref{sec:sho} as a first simple example, with particular 
attention paid to autocorrelations.  Next, Sec.~\ref{sec:phi4} is
dedicated to Monte Carlo calculations in one of the simplest quantum
field theories: a real scalar field in two spatial dimensions (three 
space-time dimensions).  The theory is first formulated on a space-time 
lattice, then a simple Metropolis updating scheme is described.  The 
Metropolis method is seen to be plagued by strong autocorrelations.
An action-preserving (microcanonical) updating method is then described,
and its effectiveness in reducing autocorrelations is demonstrated.
Monte Carlo estimates in the free scalar field theory are compared
with exactly known results, then a $\phi^4$ interaction term is
included in the action.  This section introduces correlated-$\chi^2$
fitting, as well as jackknife and bootstrap error estimates.

There is insufficient time in these six introductory lectures to 
describe lattice QCD in any detail.  Only very brief comments about
lattice QCD are made in Sec.~\ref{sec:qcd} before concluding 
remarks are given in Sec.~\ref{sec:conclude}.   The goal of these 
lectures is 
to provide the beginner with the basic skills needed to start carrying 
out Monte Carlo studies in quantum field theories, as well as to 
present the underlying theoretical foundations of the method. 
References for further reading are given at the end for
those interesting in pursuing studies in lattice QCD.

\section{Path integrals in quantum mechanics}
\label{sec:paths}

\subsection{Correlation functions and imaginary time}
Consider a small particle of mass $m$ constrained to move only along the
$x$-axis.  Its trajectory is described by its $x$ location as a function of
time, which we write as $x(t)$.  A key quantity in the quantum mechanics of 
such a system is the \textit{transition amplitude}
     \beq Z(b,a)\equiv \langle x_b(t_b)\ \vert \ x_a(t_a)\rangle, \eeq
where $Z(b,a)$ is the probability amplitude for a particle to go from point
$x_a$ at time $t_a$ to point $x_b$ at time $t_b$.  Here, we will work in 
the Heisenberg picture in which state vectors $\vert\Psi\rangle$ are 
stationary and operators and their eigenvectors evolve with time 
       \beqs x(t) &=& e^{iHt/\hbar}\ x(0)\ e^{-iHt/\hbar},\\
       \vert x(t)\rangle &=& e^{iHt/\hbar}\ \vert x(0)\rangle. \eeqs
We often will shift the Hamiltonian so the ground state energy is zero:
     \beqs H\ \vert\phi_n(t)\rangle &=&  E_n\ \vert\phi_n(t)\rangle,
      \qquad E_0=0,\\
      \vert\phi_0(t)\rangle&=&\vert\phi_0(0)\rangle\equiv\vert 0\rangle.
      \eeqs
The transition amplitude contains information about all energy levels and 
all wavefunctions, as can be seen from its spectral representation.
Insert a complete and discrete set of Heisenberg-picture eigenstates
  $\vert\phi_n(t)\rangle$ of the Hamiltonian $H$ into the transition amplitude,
     \beq Z(b,a)\equiv \langle x_b(t_b)\ \vert \ x_a(t_a)\rangle
          =  \sum_n\langle x_b(t_b)\ \vert\phi_n(t_b)\rangle
        \langle\phi_n(t_b)\vert \ x_a(t_a)\rangle, \eeq
then use $\vert\phi_n(t)\rangle=e^{iHt/\hbar}\vert\phi_n(0)\rangle
         =e^{iE_nt/\hbar}\vert\phi_n(0)\rangle$ to obtain
     \beq Z(b,a) =  \sum_n e^{-iE_n(t_b-t_a)/\hbar}
         \langle x_b(t_b)\ \vert\phi_n(t_b)\rangle
        \langle\phi_n(t_a)\vert \ x_a(t_a)\rangle. \eeq
Since $\langle x(t)\vert \phi_n(t)\rangle\equiv\varphi_n(x)$
is the wavefunction in coordinate space of the $n$-th stationary
state, one sees how the transition amplitude provides information about
both the stationary state energies and their wavefunctions:
     \beq Z(b,a) =  \sum_n \varphi_n(x_b)\varphi_n^\ast(x_a)
              \ e^{-iE_n(t_b-t_a)/\hbar}.\eeq

Often, one is interested in evaluating the expectation value
of observables in the ground state, or vacuum.  The above transition
amplitude can yield this information by taking
$t_a=-T$ and $t_b=T$ in the limit $T\rightarrow (1-i\epsilon)\infty$:
    \beqs
     \langle \xb{T} \vert \xa{-T}\rangle &=&
     \langle \xb{0}\vert e^{-iHT/\hbar} \ e^{iH(-T)/\hbar}\vert \xa{0}\rangle\\
      &=&\sum_{n=0}^\infty \langle \xb{0}\vert \phi_n(0)\rangle
      \langle\phi_n(0)\vert \xa{0}\rangle\ e^{-2iE_nT/\hbar}\\
     &\rightarrow&  \langle \xb{0}\vert 0\rangle
     \langle 0\vert \xa{0}\rangle,
     \eeqs
which follows from inserting a complete set of energy eigenstates, using
$E_{n+1}\geq E_n,\ E_0=0$, and assuming a nondegenerate vacuum.
This vacuum saturation trick allows the possibility of probing ground 
state (vacuum) properties.  Now apply the limit 
$T\rightarrow(1-i\epsilon)\infty$ to a more complicated amplitude
   \beqs &&
   \langle \xb{T}\vert \xt \vert \xa{-T}\rangle\\ &=&
   \langle \xb{0}\vert e^{-iHT/\hbar}\ \xt \ e^{-iHT/\hbar}\vert 
   \xa{0}\rangle\\
   &=&\sum_{n,\textcolor{purple}{m}} \langle \xb{0}\vert \phin\rangle
   \langle\phin\vert \xt \vert \phim\rangle
   \langle \phim\vert \xa{0}\rangle\\[-5mm] &&
   \qquad\qquad\qquad\qquad\qquad\qquad\times
   e^{-i(E_n+\textcolor{purple}{E_m})T/\hbar}\\
   &\rightarrow&  \langle \xb{0}\vert 0\rangle
   \langle 0\vert \xt \vert 0\rangle
   \langle 0\vert \xa{0}\rangle.
   \eeqs
Hence, the vacuum expectation value of $x(t_2)x(t_1)$ is obtained from
    \beq\langle 0\vert \xt \vert 0\rangle
     = \lim_{T\rightarrow(1-i\epsilon)\infty}
    \frac{
    \langle \xb{T}\vert \xt \vert \xa{-T}\rangle}
    {\langle \xb{T} \vert \xa{-T}\rangle}.
    \eeq
This result generalizes to higher products of the position operator.

A key point to keep in mind is that all observables can be extracted 
from the correlation functions (vacuum expectation values) of the
position operator $x(t)$.  For example, the energies of the stationary
states can be obtained from
    \beqs \langle 0\vert x(t)x(0)\vert 0\rangle &=&
    \langle 0 \vert e^{iHt/\hbar} x(0) e^{-iHt/\hbar} x(0)\vert 0\rangle\\
    &=& \sum_n \langle 0\vert x(0) e^{-iHt/\hbar}\vert\phi_n(0)\rangle
    \langle\phi_n(0)\vert x(0)\vert 0\rangle \\
    &=& \sum_n \vert\langle 0\vert x(0)\vert\phi_n(0)\rangle\vert^2
    e^{-iE_nt/\hbar},
    \eeqs
and similarly for more complicated correlation functions:
    \beqs \langle 0\vert x^2(t)x^2(0)\vert 0\rangle &=&
    \langle 0 \vert e^{iHt/\hbar} x^2(0) e^{-iHt/\hbar} x^2(0)\vert 0\rangle\\
    &=& \sum_n \vert\langle 0\vert x^2(0)\vert\phi_n(0)\rangle\vert^2
    e^{-iE_nt/\hbar}.
    \eeqs
But it is difficult to extract the energies $E_n$ from such oscillatory 
functions.  It would be much easier if we had \alert{decaying} exponentials.
We can get decaying exponentials if we rotate from the \alert{real}
to the \alert{imaginary} axis in time (Wick rotation) 
         $t\rightarrow -i\tau$
    \beqs \langle 0\vert x(t)x(0)\vert 0\rangle
     &=& \sum_n \vert\langle 0\vert x(0)\vert\phi_n(0)\rangle\vert^2
     e^{-E_n\tau/\hbar}\\
     &\stackrel{\tau\rightarrow \infty}{\longrightarrow}& \vert\langle 0\vert 
     x(0)\vert 0\rangle\vert^2
     + \vert \langle 0\vert x(0)\vert \phi_1(0)\rangle\vert^2 e^{-E_1\tau/\hbar}.
     \eeqs
Later, we will see that this imaginary time formalism provides another 
important advantage for Monte Carlo applications.

\subsection{Path integrals}
\begin{wrapfigure}{r}{1.2in}
\vspace{-5mm}
\includegraphics[height=1.6in,bb=14 14 197 275]{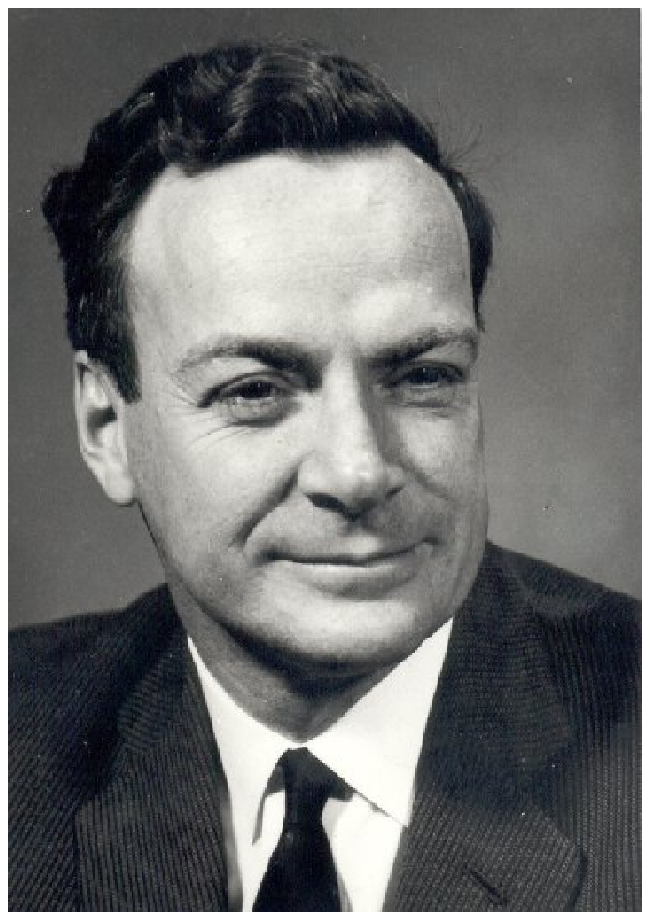}
\end{wrapfigure}
The evaluation of the quantum-mechanical transition amplitude can be 
accomplished in several ways.  In the 1940s, Richard Feynman developed 
an alternative formulation\cite{feynman1948} of quantum mechanics 
as the topic of his Ph.D.\ thesis.  In his formulation, the quantum 
mechanical law of motion expresses the transition amplitude as a
\alert{sum over histories} or a \alert{path integral}:
     \beq Z(b,a) \sim \sum_{\stackrel{\mbox{all paths } \displaystyle 
      x(t)}{\mbox{from $a$ to $b$}}}
      \exp\left( iS[x(t)]/\hbar\right). \eeq
All paths contribute to the probability amplitude, but with different 
\alert{phases} determined by the action $S[x(t)]$.  Evaluating the
transition amplitude in this formalism requires computing a 
multi-dimensional integral, but no differential equations need to be
solved and no large matrices need to be diagonalized.  His approach
also has a conceptual advantage:  the classical limit clearly emerges
when small changes in the path yield changes in the action large 
compared to $\hbar$, causing the phases to cancel out so that only
the path of least action $\delta S=0$ dominates the sum over histories.

\begin{figure}[t]
\begin{center}
\includegraphics[height=3in,bb=0 0 495 430]{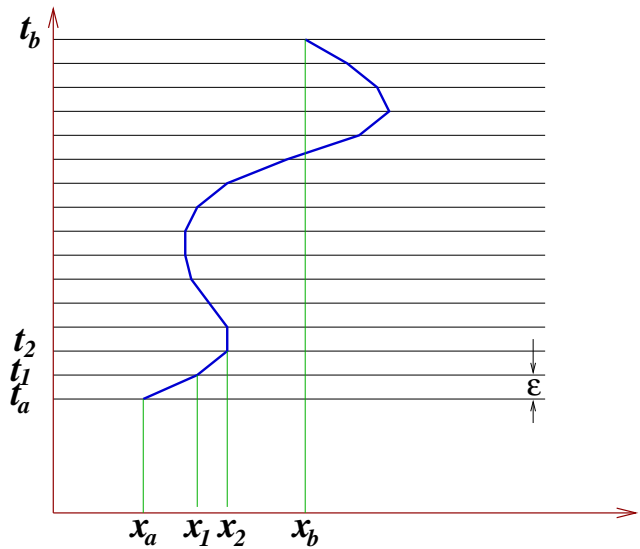}
\end{center}
\caption{A typical path in the path integral for a nonrelativistic
particle moving in one dimension.
\label{fig:path}}
\end{figure}

For a single particle constrained to move only along the $x$-axis, 
the action, being the time integral of the Lagrangian (kinetic minus 
potential energy), is given by
    \beq S = \int\! dt\ L(x,\dot{x}) = \int\! dt\ \Bigl(K-U\Bigr). \eeq
To define the path integral needed to evaluate the transition amplitude,
one first divides time into small steps of width $\varepsilon$, where
        $N\varepsilon=t_b-t_a$ for large integer $N$.
The path integral is defined as
    \beq  Z(b,a)=\lim_{N\rightarrow \infty}\frac{1}{A}
    \int_{-\infty}^\infty
    \frac{dx_1}{A}\frac{dx_2}{A}\cdots\frac{dx_{N-1}}{A}
    \ e^{iS[x(t)]/\hbar}, \eeq
where $A$ is a normalization factor depending on $\varepsilon=(t_b-t_a)/N$ and
chosen so that the path integral is well-defined (see later).
In a nonrelativistic theory, paths cannot double-back in time,
so a typical path looks like the one shown in Fig.~\ref{fig:path}.

\subsection{Relationship to the Schr\"odinger equation}
It is interesting to show how the above expression is equivalent to
the familiar Schr\"odinger equation.  The probability amplitude 
$\psi(x_b,t_b)$ at time $t_b$, assuming an amplitude $\psi(x_a,t_a)$ at 
an earlier time $t_a$, is given by
      \beq \psi(x_b,t_b)=\int Z(b,a)\ \psi(x_a,t_a)\ dx_a. \eeq
Take $t_a=t$ and $t_b=t+\varepsilon$ one time slice away, then
    \beq \psi(x_b,t+\varepsilon)=\frac{1}{A}\int_{-\infty}^\infty
      \exp\left[ \frac{i\varepsilon}{\hbar}
   L\left(\frac{x_b+x_a}{2},\frac{x_b-x_a}{\varepsilon}\right)\right]
     \psi(x_a,t)\ dx_a,\eeq
where in $L(x,\dot{x})$, the speed is $\dot{x}=(x_b-x_a)/\varepsilon$ and 
the mid-point prescription $x\rightarrow (x_b+x_a)/2$ is used.
If the particle is subject to a potential energy, then
$L=\frac{1}{2}m\dot{x}^2\!-\!V(x,t)$,
and it is convenient to write $x_b=x,\ x_a=x+\eta$ so that
    \beq \psi(x,t+\textcolor{red}{\varepsilon})=\frac{1}{A}\int_{-\infty}^\infty
     e^{im\textcolor{darkgreen}{\eta}^2/(2\hbar\textcolor{red}{\varepsilon})}e^{-i\textcolor{red}{\varepsilon} V(x+\textcolor{darkgreen}{\eta}/2,t)/\hbar}
     \psi(x+\textcolor{darkgreen}{\eta},t)\ d\textcolor{darkgreen}{\eta}.\eeq
The rapid oscillation of $e^{im\eta^2/(2\hbar\varepsilon)}$ except when 
$\eta\sim O(\sqrt{\varepsilon})$ leads to the fact that the integral is
dominated by contributions from $\eta$ having values of this order.
Given this, we can expand all expressions to $O(\textcolor{red}{\varepsilon})$ 
and $O(\textcolor{darkgreen}{\eta}^2)$, except 
     $e^{im\eta^2/(2\hbar\textcolor{red}{\varepsilon})}$ 
($\psi$ refers to $\psi(x,t)$), yielding
 \beqs \psi+\textcolor{red}{\varepsilon}\frac{\partial \psi}{\partial t}
     &=&\frac{1}{A}\int_{-\infty}^\infty\!\!
     e^{im\textcolor{darkgreen}{\eta}^2/(2\hbar\textcolor{red}{\varepsilon})}
     \Bigl[1\!-\!\frac{i\textcolor{red}{\varepsilon}}{\hbar}
        V(x,t)\Bigr]
      \Bigl[\psi\!+\!\textcolor{darkgreen}{\eta}\frac{\partial\psi}{\partial x}
        +\frac{\textcolor{darkgreen}{\eta}^2}{2}\frac{\partial^2\psi}{\partial x^2}\Bigl]
          d\textcolor{darkgreen}{\eta},\\
     &=&\frac{1}{A}\int_{-\infty}^\infty\!\!
     e^{im\textcolor{darkgreen}{\eta}^2/(2\hbar\textcolor{red}{\varepsilon})}\Bigl[\psi-\frac{i\textcolor{red}{\varepsilon}}{\hbar}
        V(x,t)\psi +\textcolor{darkgreen}{\eta}\frac{\partial\psi}{\partial x}
        +\frac{\textcolor{darkgreen}{\eta}^2}{2}\frac{\partial^2\psi}{\partial x^2}\Bigl]
          d\textcolor{darkgreen}{\eta}.\eeqs
 Matching the leading terms on both sides determines $A$ (using analytic
           continuation to evaluate the integral):
   \beq 1=\frac{1}{A}\int_{-\infty}^\infty
     e^{im\eta^2/(2\hbar\varepsilon)}d\eta= \frac{1}{A}\left(
      \frac{2\pi i\hbar\varepsilon}{m}\right)^{1/2}\quad\Rightarrow\quad
       A= \left(\frac{2\pi i\hbar\varepsilon}{m}\right)^{1/2}.\eeq
  Given the following integrals,
    \beq \frac{1}{A}\int_{-\infty}^\infty\!\!
     e^{im\eta^2/(2\hbar\varepsilon)}\, \eta\ d\eta =0,\qquad
     \frac{1}{A}\int_{-\infty}^\infty\!\!
     e^{im\eta^2/(2\hbar\varepsilon)}\, \eta^2 d\eta
  =\frac{i\hbar\varepsilon}{m},
     \eeq
 then the $O(\varepsilon)$ part of the equation yields
     \beq -\frac{\hbar}{i}\frac{\partial\psi}{\partial t}
       =-\frac{\hbar^2}{2m}\frac{\partial^2\psi}{\partial x^2}+V(x,t)\psi.\eeq
This is the Schr\"odinger equation!

\subsection{Example: free particle in one dimension}
Now let us explicitly evaluate the path integrals for several simple
examples.  First, consider a free particle of mass $m$ in one dimension.
The Lagrangian of a free particle in one dimension is
  \beq L=\frac{1}{2}m\dot{x}^2,\eeq
so the amplitude for the particle to travel from $x_a$ at time $t_a$
to location $x_b$ at later time $t_b$ is
   \beq\ip{x_b(t_b)}{x_a(t_a)}=\int_a^b{\cal D}x(t) \exp(iS[b,a]/\hbar),
   \eeq
summing over all allowed paths with $x(t_a)=x_a$ and $x(t_b)=x_b$.
The classical path $x_{\rm cl}(t)$ is obtained from $\delta S=0$
and boundary conditions: 
       \beq \ddot{x}_{\rm cl}(t)=0, \qquad x_{\rm cl}(t)=x_a + (x_b-x_a)
           \frac{(t-t_a)}{(t_b-t_a)},\eeq
and the classical action is
       \beq S_{\rm cl}[b,a]=\int_{t_a}^{t_b}\!\!dt\ \textstyle\frac{1}{2}
       m\dot{x}_{\rm cl}^2 = \displaystyle\frac{m(x_b-x_a)^2}{2(t_b-t_a)}.
       \eeq
Write $x(t)=x_{\rm cl}(t)+\chi(t)$ with $\chi(t_a)=\chi(t_b)=0$, 
then
   \beq S[b,a]=S_{\rm cl}[b,a]+\int_{t_a}^{t_b}\!\!dt\ \textstyle\frac{1}{2}
    m\dot{\chi}^2, \eeq
where $S_{\rm cl}[b,a]$ is the classical action.  Notice that there
are no terms linear in $\chi(t)$ since $S_{\rm cl}$ is an extremum.
The transition amplitude becomes
   \beqs Z(b,a)&=&F(T)\exp(iS_{\rm cl}/\hbar),\\
       F(T)&=&\int_0^0\!\!{\cal D}\chi\ \exp\left\{\frac{im}{2\hbar}
    \int_0^T\!dt\ \dot{\chi}^2\right\},
    \eeqs
where $T=t_b-t_a$.
Partition time into discrete steps of length $\varepsilon$,
use the midpoint prescription, and note that $\chi_0=\chi_N=0$:
    \beqs \int_0^0{\cal D}\chi  &=& \frac{1}{A}
    \int_{-\infty}^\infty\!\!\left(\prod_{l=1}^{N-1}\frac{d\chi_l}{A}\right),
    \qquad  A = \left(\frac{2\pi i\hbar\varepsilon}{m}\right)^{1/2},\\
    \int_0^T\!\!\!\!dt\ \dot{\chi}^2 &=&  
    \frac{1}{\varepsilon}\!\sum_{j=0}^{N-1}
       (\chi_{j+1}\!-\!\chi_j)^2,\\
       F(T)&=& \left(\frac{m}{2\pi i\hbar\varepsilon}\right)^{N/2}
      \!\! \int_{-\infty}^\infty\!\!\left(\prod_{l=1}^{N-1}d\chi_l\right)
       \exp\left\{\frac{im}{2\hbar\varepsilon} \chi_j M_{jk}\chi_k 
    \right\}.   \eeqs
A multivariate Gaussian integral remains:
   \beq F(T) = \left(\frac{m}{2\pi i\hbar\varepsilon}\right)^{N/2}
      \int_{-\infty}^\infty\!\!\left(\prod_{l=1}^{N-1}d\chi_l\right)
       \exp\left\{\frac{im}{2\hbar\varepsilon} \chi_j M_{jk}\chi_k 
     \right\}, \eeq
  where $M$ is a symmetric $(N-1)\times(N-1)$ matrix
  \beq  M = \left[\begin{array}{rrrrr}
         2 & -1 & 0 & 0 & \cdots \\ -1 & 2 & -1 & 0 &\cdots \\
       0 & -1 & 2 & -1 & \cdots \\ \vdots & \vdots & \vdots & \vdots
       & \ddots  \end{array} \right].\eeq
  Gaussian integrals of a symmetric matrix $A$ are easily evaluated,
        \beq \int_{-\infty}^\infty\!\! \left(\prod_{i=1}^n d\chi_i\right)
          \exp\Bigl(-\chi_j A_{jk} \chi_k \Bigr)
          = \left(\frac{\pi^n}{\det A}\right)^{1/2},\eeq
so the result for $F(T)$ is
   \beq F(T) = \left(\frac{m}{2\pi i\hbar\varepsilon\det M}\right)^{1/2}.
 \eeq
We now need to compute $\det(M)$.
Consider an $n\times n$ matrix $B_n$ of form
       \beq B_n=\left(\begin{array}{rrrrr} 2b & -b & 0 & 0 &\cdots\\
          -b & 2b & -b & 0 & \cdots\\ 0 & -b 
           & 2b & -b & \cdots\\
          \vdots & \vdots & \vdots & \vdots& \ddots\end{array} \right)_{n,n}.
      \eeq
Notice that
       \beqs \det B_n &=& 2b\det B_{n-1}
        +b\det\left(\begin{array}{c|ccc} -b & -b & 0 &\cdots\\ \hline
          \begin{array}{c}0\\ \vdots\end{array}
        & \multicolumn{3}{c}{B_{n-2}}\end{array} \right),\\&=&2b
           \det B_{n-1}
           -b^2\det B_{n-2}.
       \eeqs
Define $I_n=\det B_n$, then we have the recursion relation
      \beq I_{n+1}=2b I_n - b^2 I_{n-1},\qquad I_{-1}=0,\quad I_0=1,\qquad
         n=0,1,2,\dots.\eeq
Rewrite $I_{n+1}=2b I_n - b^2 I_{n-1},\  I_{-1}=0,\  I_0=1$ as
     \beq  \left(\begin{array}{c} I_{n+1}\\ I_n\end{array}\right)
          = \left(\begin{array}{cc} 2b & -b^2\\ 1 & 0 \end{array}\right)
       \left(\begin{array}{c} I_{n}\\ I_{n-1}\end{array}\right)
      =  \left(\begin{array}{cc} 2b & -b^2\\ 1 & 0 \end{array}\right)^{n}
       \left(\begin{array}{c} I_{1}\\ I_{0}\end{array}\right). \eeq
It is then straightforward to show that
      \beq 
      \left(\begin{array}{cc} 2b & -b^2\\ 1 & 0 
        \end{array}\right)^n = 
       \left(\begin{array}{cc} (n+1)b^n &  -nb^{n+1} \\ 
         nb^{n-1} & -(n-1)b^n\end{array}\right),
       \eeq
   so that
      \beq  \left(\begin{array}{c} I_{n+1}\\ I_n\end{array}\right)
      = \left(\begin{array}{cc} (n+1)b^n &  -nb^{n+1} \\ 
         nb^{n-1} & -(n-1)b^n\end{array}\right)
       \left(\begin{array}{c}  2b\\ 1\end{array}\right),
      \eeq
  and thus, $I_n = \det B_n = (n+1) b^n$.
  Here, $b=1$ and $n=N-1$ so $\det M = N$, and using $N\varepsilon=t_b-t_a$,
  we obtain
     \beq F(t_b,t_a)= \left(\frac{m}{2\pi i\hbar(t_b\!-\!t_a)}\right)^{1/2}. \eeq
  The final result for the transition amplitude for a free particle
in one dimension is
       \beq \ip{x_b(t_b)}{x_a(t_a)}=
         \left(\frac{m}{2\pi i\hbar(t_b\!-\!t_a)}\right)^{1/2}
         \ \exp\left\{ \frac{im(x_b-x_a)^2}{2\hbar(t_b-t_a)}\right\}. \eeq

\subsection{Infinite square well}
As a second example, consider one of the first systems usually studied 
when learning quantum mechanics: the infinite square well.  This is
a particle moving in one dimension under the influence of a potential
given by
     \beq V(x) = \left\{ \begin{array}{ll} 0 & \mbox{for $0<x<L$},\\
        \infty & \mbox{for $x\leq 0$ and $x\geq L$}.\end{array}\right.
      \eeq
The path integral for the transition amplitude in this case is given by
    \beq  Z(b,a)=\lim_{N\rightarrow \infty}\frac{1}{A}
    \int_{0}^L\!\!\!\frac{dx_1}{A}
     \cdots\int_{0}^L\!\!\!\frac{dx_{N-1}}{A}
     \ \exp\left\{\frac{im}{2\varepsilon\hbar}
      \sum_{j=0}^{N-1}(x_{j+1}\!-\!x_j)^2\right\},
     \eeq
where the paths are limited to $0<x<L$.
Gaussian integrals over bounded domains produce error functions ($\mathrm{erf}$),
making direct evaluation in closed form difficult.  A simple 
trick\cite{goodman1981}
to evaluate the path integral in this case is to extend the regions of 
integration to $-\infty<x<\infty$, but subtract off all forbidden paths.
In so doing, we express the square well amplitude as an
infinite sum of free particle amplitudes.

\begin{figure}[t]
\begin{center}
\includegraphics[width=4.25in,bb=0 0 733 433]{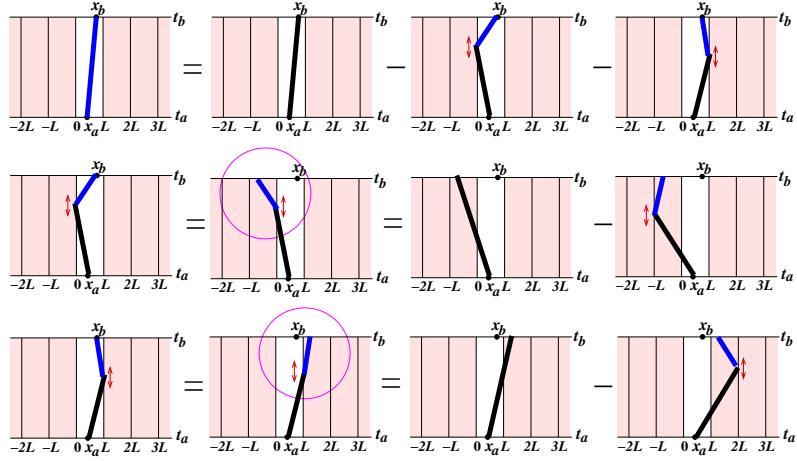}
\end{center}
\caption{Expressing the set of confined paths (solid lines) of the infinite 
square well in terms of sets of free paths (dashed lines).  The minus
signs are set difference operators.  Details
are described in the text.  The circles indicate the applications
of reflections which preserve the free-particle action.
\label{fig:sqwell1}}
\end{figure}

To help describe these path cancellations, let us refer to unbounded
paths which can visit any value of $x$ as \alert{free paths},
and we shall refer to paths which never cross an $x=nL$ boundary,
for integer $n$, as \alert{confined paths}.  The set $S_C$ of
all confined paths from $x_a$ at time $t_a$ to $x_b$ at later time $t_b$
is given by the set $S_F$ of all free paths from $x_a$ to $x_b$, excluding
all free paths which cross the $x=0$ or $x=L$ boundary at least once.
The set of free paths which cross the $x=0$ or $x=L$ boundary at least
once can be partitioned into two non-intersecting subsets: the set $S_{1L}$
of paths whose last boundary crossing occurs at the $x=0$ boundary at time 
$t_1$, and the set $S_{1R}$ of paths whose last boundary crossing occurs 
at the $x=L$ boundary at time $t_1$, for all possible values of $t_1$.
For a particular $t_1$, each subset is
the set of all free paths from $x_a$ at $t_a$ to $x=0$ (or $x=L$)
at time $t_1$ with all \alert{confined} paths from $x=0$ (or $x=L)$ at $t_1$
to $x_b$ at $t_b$.  The set expression $S_C=S_F-S_{1R}-S_{1L}$ is illustrated 
graphically in the top row of Fig.~\ref{fig:sqwell1}.  In this figure, the 
solid lines represent all 
confined paths between the end points of the line (those that do 
\alert{not cross} an $nL$ boundary, for integer $n$), and the dashed lines
represent all free paths between the end points of the line.  Remember 
that there is no doubling back in time.  The minus signs are set
difference operators.

Next, consider a particular path in set $S_{1L}$.  The section of the
path from $x=0$ at $t_1$ to $x_b$ at $t_b$ can be reflected 
$x\rightarrow -x$ without changing the free-particle action.  This is because
the free-particle Lagrangian depends only on the square of the speed
which is left unchanged except at a finite number of points, a set of
measure zero. Hence, as far as
the path integral is concerned, the set $S_{1L}$ can be replaced
by the set $S_{1L}^\prime$ of all free paths from $x_a$ at $t_a$
to $x=0$ at $t_1$ with all confined paths from $x=0$ at $t_1$ to
$-x_b$ at $t_b$, for all possible $t_1$.  With a little thought,
one can see that $S_{1L}^\prime$ is the set $S_{F1L}$ of all free paths
from $x_a$ at $t_a$ to $-x_b$ at $t_b$, excluding the set $S_{2L}$
of all free paths from $x_a$ at $t_a$ to $x=-L$ at $t_1$ with all
confined paths from $x=-L$ at $t_1$ to $-x_b$ at $t_b$.  The set
expression $S_{1L}^\prime=S_{F1L}-S_{2L}$ is 
illustrated in the second row of Fig.~\ref{fig:sqwell1}.

Consider a particular path in the set $S_{1R}$.  The section of
the path from $x=L$ at $t_1$ to $x_b$ at $t_b$ can be reflected
$x\rightarrow 2L-x$ without changing the free-particle action, so $S_{1R}$
can be replaced by the set $S_{1R}^\prime$ of all free paths from 
$x_a$ at $t_a$ to $x=L$ at $t_1$ with all confined paths from $x=L$ 
at $t_1$ to $2L-x_b$ at $t_b$, for all possible $t_1$.  Again,
it is not difficult to see that $S_{1R}^\prime$ is the set $S_{F1R}$ of 
all free paths from $x_a$ at $t_a$ to $2L-x_b$ at $t_b$, excluding the 
set $S_{2R}$ of all free paths from $x_a$ at $t_a$ to $x=2L$ at $t_1$ 
with all confined paths from $x=2L$ at $t_1$ to $2L-x_b$ at $t_b$, as
illustrated in the third row of Fig.~\ref{fig:sqwell1}.

This procedure can be iterated again and again until one obtains the
final result as a sum of free propagators to an infinite number of
mirror points:
   \beqs &&\ip{x_b,t_b}{x_a,t_a}_{\rm conf} 
  = \ip{x_b,t_b}{x_a,t_a}_{\rm free}\\
    &&\qquad- \ip{-x_b,t_b}{x_a,t_a}_{\rm free} 
  - \ip{2L-x_b,t_b}{x_a,t_a}_{\rm free}\\
    &&\qquad+ \ip{-2L+x_b,t_b}{x_a,t_a}_{\rm free} 
   + \ip{2L+x_b,t_b}{x_a,t_a}_{\rm free}
    + \cdots,\\
   &=&\sum_{n=-\infty}^\infty\Bigl\{
      \ip{2nL+x_b,t_b}{x_a,t_a}_{\rm free}
     -\ip{2nL-x_b,t_b}{x_a,t_a}_{\rm free}\Bigr\}.
    \eeqs
Substituting the amplitude for a free particle into this expression yields
       \beqs &&\ip{x_b(t_b)}{x_a(t_a)}_{\rm conf}
    =\left(\frac{m}{2\pi i\hbar (t_b-t_a)}\right)^{1/2}\\
   &&\times \sum_{n=-\infty}^\infty\!\!\left(
      \!  \exp\left\{ \frac{im(2nL\!+\!x_b\!-\!x_a)^2}{2\hbar(t_b-t_a)}\right\} 
  \!-\! \ \exp\left\{ \frac{im(2nL\!-\!x_b\!-\!x_a)^2}{2\hbar(t_b-t_a)}\right\}
    \right). \eeqs
Apply Poisson summation and integrate the Gaussian
      \beqs &&\sum_{n=-\infty}^\infty f(n)
        =\sum_{j=-\infty}^\infty\int_{-\infty}^\infty 
          \! ds\ f(s) e^{2\pi ijs},\\
      && \int_{-\infty}^\infty\!\! ds \exp\Bigl(-i\alpha s^2\pm i\beta s\Bigr)
         = \sqrt{\frac{\pi}{i\alpha}}\exp\left(\frac{i\beta^2}{4\alpha}\right),
   \eeqs
to finally obtain the spectral representation of the transition amplitude
for an infinite square well: 
     \beqs &&\ip{x_b(t_b)}{x_a(t_a)}_{\rm well}=\sum_{n=1}^\infty
          \varphi_n(x_b)\varphi^\ast_n(x_a)e^{-iE_n(t_b-t_a)/\hbar},\\
       &&E_n=\frac{n^2\pi^2\hbar^2}{2mL^2},\qquad \varphi_n(x)=\sqrt{\frac{2}{L}}
           \sin\left(\frac{n\pi x}{L}\right).
     \eeqs
The familiar energy levels and wavefunctions have been obtained
using only path integrals.

\subsection{Free particle in 1D periodic box}

\begin{figure}[t]
\begin{center}
\includegraphics[width=3.0in,bb=0 0 500 391]{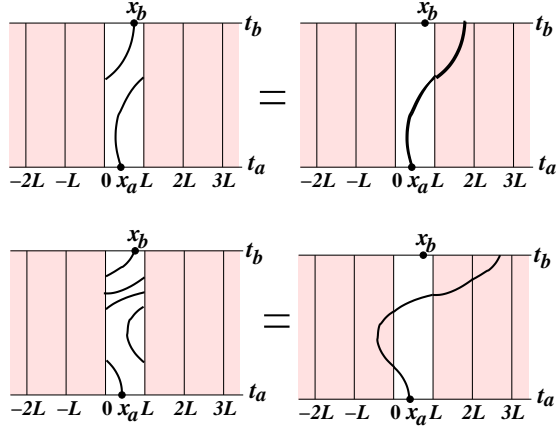}
\end{center}
\caption[capthree]{Each path in the periodic box is equivalent
 to a free path leading to an appropriate mirror point.  The
 equivalent free path is found by horizontally translating sections 
 of the periodic path to form a continuous free path.
\label{fig:periodic}}
\end{figure}

For our third example, consider a particle moving in one dimension 
with periodic boundary conditions at $x=0$ and $x=L$.  Once again,
directly enforcing boundary conditions on the path integrals is difficult,
so it is best to proceed by using a trick similar to that used for the 
infinite square well, that is, to express the set of allowed paths in 
terms of an equivalent set of unrestricted paths.  Each path in the periodic 
box is equivalent to a free path leading to an appropriate mirror point.  
The equivalent free path is found by horizontally translating sections 
of the periodic path to form a continuous free path, as shown in
Fig.~\ref{fig:periodic}.  The resulting amplitude is a sum of
free amplitudes to an infinite number of mirror points:
   \beq \ip{x_b,t_b}{x_a,t_a}_{\rm periodic} 
  = \sum_{n=-\infty}^\infty\ip{x_b+nL,t_b}{x_a,t_a}_{\rm free}.
    \eeq
Substitute the amplitude for a free particle,
       \beq\ip{x_b(t_b)}{x_a(t_a)}
    =\left(\frac{m}{2\pi i\hbar (t_b-t_a)}\right)^{1/2}
    \sum_{n=-\infty}^\infty\!\!
      \exp\left\{ \frac{im(nL\!+\!x_b\!-\!x_a)^2}{2\hbar(t_b-t_a)}\right\} ,
   \eeq
apply Poisson summation, and integrate the Gaussian,
      \beqs &&\sum_{n=-\infty}^\infty f(n)
        =\sum_{j=-\infty}^\infty\int_{-\infty}^\infty 
          \! ds\ f(s) e^{2\pi ijs},\\[-2mm]
      && \int_{-\infty}^\infty\!\! ds \exp\Bigl(-i\alpha s^2\pm i\beta s\Bigr)
         = \sqrt{\frac{\pi}{i\alpha}}\exp\left(\frac{i\beta^2}{4\alpha}\right),\eeqs
to obtain the spectral representation of the transition amplitude:
     \beqs &&\ip{x_b(t_b)}{x_a(t_a)}=\sum_{n=-\infty}^\infty
          \varphi_n(x_b)\varphi^\ast_n(x_a)e^{-iE_n(t_b-t_a)/\hbar},\\
       &&E_n=\frac{p_n^2}{2m},\qquad p_n=\frac{2\pi n\hbar}{L},\qquad
            \varphi_n(x)=\frac{1}{\sqrt{L}} e^{ip_nx/\hbar}.
     \eeqs
The quantization of the momentum, and the familiar energy levels and 
wavefunctions have once again emerged using only path integrals.

\subsection{Simple harmonic oscillator in 1D}
The one-dimensional simple harmonic oscillator is our last example.
The kinetic and potential energies of a simple harmonic oscillator of 
mass $m$ and frequency $\omega$ are given by
    \beq K = \textstyle\frac{1}{2}m \dot{x}^2,\qquad
         U=\frac{1}{2}m\omega^2 x^2, \eeq
so the action is 
    \beq  S[x(t)]=\int_{t_a}^{t_b}\! dt\ \left(
         \textstyle\frac{1}{2}m\dot{x}^2-\frac{1}{2}m\omega^2 x^2\right).
    \eeq
The classical equations of motion are
     \beq \delta S=0 \quad\Rightarrow\quad \ddot{x}_{\rm cl} 
         + \omega^2  x_{\rm cl} = 0, \eeq
and the value of action for the classical path is
     \beq  S_{\rm cl} = \frac{m\omega}{2\sin(\omega T)}
     \Biggl[ (x_a^2+x_b^2)\cos(\omega T)
      - 2x_a x_b\Biggr], \eeq
where $T=t_b-t_a$.
To calculate the amplitude $Z(b,a)=\ip{x_b(t_b)}{x_a(t_a)}_{\rm sho}$,
write the path as a deviation from the classical path:    
    \beq x(t)=x_{\rm cl}(t)+\chi(t),\qquad  \chi(t_a)=\chi(t_b)=0.\eeq
The amplitude can then be written as
   \beqs Z(b,a)&=&F(T)\exp(iS_{\rm cl}/\hbar),\\
       F(T)&=&\int_0^0\!\!{\cal D}\chi\ \exp\left\{\frac{im}{2\hbar}
    \int_0^T\!dt\ (\dot{\chi}^2-\omega^2\chi^2)\right\}.
    \eeqs
Partition time into discrete steps of length $\varepsilon$
and use a midpoint prescription:
    \beqs \int_0^0{\cal D}\chi\!\! &=& \frac{1}{A}
    \int_{-\infty}^\infty\!\!\left(\prod_{l=1}^{N-1}\frac{d\chi_l}{A}\right),
    \qquad  A = \left(\frac{2\pi i\hbar\varepsilon}{m}\right)^{1/2},\\
    \int_0^T\!\!dt(\dot{\chi}^2\!-\!\omega^2\chi^2)&=&   
    \frac{1}{\varepsilon}\!\sum_{j=0}^{N-1}\left[
       (\chi_{j+1}\!-\!\chi_j)^2\!-\!\frac{\varepsilon^2\omega^2}{4}
       (\chi_{j+1}\!+\!\chi_j)^2\right],\\
       F(T)&=& \left(\frac{m}{2\pi i\hbar\varepsilon}\right)^{N/2}
      \!\!\!\int_{-\infty}^\infty\!\!\left(\prod_{l=1}^{N-1}d\chi_l\right)
       \exp\left\{\frac{im}{2\hbar\varepsilon} \chi_j M_{jk}\chi_k 
    \right\}.   \eeqs
A multivariate Gaussian integral remains:
   \beq F(T) = \left(\frac{m}{2\pi i\hbar\varepsilon}\right)^{N/2}
      \int_{-\infty}^\infty\!\!\left(\prod_{l=1}^{N-1}d\chi_l\right)
       \exp\left\{\frac{im}{2\hbar\varepsilon} \chi_j M_{jk}\chi_k 
     \right\} ,\eeq
  where $M$ is a symmetric $(N-1)\times(N-1)$ matrix
  \beq  M = \left[\begin{array}{rrrrr}
         2 & -1 & 0 & 0 & \cdots \\ -1 & 2 & -1 & 0 &\cdots \\
       0 & -1 & 2 & -1 & \cdots \\ \vdots & \vdots & \vdots & \vdots
       & \ddots  \end{array} \right]
        -\frac{\varepsilon^2\omega^2}{4}
        \left[\begin{array}{rrrrr}
         2 & 1 & 0 & 0 & \cdots \\ 1 & 2 & 1 & 0 &\cdots \\
       0 & 1 & 2 & 1 & \cdots \\ \vdots & \vdots & \vdots & \vdots
       & \ddots  \end{array} \right]. \eeq
Such Gaussian integrals are easily evaluated:
   \beq F(T) = \left(\frac{m}{2\pi i\hbar\varepsilon\det M}\right)^{1/2}.
   \eeq
Now we must compute $\det M$.  Consider $\det(B_n)$, where the $n\times n$ 
matrix $B_n$ has the form
    \beq B_n=\left(\begin{array}{ccccc} a & b & 0 & 0 &\cdots\\
       b & a & b & 0 & \cdots\\ 0 & b & a & b & \cdots\\
       \vdots & \vdots & \vdots & \vdots& \ddots\end{array} \right)_{n,n}
     .\eeq
$B_n$ matches $M$ for $n=N-1$, $a=2(1-\epsilon^2\omega^2/4)$,
    and    $b=-(1+\epsilon^2\omega^2/4)$.
Notice that
        \beqs \det B_n &=& a\det B_{n-1}
         -b\det\left(\begin{array}{c|ccc} b & b & 0 &\cdots\\ \hline
           \begin{array}{c}0\\ \vdots\end{array}
           & \multicolumn{3}{c}{B_{n-2}}\end{array} \right),\\
            &=&a\det B_{n-1}
            -b^2\det B_{n-2}.
        \eeqs
Define $I_n=\det B_n$ to obtain the recursion relation
      \beq I_{n+1}=a I_n - b^2 I_{n-1},\qquad I_{-1}=0,\quad I_0=1,\qquad
          n=0,1,2,\dots.\eeq
Rewrite this recursion relation as
       \beq  \left(\begin{array}{c} I_{n+1}\\ I_n\end{array}\right)
           = \left(\begin{array}{cc} a & -b^2\\ 1 & 0 \end{array}\right)
        \left(\begin{array}{c} I_{n}\\ I_{n-1}\end{array}\right)
       =  \left(\begin{array}{cc} a & -b^2\\ 1 & 0 \end{array}\right)^{n}
        \left(\begin{array}{c} I_{1}\\ I_{0}\end{array}\right), \eeq
 diagonalize as follows,
      \beqs
      &&\left(\begin{array}{cc} a & -b^2\\ 1 & 0 \end{array}\right)
      = {\cal S}\ \left(\begin{array}{cc}\lambda_+ & 0 \\ 0 & 
      \lambda_-\end{array}\right)\ {\cal S}^{-1}, \\
      && \lambda_\pm = \frac{1}{2}\Bigl(a\pm\sqrt{a^2-4b^2}\Bigr),\\
      && {\cal S}=\left(\begin{array}{cc}\lambda_+ & \lambda_-\\1&1
       \end{array}\right),\quad
       {\cal S}^{-1}=\frac{1}{\lambda_+-\lambda_-}
       \left(\begin{array}{cc} 1 & -\lambda_-\\-1 & \lambda_+
       \end{array}\right),
       \eeqs
then we have
      \beq  \left(\begin{array}{c} I_{n+1}\\ I_n\end{array}\right)
      =  {\cal S}\left(\begin{array}{cc} \lambda_+^n & 0\\ 0 & \lambda_-^n
         \end{array}\right) {\cal S}^{-1}
       \left(\begin{array}{c} a\\ 1\end{array}\right).
      \eeq
Thus,
     \beq I_n =\det B_n= \frac{\lambda_+^{n+1}-\lambda_-^{n+1}}{
       \lambda_+-\lambda_-},  \qquad (\lambda_+\neq\lambda_-).
     \eeq
Using $\lambda_\pm = 1\pm i\omega\epsilon + O(\varepsilon^2)$
         yields
        \beqs  \lim_{\genfrac{}{}{0pt}{}{\varepsilon\rightarrow 0}{N\rightarrow
        \infty}}\varepsilon \det M
         &=& \lim_{\genfrac{}{}{0pt}{}{\varepsilon\rightarrow 0}{
             N\rightarrow  \infty}}
        \varepsilon 
        \frac{1}{2i\omega\varepsilon}\Bigl((1+i\omega\varepsilon)^N
           -(1-i\omega\varepsilon)^N \Bigr),\\
         &=& \lim_{\genfrac{}{}{0pt}{}{\varepsilon\rightarrow 0}{
                N\rightarrow  \infty}}
         \frac{1}{2i\omega}\left(\left(1+\frac{i\omega T}{N}\right)^N
           -\left(1-\frac{i\omega T}{N}\right)^N \right),\\
         &=& \frac{1}{2i\omega}\left(e^{i\omega T}-e^{-i\omega T}\right)
         = \frac{\sin\omega T}{\omega}.
      \eeqs
The final result for the path integral is
     \beq  \ip{x_b(t_b)}{x_a(t_a)}_{\rm sho}
         =\left(\frac{m\omega}{2\pi i\hbar\sin\bigl(\omega(t_b\!-\!t_a)\bigr)}
      \right)^{1/2}\ \exp\Bigl\{iS_{\rm cl}/\hbar\Bigr\}.
     \eeq

\begin{figure}[t]
\begin{center}
\includegraphics[width=4.5in,bb=49 46 603 413]{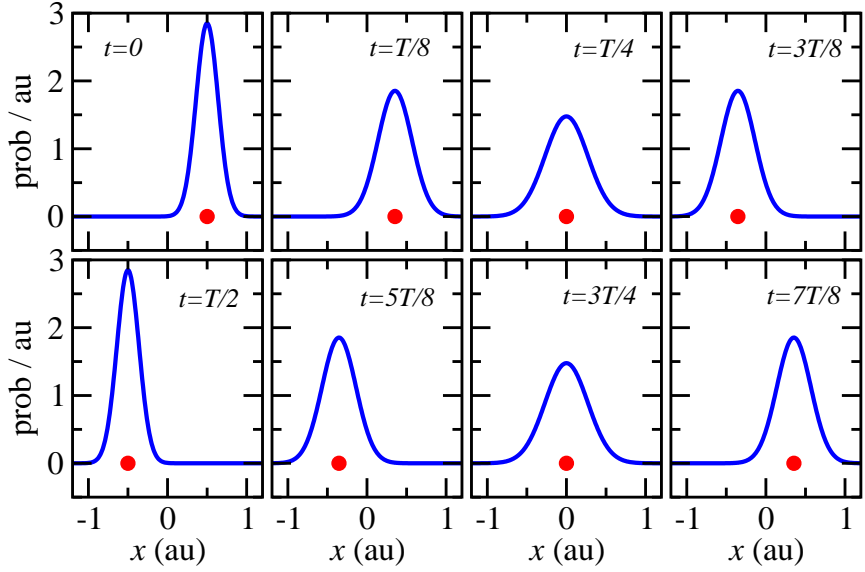}
\end{center}
\caption[capsho]{Time evolution of a Gaussian wave packet
for a simple harmonic oscillator of mass  
 $m=1\mathrm{g/mol}= 1.66\times 10^{-27} \mathrm{kg}$
 and frequency $\omega=3\times 10^{14} \mathrm{radians/sec}$.
 The initial wave packet at $t=0$ is centered at 0.5 au
 and has initial width $\sigma=$0.14 au.  Note that
 1 au (atomic unit) = 0.529 angstrom.  The dot indicates
 the location of a classical oscillator of the
 same mass and frequency.  The probabilities are shown
 for times $t=nT/8$ for $n=0,1,2,\dots,7$, where $T$ is the
 period.
\label{fig:sho_packet}}
\end{figure}

Consider the temporal evolution of a Gaussian wave packet for this
system.  If the probability distribution corresponding to the initial 
wave packet at time $t_a=0$ is a Gaussian:
    \beq \vert\phi(x_a,t_a)\vert^2=\frac{1}{\sigma \sqrt{2\pi}}
     \exp\left(-\frac{(x_a-\bar{x})^2}{2\sigma^2}\right),\eeq
then the probability amplitude at a later time $t_b$ is
    \beqs \phi(x_b,t_b)&=& \int_{-\infty}^\infty\!\!\!\!\!dx_a\ Z(b,a)
      \ \phi(x_a,0),\\
     &=& \left(\frac{-im\omega(2\pi)^{-3/2}}{\hbar\sigma
     \sin(\omega t_b)}\right)^{1/2}\!\!\!\!
     \int_{-\infty}^\infty
     \!\!\!\!\! dx_a\ e^{iS_{\rm cl}/\hbar}
    \ e^{-(x_a-\bar{x})^2/(4\sigma^2)},
     \eeqs
so the probability distribution remains a Gaussian but with a
varying width $s(t)$:
    \beq \vert\phi(x_b,t_b)\vert^2=\frac{1}{s(t_b) \sqrt{2\pi}}
    \exp\left(-\frac{(x_b-\bar{x}\cos(\omega t_b))^2}{2s^2(t_b)}\right),\eeq
where the width is given by
    \beq s(t_b) = \sigma\left\{ \cos^2(\omega t_b)+\frac{
    \hbar^2}{4m^2\omega^2\sigma^4} \sin^2(\omega t_b)
    \right\}^{1/2}.\eeq
The time evolution of such a Gaussian wave packet for a 
simple harmonic oscillator is shown in Fig.~\ref{fig:sho_packet}.
Note that this evolution was completely calculated above using path 
integrals; the Schr\"odinger equation was not used.

\subsection{Correlation functions and observables}
We have so far seen that path integrals give us simple transition amplitudes,
such as
   \beq \langle x_b(t_b)\vert x_a(t_a)\rangle
     = \int_a^b\!\! {\cal D}x\ \exp\left\{\frac{i}{\hbar}
    \int_{t_a}^{t_b}\!\! dt\ L(x,\dot{x})\right\},\eeq
but this important result generalizes to more complicated amplitudes:
   \beqs
    && \langle x_b(t_b)\vert \ x(t_2)\ x(t_1)\ \vert x_a(t_a)\rangle,\\
    &=& \int_a^b\!\! {\cal D}x\ x(t_2)x(t_1)\ \exp\left\{\frac{i}{\hbar}
    \int_{t_a}^{t_b}\!\! dt\ L(x,\dot{x})\right\},\eeqs
for $t_a < t_1 < t_2 < t_b $.
In the imaginary time formalism, paths contribute to the sum over histories 
with real exponential weights (not phases):
  \beqs && \langle x_b(\tau_b)\vert \ x(\tau_2)\ x(\tau_1)\ \vert 
    x_a(\tau_a)\rangle\\
     &=& \int_a^b\!\! {\cal D}x\ x(\tau_2)x(\tau_1)\ \exp\left\{-\frac{1}{\hbar}
    \int_{\tau_a}^{\tau_b}\!\! d\tau\ L(x,\dot{x})\right\}.\eeqs
Now the classical path gets the highest weighting. Note that weights are all 
\alert{real} and \alert{positive} since the action is real.  This fact will 
be crucial for the Monte Carlo method.

Another important fact is that correlation functions (vacuum expectation values) 
can be obtained from ratios of path integrals.  For example, a two-point
function can be obtained from
  \beqs  &&\langle 0\vert \xt \vert 0\rangle
    = \lim_{T\rightarrow \infty}
   \frac{
   \langle \xb{T}\vert \xt \vert \xa{-T}\rangle}
   {\langle \xb{T} \vert \xa{-T}\rangle},\\
   &=& \frac{
    \displaystyle\int_a^b {\cal D}x\  \xt 
   \exp\left\{-\frac{1}{\hbar}\int_{-\infty}^\infty\! d\tau 
   L(x,\dot{x})\right\}}
   { \displaystyle\int_a^b {\cal D}x\  
   \exp\left\{-\frac{1}{\hbar}\int_{-\infty}^\infty\! d\tau 
   L(x,\dot{x})\right\}}, \eeqs
and more complicated correlation functions can similarly be obtained.
In fact, \alert{any} correlation function can be computed using
path integrals.

For example, consider the simple harmonic oscillator.
Evaluating path integrals as before, the following 
   correlation functions can be obtained:
   \beqs \langle 0\vert x(\tau_1)\vert 0\rangle &=& 0,\\
   \langle 0\vert x(\tau_2) x(\tau_1)\vert 0\rangle &=&
   \frac{\hbar}{2m\omega}e^{-\omega(\tau_2\!-\!\tau_1)},\\
   \langle 0 \vert x(\tau_4) x(\tau_3) x(\tau_2) x(\tau_1)\vert 0\rangle &=&
   \left(\frac{\hbar}{2m\omega}\right)^2e^{-\omega(\tau_4\!-\!\tau_1)}
   \Bigl[ e^{-\omega(\tau_2\!-\!\tau_3)}
  +2e^{-\omega(\tau_3\!-\!\tau_2)}\Bigr],\eeqs
where $\tau_1 \leq \tau_2 \leq \tau_3\leq \tau_4$.
Comparison with the spectral representations tells us
  \beq \langle 0\vert x(\tau) x(0)\vert 0\rangle =
   \frac{\hbar}{2m\omega}e^{-\omega\tau} 
    \ \Rightarrow 
    \ E_1-E_0=\hbar \omega, \quad \vert \langle 1
   \vert x(0)\vert 0\rangle\vert^2 = \frac{\hbar}{2m\omega}.\eeq 
As another example in the SHO case, consider exciting the vacuum with 
the $x(\tau)^2$ operator:
  \beq  \langle 0\vert x^2(\tau)x^2(0)\vert 0\rangle
   = \left(\frac{\hbar}{2m\omega}\right)^2\biggl(1+2e^{-2\omega
   \tau}\biggr).\eeq
Compare with the spectral representation at large time separations,
  \beqs \lim_{\tau\rightarrow\infty}
  \langle 0\vert x^2(\tau)x^2(0)\vert 0\rangle &=&
  \vert \langle 0\vert x^2(0)\vert 0\rangle\vert^2 
  +  \vert \langle 2\vert x^2(0)\vert 0\rangle\vert^2\ e^{-(E_2-E_0)t/\hbar}
  +\dots,\\
  &=& \left(\frac{\hbar}{2m\omega}\right)^2\biggl(1+2e^{-2\omega
  \tau}\biggr),\eeqs
to arrive at the following interpretation:
  \beq E_2-E_0=2\hbar\omega, \eeq 
  \beq \vert \langle 0\vert x^2(0)\vert 0\rangle\vert^2
 =\left(\frac{\hbar}{2m\omega}\right)^2, \qquad
 \vert \langle 2\vert x^2(0)\vert 0\rangle\vert^2
 =2\left(\frac{\hbar}{2m\omega}\right)^2.\eeq
One last example in the SHO:
to determine the expectation value of $x(0)^2$ in first-excited state,
evaluate
  \beq \langle 0\vert x(\tau)\ x^2({\textstyle\frac{1}{2}}\tau)\ x(0)\vert 0\rangle
   = 3\left(\frac{\hbar}{2m\omega}\right)^2\ e^{-\omega \tau},\eeq
and compare with its spectral interpretation at large times:
  \beqs &&\lim_{\tau\rightarrow\infty}
   \langle 0\vert x(\tau) x^2(\textstyle\frac{1}{2}\tau) x(0)\vert 0\rangle\\
   && \qquad = \vert\langle 0\vert x(0)\vert 1\rangle\vert^2
   \langle 1\vert x^2(0)\vert 1\rangle
   \ e^{-(E_1-E_0)\tau/\hbar}+\cdots, \eeqs
 since $\langle 0\vert x(0)\vert 0\rangle = \langle 0\vert x(\tau)\vert 
   0\rangle =0$.
 By inspection and using previously derived results, one concludes that
  \beq \langle 1\vert x^2(0)\vert 1\rangle = \frac{3\hbar}{2m\omega}.\eeq

\begin{wrapfigure}{r}{0.9in}
\vspace{0mm}
\includegraphics[height=1.25in,bb=14 14 136 195]{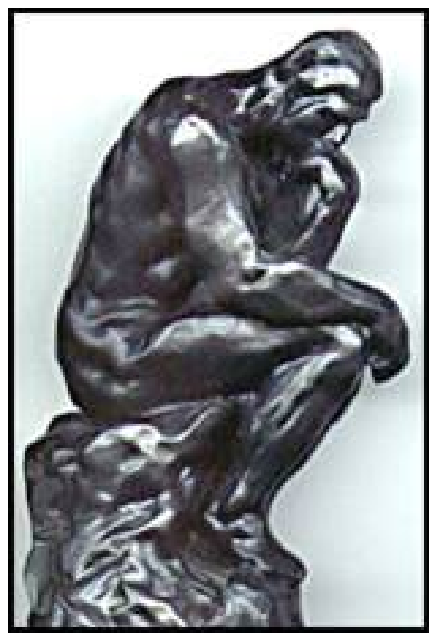}
\end{wrapfigure}

Now pause for reflection.
We have seen that observables in quantum mechanics can be extracted from
correlation functions (vacuum expectation values), and that
the imaginary time formalism is a great trick for assisting in 
such extractions.  Correlation functions can be computed via ratios of
path integrals
  \beqs &&\langle 0\vert \xt \vert 0\rangle\\
   &=& \frac{
   \displaystyle\int_a^b {\cal D}x\  \xt 
   \exp\left\{-\frac{1}{\hbar}\int_{-\infty}^\infty\! d\tau 
   L(x,\dot{x})\right\}}
   { \displaystyle\int_a^b {\cal D}x\  
   \exp\left\{-\frac{1}{\hbar}\int_{-\infty}^\infty\! d\tau 
   L(x,\dot{x})\right\}}.
  \eeqs

\section{Monte Carlo integration and Markov chains}
\label{sec:montecarlo}

In rare situations, the path integrals in transition amplitudes
can be computed exactly, such as the simple harmonic oscillator
and a free particle.  Sometimes the action can be written  $S=S_0+gS_I$,
where $S_0$ describes the free motion of the particles and
$S_I$ describes the interactions of the particles, but the coupling $g$
is small.  Typically, the path integrals using $S_0$ are Gaussian and can 
be exactly computed, and the interactions can be taken into account
using perturbation theory as an expansion in $g$.  However, if the
interactions are \alert{not weak}, such as in quantum chromodynamics,
one must somehow numerically evaluate the needed path integrals with
powerful computers.

The trapezoidal rule and Simpson's rule are not feasible for
integrals of very large dimension.  These methods require far too many
function evaluations.  One of the most productive ways of proceeding
is to start \alert{gambling}!  The Monte Carlo method comes to our
rescue.  The basic theorem of Monte Carlo integration is
    \beqs && \int_{\cal V} f(\vec{x})\ d^Dx
    \approx {\cal V}\langle f\rangle \pm {\cal V}\sqrt{\frac{\langle f^2\rangle
    -\langle f\rangle^2}{N}},\\
    &&\langle f\rangle \equiv \frac{1}{N}\sum_{i=1}^N
     f(\vec{x}_i),\qquad\qquad
    \langle f^2\rangle \equiv \frac{1}{N}\sum_{i=1}^N
    f(\vec{x}_i)^2, \eeqs
where the $N$ points $\vec{x}_1,\dots,\vec{x}_N$ are chosen \alert{independently}
and \alert{randomly} with a uniform probability distribution throughout
the $D$-dimensional volume ${\cal V}$.  The method is justified by the law of 
large numbers and the central limit theorem.  In the limit 
$N\rightarrow\infty$, the above Monte Carlo estimate tends to a normal 
distribution and the uncertainty tends to a standard deviation.

The above method sounds too good to be true.  Although the above
method should work in principle, it is impractical for
evaluating quantum mechanical path integrals, unless suitably modified
to incorporate importance sampling.  Before discussing this, a closer
look at the simple Monte Carlo method is warranted, and this can be
facilitated with a quick review of probability theory.

\subsection{Quick review of probabilities}
\label{sec:prob}

Consider an experiment whose outcome depends on chance.
Represent an outcome by $X$ called a \alert{random variable},
and the \alert{sample space} $\Omega$ of the experiment is the set of all 
possible outcomes. $X$ is called \alert{discrete} if $\Omega$ is finite 
or countably infinite, and \alert{continuous} otherwise.
The probability distribution for discrete $X$ is a real-valued function $p_X$
on the domain $\Omega$ satisfying $p_X(x)\geq 0$ for all $x\in\Omega$ and
$\sum_{x\in\Omega}p_X(x)=1$. For any subset $E$ of $\Omega$, the
\bfalert{probability} of $E$ is $P(E)=\sum_{x\in E}p_X(x)$.
A sequence of random variables $X_1,X_2,\dots,X_N$ that are
mutually independent and have the same probability distribution is called 
an \alert{independent trials process}.

For a continuous real-valued $X$, the real-valued function $p_X$ is a 
probability \alert{density} and the probability of an outcome between
real values $a$ and $b$ is $P(a\leq X\leq b)=\int_a^bp_X(s)ds$.
The \alert{cumulative} distribution is 
$F_X(x)=P(X\leq x)=\int_{-\infty}^x p_X(s)ds$.
A common probability density is the \alert{normal} 
distribution $p_X(x)=\dd\frac{1}{\sqrt{2\pi}\sigma}
        e^{-(x-\mu)^2/(2\sigma^2)}$.

The \bfalert{expected value} of $X$ is 
    \beq E(X)=\sum_{x\in\Omega} x\ p_X(x)
    \qquad\left(=\int_{-\infty}^\infty s\ p_X(s)ds\right). \eeq
The expected value satisfies $E(X+Y)=E(X)+E(Y)$ and $E(cX)=cE(X)$,
and for \underline{in}dependent random variables $X,Y$ one has
          $E(XY)=E(X)E(Y)$.
One can show that $E(X)$ is the average of outcomes if repeated many 
times.  For a continuous real-valued function $f$, one can also show that
        \beq E(f(X))=\sum_{x\in\Omega}f(x)\ p_X(x)
          \qquad\left(=\int_{-\infty}^\infty f(s)\ p_X(s)\,ds\right) .
        \eeq
To see this,  group together terms in $\sum_x\! f(x)p_X(x)$ 
having same $f(x)$ value.  Denote the set of different $f(x)$ 
values by ${\cal F}$, and the subset of $\Omega$ leading to same value 
of $f(x)$ by $\Omega_{f(x)}$, then
     \beqs&& \sum_{x\in\Omega}f(x)p_X(x)=\sum_{y\in{\cal F}}
         \sum_{x\in\Omega_{f(x)}}\!\! f(x)p_X(x)=\sum_{y\in{\cal F}} y
         \,(\!\!\sum_{x\in\Omega_{f(x)}}\!\!\!p_X(x))\\
         &&=\sum_{y\in{\cal F}}yp(y)
          =E(f(x)).
          \eeqs

The \bfalert{variance} of $X$ is $V(X)=E(\ (X-E(X))^2\ )$,
and the \bfalert{standard deviation} of $X$ is $\sigma(X)=\sqrt{V(X)}$.
The variance satisfies $V(cX)=c^2V(X)$ and $V(X+c)=V(X)$,
and for \underline{in}dependent random variables $X,Y$, one has
 $V(X+Y)=V(X)+V(Y)$. Let $X_1,\dots,X_N$ be an independent trials process 
with $E(X_j)=\mu$ and $V(X_j)=\sigma^2$, and define
  $A_N=(X_1+X_2+\dots +X_N)/N$, then one can easily show 
      \beq  E(A_N)=\mu,\qquad V(A_N)=\sigma^2/N. \eeq

An important theorem in probability and statistics is known
as \bfalert{Chebyshev's inequality}:  Let $X$ be a random variable (discrete
or continuous)
with $E(X)=\mu$ and let $\epsilon>0$ be any positive real number, then
  \beq P(\vert X-\mu\vert\geq \epsilon)\leq \frac{V(X)}{\epsilon^2}.\eeq
\begin{proof}
 Let $p_X(x)$ denote the probability distribution of $X$, then the
 probability that $X$ differs from $\mu$ by at least $\epsilon$ is
  \beq \qquad \qquad P(\vert X-\mu\vert\geq\epsilon)=\sum_{\vert x-\mu\vert\geq
         \epsilon} p_X(x).\eeq
 Considering the ranges of summation and that we have positive summands,
       \beq V(X)=\sum_x(x-\mu)^2p_X(x)\geq \sum_{\vert x-\mu\vert\geq\epsilon}
               (x-\mu)^2p_X(x)\geq \sum_{\vert x-\mu\vert\geq\epsilon}
                \epsilon^2 p_X(x), \eeq
 but the rightmost expression is
         \beq \epsilon^2\sum_{\vert x-\mu\vert\geq\epsilon}
                 p_X(x)= \epsilon^2 P(\vert X-\mu\vert\geq\epsilon).\eeq
 Thus, we have shown
            $V(x)\geq \epsilon^2 P(\vert X-\mu\vert\geq\epsilon)$.
\end{proof}

An important consequence of Chebyshev's inequality is the 
\bfalert{weak law of large numbers}: Let $X_1,X_2,\dots,X_N$ be an
independent trials process with $E(X_j)=\mu$ and $V(X_j)=\sigma^2$, 
where $\mu,\sigma$ are finite, and let $A_N=(X_1+X_2+\dots +X_N)/N$.  
Then for any $\epsilon>0$,
      \beq \lim_{N\rightarrow\infty}
           P(\vert A_N-\mu\vert\geq\epsilon)= 0,\qquad
           \lim_{N\rightarrow\infty}
           P(\vert A_N-\mu\vert < \epsilon)= 1.\eeq
\begin{proof}
We previously stated that $E(A_N)=\mu$ and $V(A_N)=\sigma^2/N$,
and from the Chebyshev inequality,
  \beq P(\vert A_N-\mu\vert\geq \epsilon)\leq \frac{V(A_N)}{\epsilon^2}
   =\frac{\sigma^2}{N\epsilon^2}
   \stackrel{N\rightarrow\infty}{\longrightarrow} 0.\eeq
\end{proof}
\noindent
This is also known as the \alert{law of averages}, and applies to 
continuous random variables as well.
 
A different version of the above law is known as 
the \bfalert{strong law of large numbers}: Let $X_1,X_2,\dots,X_N$ be an
 independent trials process with $E(X_j)=\mu$ and $V(X_j^2)=\sigma^2$, 
 where $\mu,\sigma$ are finite, then 
 \beq  P\left(\lim_{N\rightarrow\infty}
            (X_1+X_2+\dots +X_N)/N = \mu\right)= 1.\eeq
\begin{proof}
We shall assume that the random variables $X_j$
have a finite fourth moment $E(X_j^4)=K<\infty$.
The finiteness of $E(X_j^4)$ is not needed, but simplifies the proof.
For a proof without this assumption, see Ref.~\refcite{etemadi}.

Define $Y_j=X_j-\mu$ so $E(Y_j)=0$.  Define $B=E(Y_i^2)<\infty$ and
$C=E(Y_j^4)<\infty$, then define $A_N=(Y_1+Y_2+\dots +Y_N)/N$.  Consider
\[ N^4 E(A_N^4) = E(\, (Y_1+Y_2+\cdots +Y_N)^4\, ).\]
Expanding $A_N^4$ yields terms of the form
$Y_i^4$, $Y_i^3 Y_j$, $Y_i^2 Y_j^2$, $Y_i^2 Y_j Y_k$, and $Y_iY_jY_kY_l$, where
$i,j,k,l$ are all different.
Given that $E(Y_j)=0$ and all $Y_j$ are independent, then
\beqs E(Y_i^3 Y_j) &=& E(Y_i^3)E(Y_j)=0, \qquad \mbox{($i,j,k,l$ all different)}\\
 E(Y_i^2 Y_j Y_k) &=& E(Y_i^2) E(Y_j) E(Y_k)=0,\\
 E(Y_iY_jY_kY_l) &=& E(Y_i)E(Y_j)E(Y_k)E(Y_l)=0,\\
 E(Y_i^2Y_j^2) &=& E(Y_i^2)E(Y_j^2) = B^2.
\eeqs
Since the random variables are identically distributed, then
$E(Y_i^4)$ and $E(Y_i^2Y_j^2)$ are independent of $i,j$, so we have
             \beq N^4E(A_N^4)=NE(Y_j^4)
          +6\textstyle\genfrac{(}{)}{0pt}{}{N}{2}E(Y_i^2Y_j^2)
               =NC +3N(N-1)B^2.\eeq
Since $0\leq V(Y_j^2)=E((Y_j^2-E(Y_j^2)^2)=E(Y_j^4)-E(Y_j^2)^2$ then
            $B^2=E(Y_j^2)^2\leq E(Y_j^4)=C$
so $E(A_N^4)\leq C/N^3+3C/N^2$, which means
           \beq  E\left(\sum_{N=1}^\infty A_N^4\right)=\sum_{N=1}^\infty E(A_N^4)
            \leq\sum_{N=1}^\infty\left(\frac{C}{N^3}+\frac{3C}{N^2}\right)
            <\infty.\eeq
This implies $\sum_{N=1}^\infty A_N^4 < \infty$ with unit probability,
    and convergence of this series 
   implies $\lim_{N\rightarrow\infty}A_N^4=0$, which means that
               $\lim_{N\rightarrow\infty}A_N=0$.
\end{proof}
\noindent
This proves that $E(X)$ is the average of outcomes for many repetitions.

Uncertainties in Monte Carlo estimates depend upon 
the celebrated \bfalert{central limit theorem}: Let $X_1,X_2,\dots,X_N$ be independent 
         random variables with common distribution having $E(X_j)=\mu$ and 
         $V(X_j)=\sigma^2$, where $\mu,\sigma$
         are finite, and let $A_N=(X_1+X_2+\dots +X_N)/N$.  Then for
         $a<b$,
      \beq \lim_{N\rightarrow\infty}
           P\left(\frac{a\sigma}{\sqrt{N}}< (A_N-\mu)<\frac{b\sigma}{\sqrt{N}}
           \right)= \frac{1}{\sqrt{2\pi}}
           \int_a^b e^{-x^2/2}dx. \eeq
Alternatively, the distribution of $(X_1+\dots +X_N-N\mu)
         /(\sigma\sqrt{N})$ tends to the standard normal (zero mean, unit variance).
\begin{proof}
Define $S_N = N(A_N-\mu)/(\sigma\sqrt{N})$, and if $t$ is a real-valued
parameter, then
\beqs
  E\left(e^{tS_N}\right)&=&E\left(e^{t(X_1-\mu)/(\sigma\sqrt{N})}
 e^{t(X_2-\mu)/(\sigma\sqrt{N})}\cdots e^{t(X_N-\mu)/(\sigma\sqrt{N})}\right),\\
&=&E\left(\!e^{t(X_1-\mu)/(\sigma\sqrt{N})}\!\right)
 E\left(\!e^{t(X_2-\mu)/(\sigma\sqrt{N})}\!\right)\cdots 
  E\left(\!e^{t(X_N-\mu)/(\sigma\sqrt{N})}\!\right),\\
 &=& \left\lbrack E\left(e^{t(X_1-\mu)/(\sigma\sqrt{N})}\right) \right\rbrack^N,
\eeqs
where, in the last two steps above, we used, respectively, the facts that the
$X_j$ are independent and are identically distributed.  Now carry out a
Taylor series expansion about $t=0$:
\beqs
  E\left(\!e^{t(X_1\!-\!\mu)/(\sigma\sqrt{N})}\!\right)
 &=& E\left( 1 + \frac{t(X_1-\mu)}{\sigma\sqrt{N}}\
     + \frac{t^2(X_1-\mu)^2}{2\sigma^2 N} + \cdots\right),\\
 &=& E(1) + \frac{t}{\sigma\sqrt{N}}E(X_1\!-\!\mu)
     + \frac{t^2}{2\sigma^2 N}E\left((X_1\!-\!\mu)^2\right) + \cdots,\\
 &=& 1 + \frac{t}{\sigma\sqrt{N}}(0)
     + \frac{t^2}{2\sigma^2 N}\sigma^2 + \cdots
  = 1 + \frac{t^2}{2N} +\cdots .
\eeqs
From this, one sees that
\beq
   \lim_{N\rightarrow\infty}
 E\left(\!e^{t(X_1\!-\!\mu)/(\sigma\sqrt{N})}\!\right)
  =\lim_{N\rightarrow\infty}\left(1 + \frac{t^2}{2N} +\cdots\right)^N
 = e^{t^2/2}.
\eeq
This is the \alert{moment generating function} of the standardized
normal distribution.  The moment generating function of a random
variable $X$ is defined by $M_X(t)=E(e^{tX})$.  If $X$ and $Y$ are
random variables having moment generating functions $M_X(t)$ and
$M_Y(t)$, respectively, then there is a uniqueness theorem that states that
$X$ and $Y$ have the same probability distribution if and only if
$M_X(t)=M_Y(t)$ identically.  The use of this theorem completes
the proof of the central limit theorem.
\end{proof}

\subsection{Simple Monte Carlo integration}
Recall that for a continuous real-valued function $f(X)$ of a 
continuous random variable $X$ having probability distribution
$p_X(s)$, the expected value of $f(X)$ is
\[ E(\, f(X)\,) = \int_{-\infty}^\infty\!\! f(s)\, p_X(s)\, ds.\]
Now consider a uniform probability density
\[ p_X(x)=\left\{ \begin{array}{l@{\hspace{4mm}}l}1/(b-a), & a\leq x\leq b,\\
              0, & \mbox{otherwise}.\end{array}\right.\]
If one uses this probability density to obtain $N$ outcomes $X_1,X_2,\dots,X_N$,
and applies the function $f$ to obtain random variables $Y_j=f(X_j)$, then the
law of large numbers tell us that
          \beq \frac{1}{N}\sum_{j=1}^N Y_j 
        \stackrel{N\rightarrow\infty}{\longrightarrow} E(Y)
         =E(f(X))=\frac{1}{(b-a)}\int_a^b f(s)ds. \eeq
Define \[\dd\langle f\rangle\equiv \frac{1}{N}\sum_{j=1}^N f(X_j),\]
then
\[ \dd(b-a)\lim_{N\rightarrow\infty}\langle f\rangle =\int_a^b f(s)ds. \]
It is straightforward to generalize this result to multiple dimensions.
Naturally, a key question is then: how good is such an estimate for finite $N$?

For large $N$, the central limit theorem tells us that the error one 
        makes in approximating $E(X)$ by $A_N$ is $\sigma/\sqrt{N}=\sqrt{V(X)/N}$.
For $Y=f(X)$ as before, the error in approximating
        $E(f(X))$ by $\sum_j f(X_j)/N$ is $\sqrt{V(f(X))/N}$.
One can then use the Monte Carlo method to estimate the variance $V(f(X))$:
        \beq V(Y)=E((Y-E(Y))^2)\approx \langle (f-\langle f\rangle)^2\rangle
          = \langle f^2\rangle-\langle f\rangle^2.\eeq

If $p_X(x)$ is not uniform but can be easily sampled, then one can use $p_X(x)$ 
to obtain  $N$ outcomes $X_1,X_2,\dots,X_N$, then
apply the function $f$ to obtain random variables $Y_j=f(X_j)$, and
the law of large numbers tell us that
          \beq \frac{1}{N}\sum_{j=1}^N Y_j 
        \stackrel{N\rightarrow\infty}{\longrightarrow} E(Y)
         =E(f(X))=\int_a^b \!\!p_X(s)\ f(s)ds. \eeq

To summarize, simple Monte Carlo integration is accomplished using
    \beqs && \int_{\cal V} p(\vec{x})\ f(\vec{x})\ d^Dx
    \approx \langle f\rangle \pm \sqrt{\frac{\langle f^2\rangle
    -\langle f\rangle^2}{N}},\\
    &&\langle f\rangle \equiv \frac{1}{N}\sum_{i=1}^N
     f(\vec{x}_i),\qquad\qquad
    \langle f^2\rangle \equiv \frac{1}{N}\sum_{i=1}^N
    f(\vec{x}_i)^2, \eeqs
where the $N$ points $\vec{x}_1,\dots,\vec{x}_N$ are chosen \alert{independently}
and \alert{randomly} with probability distribution  $p(\vec{x})$ throughout the
$D$-dimensional volume ${\cal V}$, and this density satisfies
the normalization condition $\int_{\cal V} p(\vec{x}) d^Dx=1$.
The law of large numbers justifies the correctness of this estimate,
and the central limit theorem gives an estimate of the statistical uncertainty
in the estimate.  In the limit $N\rightarrow\infty$, the Monte Carlo estimate
will tend to be gaussian distributed and the uncertainty tends to a standard
deviation.

Monte Carlo integration requires random numbers, but computers are 
deterministic.  However, clever algorithms can produce sequences of numbers
which \alert{appear} to be random; such numbers are called pseudorandom.  
Devising a good random number generator is a science in itself, which will not 
be discussed here.  Random number generators often utilize the modulus function,
bit shifting, and shuffling to produce random 32-bit or 64-bit integers which 
can be converted to approximate uniform deviates between 0 and 1.  The
\alert{Mersenne twister}\cite{mersenne} is an example of a very good random number
generator.  It is very fast, passes all standard tests, such as the Diehard 
suite, and has an amazingly long period of $2^{19937}-1$.

\begin{figure}[t]
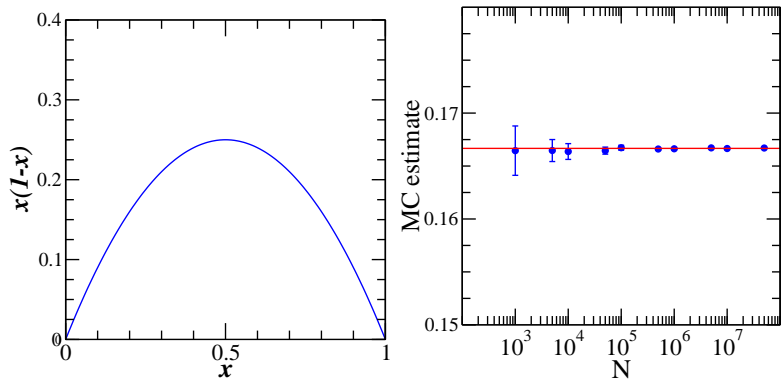

\begin{center}
\includegraphics[width=2.0in,bb=16 39 528 531]{mcint1}
\includegraphics[width=2.0in,bb=8 19 523 523]{mcint2}
\end{center}
\caption{A simple one-dimensional Monte Carlo integration.  The integrand
$x(1-x)$ is shown on the left, while Monte Carlo estimates, with
error bars, are shown on the right for several values of $N$, the number of 
random points used. 
\label{fig:mcint1}}
\end{figure}
To see the method in action, consider a simple one-dimensional example:
 \beq \int_0^1 x(1-x)\ dx = \frac{1}{6} = 0.166666\cdots.\eeq
The integrand is plotted in Fig.~\ref{fig:mcint1}.  Various Monte Carlo 
estimates, including error bars, for different values of $N$, the number 
of random points used, are also shown in this figure.  One sees the error
decreasing as $N$ increases, but notice that the method is not
particularly efficient in this simple case.

The Monte Carlo method works best for flat functions, and is most
problematic when the integrand is sharply peaked or rapidly oscillates.
  \alert{Importance sampling}
can greatly improve the efficiency of Monte Carlo integration by dramatically
reducing the variance in the estimates.  Recall that a simple Monte
Carlo integration is achieved by
\beq \int_a^b f(x)\ dx\approx\frac{(b-a)}{N}\sum_{j=1}^N f(x_j), \eeq
where the $x_j$ are chosen with uniform probability between $a$ and $b$.
Suppose that one could find a function $g(x)>0$ with $\int_a^b g(x)dx=1$ 
such that $h(x)=\dfrac{f(x)}{g(x)}$ is as close as possible to a constant.
The integral can then be evaluated by
\beq \int_a^b f(x)dx=\int_a^b h(x) g(x)dx\approx \frac{(b-a)}{N}
    \sum_{j=1}^N h(x_j), \eeq
where the $x_j$ are now chosen with probability density $g(x)$.
Since the function $h(x)$ is fairly flat, the Monte Carlo method can do
a much better job estimating the integral in this way.  The function $g(x)$
accomplishes the importance sampling, causing more points to be chosen
near peaked regions.  Of course, one must be able to sample with 
probability density $g(x)$.  Also, how can one find such a suitable function 
$g(x)$, especially for complicated multi-dimensional integrals?

Random number generators generally produce uniform deviates.  To sample
other probability densities, a transformation must be applied.
Consider a random variable $U$ with uniform density $p_U(u)=1$ for 
$0\leq x\leq 1$, and another random variable $Y=\phi(U)$, where $\phi$
is a strictly increasing function.  A strictly increasing function ensures 
that the inverse function is single-valued, and also ensures that if $u+du>u$,
then $y+dy>y$ for $y=\phi(u)$.  The probability density $p_Y$ associated with
the random variable $Y$ can be determined using the conservation of probability:
           \beq p_Y(y)dy=p_U(u)du,\qquad
                p_Y(y)=p_U(u)\frac{du}{dy}
            =p_U(\phi^{-1}(y))\frac{d\phi^{-1}(y)}{dy}.
           \eeq
Usually the desired density $p_Y$ is known, so the function $\phi$ must be
determined.  For a uniform deviate $p_U(u)=1$, then $du=p_Y(y)dy$,
and integrating yields
       \beq \int_0^u\!\! du^\prime 
     = \int_{\phi(0)}^{\phi(u)}\!\!\!\!\!\!p_Y(y)\,dy
       \quad\Rightarrow\quad u=F_Y(\phi(u))
        \quad\Rightarrow\quad \phi(u)=F_Y^{-1}(u). \eeq
$F^{-1}$ is unique since $F$ is a strictly increasing function.
In summary: a random variable $Y$ with probability density $p_Y(y)$ and
cumulative distribution $F_Y(y)=\int_{-\infty}^y p_Y(s)\,ds$
can be sampled by first choosing $U$ with uniform probability in some
interval, then applying the transformation
\beq Y=F_Y^{-1}(U).\eeq

\begin{figure}[ht]
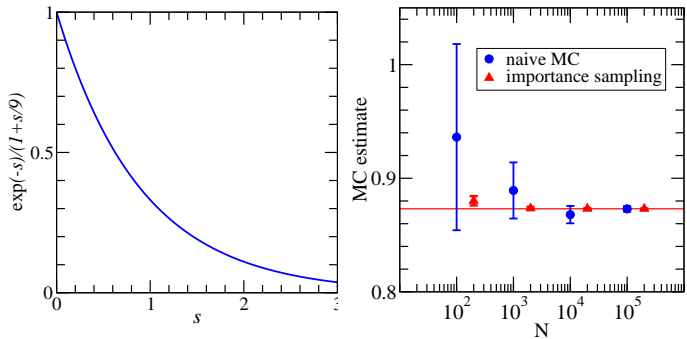

\begin{center}
 \raisebox{5pt}{\includegraphics[width=1.75in,bb=18 44 528 531]{mcint3}}
 \includegraphics[width=1.75in,bb=19 23 523 523]{mcint4}
\end{center}
\caption{Plot of the integrand $e^{-s}/(1+s/9)$ on the left,
and Monte Carlo estimates of the integral on the right.
Circles show estimates without importance sampling, whereas triangles
show estimates using importance sampling.  Statistical uncertainties
using importance sampling are dramatically smaller for a given value
of $N$, the number of random points used.
\label{fig:mcint3}}
\end{figure}

This transformation method is only applicable for probability densities whose
indefinite integral can be obtained and inverted.  Thus, the method is useful
for only a handful of density functions.  One such example is
the exponential distribution:
\beq p_Y(y)=\frac{e^{-y}}{1-e^{-b}}, \qquad \mbox{for $0\leq y\leq b$}.\eeq
The cumulative distribution and its inverse are
\beqs F_Y(y)&=&\int_0^y p_Y(s)ds = \frac{(1-e^{-y})}{(1-e^{-b})},\\
  F^{-1}_Y(u)&=&-\ln\Bigl(1-(1-e^{-b})u\Bigr).\eeqs
Now consider the integral 
\beq  \int_0^3 \frac{e^{-s}\ ds}{1+s/9} \approx 0.873109.\eeq
The integrand is shown in Fig.~\ref{fig:mcint3}, and various Monte
Carlo estimates with and without importance sampling of the integral 
are also shown in the figure.  Estimates using importance sampling
(triangles) are seen to have much smaller statistical uncertainties
for a given value of $N$, the number of random points used.

Probability densities whose cumulative distributions are not easily
calculable and invertible can be sampled using the \alert{rejection method}.
This method exploits the fact that sampling from a density $p_X(x)$ 
for $a\leq x\leq b$ is equivalent to choosing a random point in \alert{two}
dimensions with uniform probability in the \alert{area} under the curve
$p_X(x)$.  
\begin{wrapfigure}{r}{1.5in}
\vspace{0mm}
\includegraphics[width=1.35in,bb=0 0 336 212]{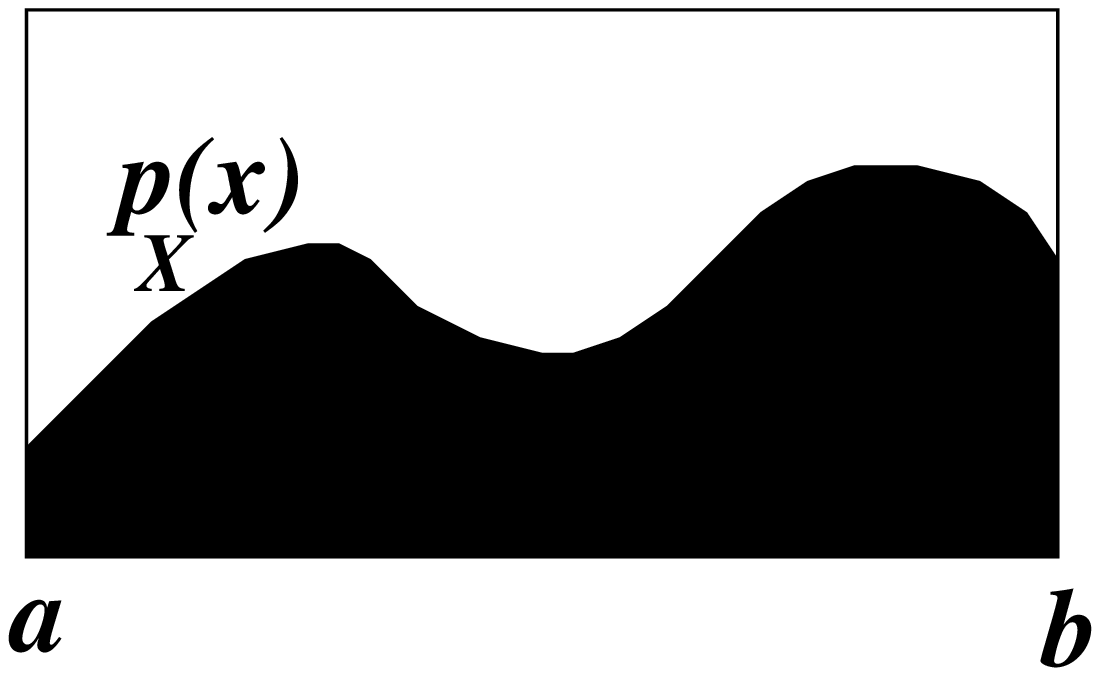}
\vspace{-3mm}
\end{wrapfigure}
For example, one can pick a random point with uniform
probability in a box $a\leq x\leq b$ horizontally and 
$0\leq y\leq \max(p_X(x))$ vertically; the result is accepted if it
lies below the curve, but if above the curve, the result is rejected
and the procedure repeated until an acceptance occurs.
If $p_X(x)$ is sharply peaked, then a more efficient implementation of
the method uses a comparison function $f(x)$ satisfying $f(x)\geq p_X(x)$ 
for all $a\leq x\leq b$ and which can be sampled by the transformation 
method.

\subsection{Monte Carlo using stationary stochastic processes}
The sampling methods described so far work well in one-dimension, but
for multi-dimensional integrals, the transformation and rejection methods 
are usually not feasible.  Fortunately, highly multi-dimensional integrals
can be handled by exploiting \alert{stationary stochastic processes}.
       
A \alert{stochastic process} is a sequence of events $X_t,\ t=0,1,2,\dots$
governed by probabilistic laws (we shall limit our attention to discrete 
``time'' $t$).  Consider a system which can be in one of $R$ discrete states
$s_1, s_2,\dots, s_R$ (generalization to a continuum of states is usually
straightforward).  The system moves or \alert{steps} successively from one 
state to another.  Given previous states of the system $X_0,X_1,\dots,X_{t-1}$, 
the conditional probability to find the system in state $X_t$ at time $t$ 
is denoted by $P(X_0,\dots,X_{t-1}\vert X_t)$ and may depend on previous
states of the system and possibly $t$.  

A stochastic process is \alert{stationary} when the probabilistic laws 
remain unchanged through shifts in time.  In other words, the joint 
probability distribution of $(X_t, X_{t+j_1},\dots, X_{t+j_n})$
is the same as that of $(X_{t+h}, X_{t+h+j_1},\dots, X_{t+h+j_n})$ for 
any $h$.  For such processes, the mean $E(X_t)=\mu$ is independent of $t$ 
(if it exists) and the variance $E((X_t-\mu)^2)=\sigma^2$ is independent
of $t$ if $E(X_t^2)$ is finite.   However, the $X_t$ are usually \alert{not}
independent random variables, but the \alert{autocovariance} 
$E((X_t-\mu)(X_s-\mu))=R(\vert t-s\vert)$ depends only on the time 
difference $\vert t-s\vert$.  The \alert{autocorrelation function} 
is defined by $\rho(t)=R(t)/R(0)$ so that $\rho(0)=1$ and 
$-1\leq \rho(t)\leq 1$ for all $t$ (from Schwartz's inequality).

The Monte Carlo method described so far requires statistically
independent random points.  In order to use points generated by a
stationary stochastic process, we must revisit the law of large
numbers and the central limit theorem for the case of dependent random 
variables.

The \bfalert{law of large numbers for stationary stochastic processes}:  
Consider a stationary stochastic process $X_1,X_2,\dots$ with $E(X_k)=\mu$ and
autocovariance $R(s)=E((X_k-\mu)(X_{k+s}-\mu))$ satisfying
$\sum_{s=0}^\infty \vert R(s)\vert <\infty$, and define
 $\overline{X}_N=(X_1+X_2+\dots +X_N)/N$, then 
\beq \lim_{N\rightarrow\infty} P(\vert \overline{X}_N-\mu\vert 
        \geq \varepsilon)=0,\quad \mbox{for any $\varepsilon>0$}.\eeq
\begin{proof}
Define $Y_n=X_n-\mu$ and $\overline{Y}_N=(Y_1+\dots+Y_N)/N$, then
 \beqs E(\overline{Y}_N^2)
  &=&\frac{1}{N^2}E\Bigl(\sum_{k=1}^NY_k^2 \!+\!2\sum_{k<l}Y_kY_l
  \Bigr)=\frac{1}{N^2}\Bigl(NR(0)\!+\!2\sum_{k<l}R(l\!-\!k)\Bigr),\\
   &=&\frac{R(0)}{N}
    +\frac{2}{N^2}\sum_{k=1}^{N-1}(N-k)\ R(k), \eeqs 
so that
   \beqs N E(\overline{Y}_N^2)&=& \Bigl\vert R(0)
       +\textstyle\sum_{k=1}^{N-1}2R(k)(N-k)/N  \Bigr\vert,\\
         &\leq& \vert R(0)\vert
       +\textstyle\sum_{k=1}^{N-1}2\vert R(k)\vert\ (N-k)/N, \\
       &\leq& \vert R(0)\vert
       +\textstyle\sum_{k=1}^{N-1}2\vert R(k)\vert. \eeqs
Since $\sum_j\vert R(j)\vert<\infty$, then $N E(\overline{Y}_N^2)<\infty$
so $\lim_{N\rightarrow\infty}E(\overline{Y}_N^2)=0$.
The Chebyshev inequality tells us that
$P(\vert \overline{X}_N-\mu\vert \geq \varepsilon)\leq
       E((\overline{X}_N-\mu)^2)/\varepsilon^2$  so
       \beq\lim_{N\rightarrow\infty} E((\overline{X}_N-\mu)^2)=0\ \mbox{implies}
        \ \lim_{N\rightarrow\infty} P(\vert \overline{X}_N-\mu\vert \geq 
     \varepsilon)=0,\eeq
which proves the weak law of large numbers for a stationary stochastic
process with an absolutely summable autocovariance.
\end{proof}
\noindent
One can show that the limiting value of the variance is
\beq\lim_{N\rightarrow\infty}NE((\overline{X}_N-\mu)^2)=
           \sum_{k=-\infty}^\infty R(k).\eeq
\begin{proof}
Since the autocovariance is absolutely summable
$\sum_k\vert R(k)\vert<\infty$, then for any $\varepsilon>0$ there exists
a $q$ such that $\sum_{k=1}^\infty 2\vert R(q+k)\vert < \varepsilon/2$.
Hence,
\beqs\Bigl\vert \!\!\!\!\!\sum_{j=-(N\!-\!1)}^{N\!-\!1}
       \!\!\!\!\!\! R(j)\!-\!NE(\overline{Y}_N^2)\Bigr\vert
      &=& \Bigl\vert R(0)\!+\!2\!\!\sum_{j=1}^{N-1}\! R(j) \!-\! \Bigl(\!
               R(0) \!+\!\sum_{k=1}^{N-1}\!\!2R(k)(N\!-\!k)/N  \Bigr)
\Bigr\vert\\
      &=& \Bigl\vert \textstyle\sum_{k=1}^{N-1}2kR(k)/N  \Bigr\vert
       \leq   \sum_{k=1}^{N-1}2k \vert R(k)\vert\ /N\\
     &=& \textstyle\sum_{k=1}^{q}2k \vert R(k)\vert\ /N + 
         \sum_{k=q+1}^{N-1}2k \vert R(k)\vert\ /N \\
     &\leq & \textstyle\sum_{k=1}^{q}2k \vert R(k)\vert\ /N + 
         \sum_{k=q+1}^{N-1}2 \vert R(k)\vert\\
     &\leq & \textstyle\sum_{k=1}^{q}2k \vert R(k)\vert\ /N + 
         \varepsilon/2.  \eeqs
Since $q$ fixed and finite, we can always increase $N$ so that
           $\sum_{k=1}^{q}2k \vert R(k)\vert\ /N < \varepsilon/2$
          which holds as $N\rightarrow\infty$.
Thus, \beq\Bigl\vert\!\!\!\!\! \sum_{j=-(N\!-\!1)}^{N\!-\!1}\!\!
       \!\!\!\! R(j)\!-\!NE(\overline{Y}_N^2)\Bigr\vert < \varepsilon,\eeq
which proves the required result in the limit as $N\rightarrow\infty$.
\end{proof}

The \bfalert{$M$-dependent central limit theorem} states the following: 
Let $X_1,X_2,\dots,X_N$ 
 be a stationary $M$-dependent sequence of random variables ($X_t$ and $X_{t+s}$
 are independent for $s>M$) such that $E(X_t)=E(X_1)=\mu$ and 
 $E((X_1\!-\!\mu)^2)<\infty$, and define 
 $\overline{X}_N=(X_1+X_2+\dots +X_N)/N$ and
 $\sigma^2=E((X_1\!-\!\mu)^2)+2\sum_{h=1}^{M}E((X_1\!-\!\mu)(X_{h+1}\!-\!\mu))$.
 Then for $a<b$,
      \beq \lim_{N\rightarrow\infty}
           P\left(\frac{a\sigma}{\sqrt{N}}< (\overline{X}_N-\mu)<\frac{b\sigma}{\sqrt{N}}
           \right)= \frac{1}{\sqrt{2\pi}}
           \int_a^b e^{-x^2/2}dx. \eeq
In other words, the distribution of $(X_1+\dots +X_N-N\mu)
         /(\sigma\sqrt{N})$ tends to a standard normal distribution
(zero mean, unit variance).
For the proof of this very important theorem, see Ref.~\refcite{Hoeffding} 
or Ref.~\refcite{anderson}.  One version of the proof relies upon splitting
the summation into blocks in such a way that the resulting variables
are essentially independent.  Note that 
\beq\sigma^2=\sum_{h=-M}^M R(h)=NE((\overline{X}_N-\mu)^2)
\ \ \mbox{for $N\gg M$},\eeq
where the autocovariance $R(h)=R(-h)=E((X_t\!-\!\mu)(X_{t+\vert h\vert}
 \!-\!\mu))$ as usual.

Monte Carlo integration using a stationary stochastic process
is summarized by the following formula:
    \beqs & \displaystyle\int_{\cal V} p(\vec{x})\ f(\vec{x})\ d^Dx
    \approx \langle f\rangle \pm \sqrt{\frac{R_0(f)+2\sum_{h\geq 1}R_h(f)
    }{N}},\\
   &\displaystyle \langle f\rangle \equiv \frac{1}{N}\!\sum_{i=1}^N
     f(\vec{x}_i),\quad
    R_h(f) \equiv \frac{1}{N\!-\!h}\sum_{i=1}^{N-h}
    \Bigl(f(\vec{x}_i)\!-\!\langle f\rangle\!\Bigr)\Bigl( f(\vec{x}_{i+h})
     \!-\!\langle f\rangle\! \Bigr), \eeqs
where the $N$ points $\vec{x}_1,\dots,\vec{x}_N$ are elements of a
\alert{stationary stochastic process} 
with stationary probability distribution $p(\vec{x})$ 
throughout the $D$-dimensional volume ${\cal V}$, which satisfies the 
normalization condition $\int_{\cal V} p(\vec{x}) d^Dx=1$.
One requires that the autocovariance is absolutely summable, that is,
$\sum_{h=0}^\infty \vert R_h(f)\vert < \infty$.
The law of large numbers justifies the correctness of the estimate,
and the $M$-dependent central limit theorem gives an estimate of 
the statistical uncertainty.

\subsection{Markov chains}
\label{sec:markov}

\begin{wrapfigure}{r}{1.5in}
\vspace{0mm}
  \includegraphics[width=1.25in,bb=14 14 420 614]{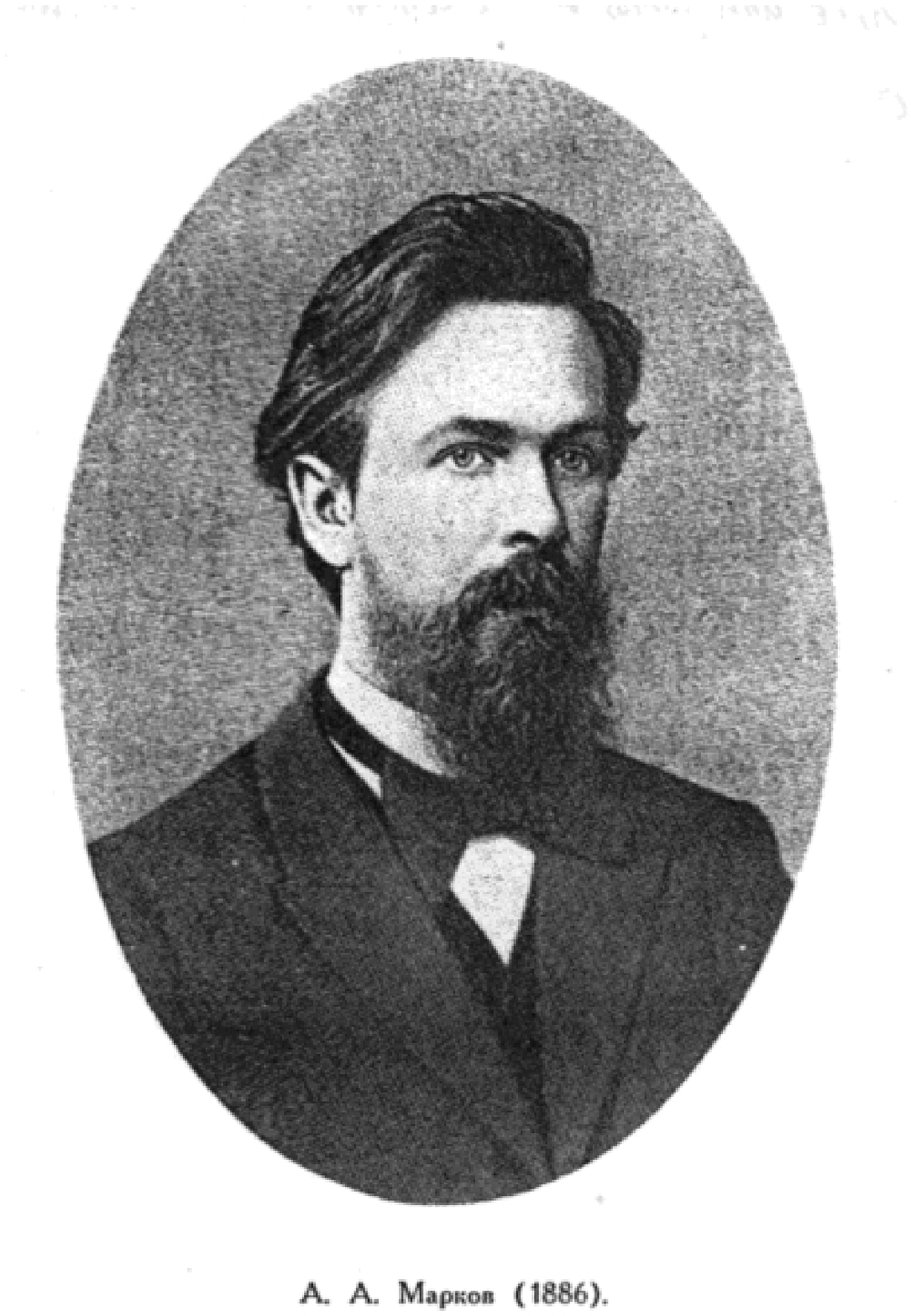}
\vspace{-3mm}
\end{wrapfigure}
Markov chains are one of the simplest types of stochastic
processes.  Markov chains were introduced by the Russian mathematician
Andrei Markov (1856-1922) in 1906.  In this section, we will discuss
Markov chains in great detail.

A \bfalert{Markov chain} is a stochastic process which generates 
a sequence of states with probabilities depending only on the current state
of the system.  Consider a system which can be in one of $R$ states
$s_1, s_2,\dots, s_R$ (again, generalization to a continuum
of states is usually straightforward).  The system moves or \alert{steps}
successively from one state to another (we shall only consider discrete 
``time'' Markov chains).  If the current state is $s_i$, then the chain 
moves to state $s_j$ at the next step with probability $p_{ij}$ which 
does \alert{not} depend on any previous states of the chain.  The
probabilities $p_{ij}$ are called the \alert{transition probabilities}.
The square $R\times R$ real-valued matrix $\mathbf{P}$ whose elements
are $p_{ij}$ is called the \alert{transition matrix} or the 
\alert{Markov matrix}.  Furthermore, we shall only deal with
\alert{time homogeneous} chains in which the transition probabilities 
$p_{ij}$ are independent of ``time'' or their position in chain.

Let us start with some basic properties of Markov chains.
\begin{itemize}
\item The transition matrix $\mathbf{P}$ has only non-negative entries 
   $p_{ij}\geq 0$.
\item  Since the probability of going from $s_i$ to \alert{any} state
       must be unity, then matrix elements must satisfy 
       $\sum_{j=1}^R p_{ij}=1$ (rows sum to unity).
\item  If the columns also sum to unity, then $\mathbf{P}$ is called a
       \alert{doubly stochastic} matrix.
\item If $\mathbf{P}_1$ and $\mathbf{P}_2$ are Markov matrices,
       then the matrix product $\mathbf{P}_1 \mathbf{P}_2$ is also
       a Markov matrix.
\item Every eigenvalue $\lambda$ of a Markov matrix satisfies
        $\vert\lambda\vert\leq 1$.
\item Every Markov matrix has at least one eigenvalue equal to unity.
\end{itemize}
It may be helpful at this point to review the properties of the
eigenvalues and eigenvectors of real square matrices.
\begin{itemize}
\item For a square matrix $\Pmat$, a nonzero column vector 
         $\vvec$ which satisfies $\Pmat\vvec=\lambda\vvec$ for
         complex scalar $\lambda$ is known as a
         \alert{right eigenvector} corresponding to \alert{eigenvalue} $\lambda$.
  Often, ``right eigenvectors" are simply called ``eigenvectors".
   \item A nonzero vector $\vvec$ satisfying
          $\vvec^T \Pmat = \lambda \vvec^T$, where $T$ indicates 
         \alert{transpose}, is known as a \alert{left eigenvector}.
   \item  Every square $R\times R$ matrix has $R$ complex eigenvalues,
         counting multiple roots according to their multiplicity.
   \item  For a real square matrix, the eigenvalues are either real or
          come in complex conjugate pairs.
   \item  Eigenvectors for distinct eigenvalues are
          linearly independent.
   \item A degenerate eigenvalue may not have distinct eigenvectors.
   \item  $R$ linearly independent eigenvectors are guaranteed only if
          all $R$ eigenvalues are distinct.
   \item A matrix $\Pmat$ and its transpose $\Pmat^T$ have the same 
   eigenvalues.
\end{itemize} 
Now let us look at the last two properties of Markov matrices above
in more detail.
\begin{itemize}\item Every eigenvalue $\lambda$ of Markov matrix $\Pmat$
         satisfies $\vert\lambda\vert \leq 1$.
\end{itemize}
\begin{proof}
  Suppose that a complex number $\lambda$ is an eigenvalue of $\Pmat$
            with corresponding eigenvector $\vvec$ so that 
 $\Pmat\vvec=\lambda\vvec$.
  Let $k$ be such that $\vert v_k\vert \geq \vert v_j\vert$ for all $j$,
  then the $k$-th component of the eigenvalue equation gives us
            $ \sum_j p_{kj}v_j = \lambda v_k$.
  Use the generalized triangle inequality for complex numbers
           $\vert \sum_k z_k\vert \leq \sum_k\vert z_k\vert$ to show
     \beq \vert\lambda v_k\vert 
          = \vert \textstyle\sum_j p_{kj}v_j\vert \leq \sum_j p_{kj}\vert v_j\vert
          \leq \sum_j p_{kj}\vert v_k\vert =\vert v_k\vert \eeq
 Thus, $\vert\lambda v_k\vert = \vert\lambda\vert \vert v_k\vert
                 \leq \vert v_k\vert$ implying
               $\vert\lambda\vert \leq 1$,
 which completes the proof.
\end{proof}
\begin{itemize}\item
 Every Markov matrix $\Pmat$ has at least one eigenvalue equal to unity.
\end{itemize}
\begin{proof}
 Let $\vvec$ be a vector satisfying $v_j=1$ for all $j$,
 then $\sum_j p_{ij}v_j=\sum_j p_{ij}=1=v_i$.
 Hence, $\vvec$ is an eigenvector corresponding to eigenvalue $1$,
 so every Markov matrix has at least one eigenvalue equal to unity. 
\end{proof}

Multi-step probabilities are determined from powers of the Markov matrix.
The $ij$-th element $p^{(n)}_{ij}$ of the matrix $\mathbf{P}^n$
is the probability that a Markov chain, starting in state $s_i$, will
be in state $s_j$ after $n$ steps; this is usually called the
\alert{$n$-step transition probability}.  For example, the probability to 
go from $s_i$ to $s_j$ in 2 steps is $\sum_{k=1}^R p_{ik}p_{kj}$.
For a starting probability vector $\mathbf{u}$, the probability that the
chain is in state $s_j$ after $n$ steps is  
$u_j^{(n)}=\sum_{i=1}^R u_i p^{(n)}_{ij}$.  Note that $u_i$ is the probability
that the starting state is $s_i$.  The previous expression can be written
in matrix form by $\uvec^{(n)T} = \uvec^T \Pmat^n$.

Another important concept is the \alert{first visit probability}, which is
the probability that a Markov chain, starting in state $s_i$, is found for the
\alert{first} time in state $s_j$ after $n$ steps.  This first visit
probability is here denoted by $f_{ij}^{(n)}$.
We define $f_{ij}^{(0)}=0$, and for one step, $f_{ij}^{(1)}=p_{ij}$.
For two steps, $f_{ij}^{(2)}=\sum_{k\neq j}p_{ik}p_{kj}$ which generalizes
to $n$-steps as
\beq f_{ij}^{(n)}=\sum_{k\neq j}p_{ik}\ f_{kj}^{(n-1)}.\eeq
An important relation for later use is
\beq p_{ij}^{(n)}=\dd\sum_{m=1}^n  f_{ij}^{(m)} p_{jj}^{(n-m)}.\eeq

The \alert{total visit probability} $f_{ij}$ is the probability that, starting
from state $s_i$, the chain will \alert{ever} visit state $s_j$:
       \beq f_{ij}=\sum_{n=1}^\infty f_{ij}^{(n)}.\eeq
The \alert{mean first passage time} $m_{ij}$ from $s_i$ to $s_j$ is the
expected number of steps to reach state $s_j$ in a Markov chain
for the first time, starting from state $s_i$ (by convention, $m_{ii}=0$):
\beq m_{ij}=\sum_{n=1}^\infty  n\ f_{ij}^{(n)}. \eeq
The \alert{mean recurrence time} $\mu_i$ of state $s_i$ is the expected number
of steps to return to state $s_i$ for the first time in a Markov chain 
starting from $s_i$:
       \beq \mu_i=\sum_{n=1}^\infty n\ f_{ii}^{(n)}.\eeq

\alert{Classes} are an important concept in studying Markov chains.
State $s_j$ is called \alert{accessible} from state $s_i$ if $p_{ij}^{(n)}>0$ 
for some finite $n$.  This is often denoted by $s_i\rightarrow s_j$.
Note that if $s_i\rightarrow s_j$ and $s_j\rightarrow s_k$, then
$s_i\rightarrow s_k$.  States $s_i$ and $s_j$ are said to \alert{communicate} 
if $s_i\rightarrow s_j$ and $s_j\rightarrow s_i$; this is denoted by 
$s_i\leftrightarrow s_j$.  Note that $s_i\leftrightarrow s_j$ and 
$s_j\leftrightarrow s_k$ implies $s_i\leftrightarrow s_k$.
A \alert{class} is a set of states that all communicate with one another.
If $C_1$ and $C_2$ are communicating classes, then either
$C_1=C_2$ or $C_1$ and $C_2$ are disjoint.  To see this, start by noting that
if $C_1$ and $C_2$ have a common state $s_i$, then 
$s_i\leftrightarrow s_{j1}$ for all $s_{j1}\in C_1$ and
$s_i\leftrightarrow s_{j2}$ for all $s_{j2}\in C_2$, so
$s_{j1}\leftrightarrow s_{j2}$, implying $C_1=C_2$.
This means that the set of all states can be partitioned into separate 
non-intersecting classes.  Also, if a transition from class $C_1$ to a
different class $C_2$ is possible, then a transition from $C_2$ to $C_1$
must not be possible, since this would imply $C_1=C_2$.

A Markov chain is called \alert{irreducible} if the probability
to go from every state to every state (not necessarily in one step)
is greater than zero.  All states in an irreducible chain are in one 
single communicating class.

States in a Markov chain can be classified according to whether they are
(a) \alert{positive recurrent} (persistent), (b) \alert{null recurrent},
or (c) \alert{transient}.  A \alert{recurrent} or \alert{persistent} state has 
$f_{ii}=\sum_{n=1}^\infty f_{ii}^{(n)}=1$, that is, there is unit probability of
returning to the state after a finite length of time in the chain.  A
\alert{transient} state has $f_{ii}=\sum_{n=1}^\infty f_{ii}^{(n)}<1$.
A recurrent state is \alert{positive} if its mean recurrence time is finite 
$\mu_i<\infty$; otherwise, it is called \alert{null}.

In addition, states in a Markov chain can be classified according to whether
they are \alert{periodic} (cyclic) or \alert{aperiodic}.
The \alert{period} of a state in a Markov chain is the greatest
common divisor of all $n\geq 0$ for which $p^{(n)}_{ii}>0$.  In other words,
the transition $s_i$ to $s_i$ is not possible except for time intervals which
are multiples of the period $d(i)$.  A \alert{periodic} state $s_i$ has period 
$d(i)>1$, whereas an \alert{aperiodic} state $s_i$ has period $d(i)=1$.

For a recurrent state, $\sum_{n=1}^\infty p^{(n)}_{ii}=\infty$,
 whereas for a transient state, $\sum_{n=1}^\infty p^{(n)}_{ii}<\infty$.
\begin{proof}
 Start with the following:
       \beq \sum_{n=1}^N p_{ij}^{(n)} = \sum_{n=1}^N \sum_{m=1}^n f_{ij}^{(m)}
         p_{jj}^{(n-m)} =\sum_{m=1}^N f_{ij}^{(m)}\sum_{n=0}^{N-m} 
         p_{jj}^{(n)} \leq \sum_{m=1}^N f_{ij}^{(m)}\sum_{n=0}^{N} 
         p_{jj}^{(n)}, \eeq
since $p_{jj}^{(n)}\geq 0$,
but for $N>N^\prime$, we also have
       \beq \sum_{n=1}^N p_{ij}^{(n)}
         =\sum_{m=1}^N f_{ij}^{(m)}\sum_{n=0}^{N-m} p_{jj}^{(n)}
         \geq \sum_{m=1}^{N^\prime} f_{ij}^{(m)}\sum_{n=0}^{N-m} 
 p_{jj}^{(n)} 
 \geq \sum_{m=1}^{N^\prime} f_{ij}^{(m)}\sum_{n=0}^{N-N^\prime} 
 p_{jj}^{(n)} ,
       \eeq
again since the $p_{jj}^{(n)}\geq 0$ and $f_{ij}^{(m)}\geq 0$.
Putting together the above results yields
     \beq \sum_{m=1}^{N^\prime} f_{ij}^{(m)}\sum_{n=0}^{N-N^\prime} p_{jj}^{(n)} 
     \leq \sum_{n=1}^N p_{ij}^{(n)}\leq \sum_{m=1}^N f_{ij}^{(m)}\sum_{n=0}^{N} 
         p_{jj}^{(n)}. \eeq
Take $N\rightarrow\infty$ first, then $N^\prime\rightarrow \infty$ to
     get
     \beq f_{ij}\sum_{n=0}^\infty p_{jj}^{(n)} 
     \leq \sum_{n=1}^\infty p_{ij}^{(n)}\leq f_{ij}\sum_{n=0}^\infty 
         p_{jj}^{(n)} \quad\Rightarrow\quad f_{ij}\sum_{n=0}^\infty p_{jj}^{(n)} 
     =\sum_{n=1}^\infty p_{ij}^{(n)}. \eeq
Set $i=j$ and use $p_{jj}^{(0)}=1$ to see that 
$f_{ii}(1+\sum_{n=1}^\infty p_{ii}^{(n)})=\sum_{n=1}^\infty p_{ii}^{(n)}$, 
so 
        \beq \sum_{n=1}^\infty p_{ii}^{(n)}=\frac{f_{ii}}{1-f_{ii}}.   \eeq
Now use the facts that $f_{ii}=1$ for a recurrent state and $f_{ii}<1$ 
for a transient state to complete the proof of the above statements.
\end{proof}
Note that the above results also imply
       \beq \sum_{n=1}^\infty p_{ij}^{(n)}=\frac{f_{ij}}{1-f_{ii}}.\eeq

A Markov chain returns to a \alert{recurrent} state infinitely often
and returns to a \alert{transient} state only a finite number of times.
\begin{proof}
 Let $g_{ij}(m)$ denote the probability that a Markov chain enters state $s_j$
 at least $m$ times, starting from $s_i$.  Clearly $g_{ij}(1)=f_{ij}$.
 One also sees that $g_{ij}(m+1)=f_{ij} g_{jj}(m)$, so $g_{ij}(m)=\left(f_{ij}\right)^m$.
 The probability of entering $s_j$ infinitely many times is
     $g_{ij}=\lim_{m\rightarrow\infty}g_{ij}(m)=\lim_{m\rightarrow\infty}(f_{ij})^m$,
 so starting in $s_j$, then
           \beq g_{jj}=\lim_{m\rightarrow\infty}
         (f_{jj})^m = \left\{ \begin{array}{l@{\hspace{4mm}}l} 1, & \mbox{for
        recurrent state since $f_{jj}=1$,}\\ 0, &\mbox{for transient state
        since $f_{jj}<1$,}\end{array}\right.\eeq
which completes the proof.
\end{proof}

Another important result for recurrent states is as follows:
if $s_i$ is recurrent and $s_i\rightarrow s_j$, then $f_{ji}=1$.
\begin{proof}
Let $\alpha$ denote the probability to reach $s_j$ from $s_i$
without previously returning to $s_i$, and since $s_i\rightarrow s_j$ we
know that $\alpha >0$.  The probability of \alert{never} 
returning to $s_i$ from $s_j$ is $1-f_{ji}$, and the probability of never returning 
to $s_i$ from $s_i$ is at least $\alpha(1-f_{ji})$.  But $s_i$ is recurrent so the
probability of no return is zero; thus, $f_{ji}=1$.  For two communicating states 
$s_i\leftrightarrow s_j$ that are each recurrent, it follows that $f_{ij}=f_{ji}=1$.
\end{proof}

All states in a class of a Markov chain are of the same type, and if periodic, all 
have the same period.
\begin{proof}
 For any two states $s_i$ and $s_j$ in a class,
 there exists integers $r$ and $s$ such that $p^{(r)}_{ij}=\alpha>0$
 and $p_{ji}^{(s)}=\beta>0$ so
             \beq p_{ii}^{(n+r+s)}=\sum_{kl} p^{(r)}_{ik}p^{(n)}_{kl}p_{li}^{(s)}
                 \geq \sum_{k} p^{(r)}_{ik}p^{(n)}_{kk}p_{ki}^{(s)}
              \geq p^{(r)}_{ij}p^{(n)}_{jj}p_{ji}^{(s)}=\alpha\beta 
              p^{(n)}_{jj}.\eeq
 If $s_i$ is transient, then the left-hand side is a term of a convergent
 series $\sum_k p_{ii}^{(k)}<\infty$, so the same must be true for 
 $p_{jj}^{(k)}$, and if
$p_{ii}^{(k)}\rightarrow 0$, then $p_{jj}^{(k)}\rightarrow 0$.
 The same statements remain true if the roles of $i$ and $j$ are
 reversed, so either both $s_i$ and $s_j$ are transient, or neither is.
 If $s_j$ is null (infinite mean recurrence time 
               $\mu_j=\sum_{n=1}^\infty n\ f_{jj}^{(n)}=\infty$), then $s_i$ must
               be null as well.
 The same statements are true if $i,j$ are reversed, so if one is a null 
               state, then so is the other.  Similarly, we can also conclude
that if either $s_i$ or $s_j$ is positive recurrent (finite mean recurrence
time), then so is the other.  Thus, we have shown that all states in a
class are of the same type (positive recurrent, null recurrent, or transient).

 Suppose $s_i$ has period $t$, then for $n=0$, the right-hand side of the above
 equation is positive, so $p_{ii}^{(r+s)}>0$, which means that $r+s$ must be a 
 multiple of $t$.  Hence, the left-hand side vanishes unless $n$ is multiple of $t$, 
 so $p_{jj}^{(n)}$ can be nonzero only if $n$ is multiple of $t$, which means
 that $s_i$ and $s_j$ have the same period.  Note that the chain is aperiodic if 
 $p_{ii}>0$ for at least \alert{one} $s_i$.
\end{proof}

States in an irreducible chain with period $d$ can be partitioned
into $d$ mutually exclusive subsets $G_0,\cdots,G_{d-1}$ such that if state
 $s_k\in G_\alpha$, then $p_{1k}^{(n)}=0$ unless $n=\alpha+\nu d$.
\begin{proof}
 Since irreducible, all states have the same period $d$ and every state
          can be reached from every other state.
 There exist for every state $s_k$ two integers $a$ and $b$ such
        that $p_{1k}^{(a)}>0$ and $p_{k1}^{(b)}>0$,
 but $p_{11}^{(a+b)}=\sum_j p_{1j}^{(a)}p_{j1}^{(b)}\geq 
        p_{1k}^{(a)}p_{k1}^{(b)}>0$, so $a+b$ divisible by $d$.
 Thus, $a+b=md$ for integer $m$, or $a=-b+md$.
 Rewrite this as $a=\alpha+\nu d$ for integer $\nu$ and $0\leq\alpha<d$.
 The parameter $\alpha$ is characteristic of state $s_k$ so all states are partitioned
 into $d$ mutually exclusive subsets $G_0,G_1,\cdots,G_{d-1}$.
\end{proof}
 With proper ordering of the $G_\alpha$ subsets, a \alert{one-step} transition
 from a state in $G_\alpha$ always leads to a state in $G_{\alpha+1}$, or from
      $G_{d-1}$ to $G_0$.  Each subset $G_\alpha$ can be considered states in an
aperiodic Markov chain with transition matrix $\Pmat^d$.

As an aside, consider the following fact concerning finite Markov chains:
in an irreducible chain having a finite number $R$ of states, there are no
null states and it is impossible that all states are transient.
\begin{proof}
 All rows of the matrix $\Pmat^n$ must add to unity.  Since each row contains a
finite number of non-negative elements, it is impossible that 
 $p_{ij}^{(n)}\rightarrow 0$ for all $i,j$ pairs.  Thus, it is impossible that
 all states are transient, so at least one state must be non-null.
 But since the chain is irreducible (one class), all states must be non-null.
\end{proof}
 In fact, in an $R$-state irreducible Markov chain, it is possible to go
 from any state to any other state in at most $R-1$ steps.

We now need to consider a very important theorem (often referred to as the
\alert{basic limit theorem} of the \alert{renewal equation}) about two sequences.
Given a sequence $f_0,f_1,f_2,\dots$ such that
 \beq  f_0=0, \qquad 0\leq f_n\leq 1, \qquad \sum_{n=0}^\infty f_n=1, \eeq
and greatest common divisor of those $n$ for which $f_n>0$ is $d\geq 1$,
and another sequence $u_0,u_1,u_2,\dots$ defined by
 \beq u_0=1, \qquad u_n=\sum_{m=1}^n f_m u_{n-m}, \qquad (n\geq 1),\eeq
then 
    \beq \lim_{n\rightarrow\infty} u_{nd} = \left\{
      \begin{array}{l@{\hspace{5mm}}l} d\mu^{-1} 
  & \mbox{if $\dd\mu=\sum_{n=1}^\infty nf_n<\infty$,}\\[5pt]
  0 & \mbox{if $\mu=\infty$.}\end{array}\right.\eeq
\begin{proof}
See Refs.~\refcite{feller, karlin} for a complete proof.  Here, we shall only 
provide a sketch of the proof of this theorem.  First, note some key properties of 
these sequences.  We know that $0\leq f_n\leq 1$ for all $n$ since $f_n\geq 0$ and 
$\sum_{n=0}^\infty f_n=1$.  Also, $0\leq u_n\leq 1$ for all $n$ can be established 
inductively.  To do this, first note that $u_0=1,\ u_1=f_1,\ u_2=f_2+f_1^2$ satisfy 
the above bounds.  Assume $0\leq u_k\leq 1$ for all $0\leq k\leq n$.  Since $f_m\geq 0$
and  $\sum_{m=1}^\infty f_m=1$, then
   $u_{n+1}=\sum_{m=1}^{n+1}f_m u_{n+1-m}\geq 0$ since it is a sum of nonnegative terms, 
   and $u_{n+1}=\sum_{m=1}^{n+1}f_m u_{n+1-m}\leq \sum_{m=1}^{n+1}f_m\leq 1$, which
completes the induction.  

Next, limit our attention to $d=1$ (the nonperiodic case).
 Since $u_n$ is a bounded sequence,
     $\lambda\equiv \limsup_{n\rightarrow\infty}u_n$ is finite,
       and there exists a subsequence $n_1<n_2<\cdots$ tending to infinity
       such that $\lim_{j\rightarrow\infty} u_{n_j}=\lambda$.
 The next step in the proof is to show that $\lim_{j\rightarrow\infty}
            u_{n_j-q}=\lambda$ for any integer $q\geq 0$ when $f_1>0$
            (we skip this here).

 Now define a new sequence $r_n=\sum_{k>n} f_k$.  Some important properties of this 
sequence are $r_n\geq 0$ for all $n$, $r_0=1$, $r_{n-1}-r_n = f_n$ for $n\geq 1$,
and $\sum_{n=0}^\infty r_n=\sum_{n=1}^\infty nf_n\equiv \mu$.
One very crucial identity is
   \beq \sum_{k=0}^N r_k u_{N-k}=1,\qquad \mbox{for all $N\geq 0$}.\eeq
To see this, define $A_N= \sum_{k=0}^N r_k u_{N-k}$.
Start with \beq u_N = \sum_{m=1}^N f_m u_{N-m}=\sum_{m=1}^N (r_{m-1}-r_m) u_{N-m},\eeq
use $r_0=1$, and rearrange to get \beq r_0u_N + \sum_{m=1}^N r_m u_{N-m}
            =\sum_{m=1}^N r_{m-1} u_{N-m}.\eeq
Take $m\rightarrow k+1$ on the right: \beq\sum_{m=0}^N r_m u_{N-m}
             =\sum_{k=0}^{N-1} r_k u_{N-1-k},\eeq
which shows $A_N=A_{N-1}$ for all $N$.  Thus, $A_N=A_{N-1}=A_{N-2}=\cdots=A_0
 =r_0u_0=1$.

Recall that $n_1<n_2<\cdots$
 is a subsequence such that $\lim_{j\rightarrow\infty} u_{n_j-q}=\lambda$
  for any integer $q\geq 0$.
  Since $\sum_{k=0}^{n_j} r_k u_{n_j-k}=1$ for all $n_j$ and
     $r_k\geq 0,\ u_k\geq 0$ for all $k$, then
       $\sum_{k=0}^N r_k u_{n_j-k} \leq 1$ for fixed $N<n_j$.
 Take the limit $j\rightarrow\infty$ so
        \beq \lim_{j\rightarrow\infty}\sum_{k=0}^N r_k u_{n_j-k}
            = \lambda\sum_{k=0}^N r_k  \leq 1.\eeq
 We already know that $\lambda\geq 0$, so take $N\rightarrow\infty$
            to have
              \beq 0\leq \lambda \leq 1/(\sum_{k=0}^\infty r_k).\eeq
 If $\sum_{k=0}^\infty r_k=\infty$ then 
             $\lim_{n\rightarrow\infty}u_n=\lambda=0$.
 If $\mu=\sum_{k=0}^\infty r_k$ is finite, $N\rightarrow\infty$
            gives  $\mu\lambda\leq 1$.
 Define $M=\sup_{n\geq 0} u_n$ so $0\leq u_k\leq M\leq 1$ for all $k$,
 and define $g(D)=\sum_{k=D+1}^\infty r_k$, noting that $g(D)\geq 0$ for all $D$
                and $\lim_{D\rightarrow\infty}g(D)=0$.
 Consider \beq\sum_{k=0}^D r_k u_{n_j-k} 
     + \sum_{k=D+1}^{n_j}r_k u_{n_j-k}=1, \quad \mbox{for $D<n_j$}.\eeq
 Thus  \beq\sum_{k=0}^D r_k u_{n_j-k} + Mg(D) \geq 1 \quad \mbox{for $D<n_j$}.\eeq
 Take $j\rightarrow\infty$ to conclude
 \beq \lambda \left(\sum_{k=0}^D r_k\right) + Mg(D) \geq 1.\eeq
 Take the limit $D\rightarrow\infty$ to obtain $\lambda\mu\geq 1$.
 We have now shown $1\leq \mu\lambda\leq 1$ so $\mu\lambda=1$.
 The proof for the nonperiodic $(d=1)$ case is now complete.

 When $d>1$, we know $f_m=0$ unless $m=nd$ for integer $n$.
 One can then show $u_m=0$ unless $m=nd$.
 Define new sequences $f^\prime_n=f_{nd}$ and $u^\prime_n=u_{nd}$ for 
               $n=0,1,2,\dots$.
 Since the new sequence is aperiodic, we know 
$ \lim_{n\rightarrow\infty}u^\prime_n
 =1/\mu^\prime$ where $\mu^\prime=\sum_{n=0}^\infty n f^\prime_n$.
 Since $f_m=0$ when $m\neq nd$, then 
  \beq \mu^\prime=\sum_{n=0}^\infty n f_{nd}=d^{-1}
  \sum_{m=0}^\infty m f_m=\mu/d.\eeq
 Thus, $\lim_{n\rightarrow\infty}u_{nd}=d\mu^{-1}$ as required.
\end{proof}

An important feature of a Markov chain is the asymptotic behavior of its 
$n$-step probabilities as $n$ becomes large.  First, consider
$p_{jj}^{(n)}$, the $n$-step probability to
go from state $s_j$ back to $s_j$ as $n$ becomes large. This
behavior can be summarized as
    \beq \lim_{n\rightarrow\infty} p_{jj}^{(dn)} = \left\{ 
  \begin{array}{l@{\hspace{5mm}}l}
        0, & \mbox{$s_j$ transient or null recurrent,}\\
        \mu_j^{-1}, & \mbox{$s_j$ aperiodic positive recurrent,}\\
        d\mu_j^{-1}, & \mbox{$s_j$ positive recurrent with period $d$.}
   \end{array}\right. \eeq
\begin{proof}
  If $s_j$ is transient, then $\sum_n p_{jj}^{(n)}$ is finite (converges),
  requiring $p_{jj}^{(n)}\rightarrow 0$.
  For a recurrent $s_j$, let $f_n=f_{jj}^{(n)}$ and $u_n=p_{jj}^{(n)}$.
  The sequences $f_n,u_n$ so defined satisfy the conditions of the basic limit 
  theorem of the renewal equation previously discussed, which tells us that
  $p_{jj}^{(dn)}\rightarrow d\mu_j^{-1}$ where
               $\mu_j=\sum_n nf_{jj}^{(n)}$ is the mean recurrence time.
  Of course, the aperiodic case applies when $d=1$.
  If $s_j$ is null recurrent, then $\mu_j=\infty$ so $p_{jj}^{(n)}
   \rightarrow \mu_j^{-1}=0$.
\end{proof}

Next, the asymptotic behavior of $p_{ij}^{(n)}$ can be summarized as
    \beq \lim_{n\rightarrow\infty} p_{ij}^{(n)} = \left\{
  \begin{array}{l@{\hspace{5mm}}l}
        0, & \mbox{$s_j$ transient or null recurrent,}\\
        f_{ij}\mu_j^{-1}, & \mbox{$s_j$ aperiodic positive recurrent.}
            \end{array}\right.\eeq
Here, we will ignore the periodic case.
\begin{proof}
 Start by noting that
   \beq p_{ij}^{(n)}=\sum_{m=1}^n f_{ij}^{(m)}p_{jj}^{(n-m)}
         = \sum_{m=1}^{n^\prime} f_{ij}^{(m)}p_{jj}^{(n-m)}
          \!\!+\!\! \sum_{m=n^\prime+1}^n f_{ij}^{(m)}p_{jj}^{(n-m)}
         \quad (n^\prime < n). \eeq
 Since $0\leq  \sum_{m=n^\prime+1}^n f_{ij}^{(m)}p_{jj}^{(n-m)}
     \leq  \sum_{m=n^\prime+1}^n f_{ij}^{(m)}$, then
       \beq\textstyle
      0\leq \Bigl(p_{ij}^{(n)}- \sum_{m=1}^{n^\prime} f_{ij}^{(m)}p_{jj}^{(n-m)}\Bigr)
          \leq \sum_{m=n^\prime+1}^n f_{ij}^{(m)}
         \quad (n^\prime < n).\eeq
 Take $n\rightarrow\infty$, then $n^\prime\rightarrow\infty$ above,
 and denote $p_{jj}=\lim_{n\rightarrow\infty}  p_{jj}^{(n)}$ to deduce
     \beq\textstyle 0 \leq \Bigl( \lim_{n\rightarrow\infty}p_{ij}^{(n)}-p_{jj}
        f_{ij}\Bigr)\leq 0 \quad\Rightarrow\quad 
         \lim_{n\rightarrow\infty}p_{ij}^{(n)}=p_{jj}\, f_{ij}.
         \eeq
 For the case of $s_j$ transient or null recurrent, $p_{jj}=0$ and $f_{ij}$ 
 finite, so $\lim_{n\rightarrow\infty}p_{ij}^{(n)}=0$.
 For $s_j$ aperiod and positive recurrent, $p_{jj}=\mu_j^{-1}$
              so $p_{ij}^{(n)}\rightarrow f_{ij}\mu_j^{-1}$.
\end{proof}

The above information will be needed in proving a very important property
of irreducible aperiodic Markov chains.  Before getting to this property,
it is convenient to introduce a few definitions and remind the reader
of two lemmas concerning sequences.

A probability vector $\wvec$ is called \alert{stationary}
    or \alert{invariant} or a \alert{fixed-point} if $\wvec^T=\wvec^T \Pmat$.
Clearly, one also has $\wvec^T=\wvec^T \Pmat^n$.  If one starts a Markov
chain with an initial probability vector that is stationary, then the
probability vector is always the same (stationary) for the chain.
When this occurs, the Markov chain is said to be in \alert{equilibrium}.

Fatou's lemma and the dominated convergence theorem will be needed
in demonstrating an important property of Markov chains, so these
theorems are briefly recapped next.

\bfalert{Fatou's lemma}: Let $a_n(t)$ for $n=1,2,\dots$ be a function
 on a discrete set $T=\{1,2,\dots\}$, assume 
  $\lim_{n\rightarrow \infty}a_n(t)$ exists for each $t$ in $T$, and suppose 
  $a_n(t)\geq 0$ for all $t,n$, then
        \beq \sum_{t\in T}\left(\lim_{n\rightarrow\infty} a_n(t)\right)
           \leq \lim_{n\rightarrow\infty} \sum_{t\in T} a_n(t).\eeq
\begin{proof}
 For any integer $M$
     \beq \sum_{t=1}^M\left(\lim_{n\rightarrow\infty}a_n(t)\right)
      = \lim_{n\rightarrow\infty}\sum_{t=1}^M a_n(t) \leq 
      \lim_{n\rightarrow\infty}\sum_{t=1}^\infty a_n(t), \eeq
 since all $a_n(t)\geq 0$.
 Take the limit $M\rightarrow\infty$ to obtain required result.
\end{proof}
 This lemma shows that taking the limit then summing the sequence is not
 the same as summing the sequence, then taking the limit. 
 For example, consider $a_n(t)=n/(n^2+t^2)$.
 For $n>t$ then $\lim_{n\rightarrow\infty}a_n(t)=0$ so
  $\sum_{t=1}^\infty\left(\lim_{n\rightarrow\infty} a_n(t)\right)=0$.
 But \beq\sum_{t=1}^\infty a_n(t)=\frac{\pi}{2}\coth(n\pi)-\frac{1}{2n} 
     \quad\mbox{so}\quad \lim_{n\rightarrow\infty}\sum_{t=1}^\infty 
   a_n(t)=\frac{\pi}{2}.\eeq

The \bfalert{dominated convergence theorem}: Let $a_n(t)$ for $n=1,2,\dots$ 
be a function on a discrete set $T=\{1,2,\dots\}$, assume 
$\lim_{n\rightarrow \infty}a_n(t)$ exists for each $t$ in $T$, and suppose 
a function $B(t)$ exists such that $\vert a_n(t)\vert \leq B(t)$ for all 
$t,n$ and $\sum_{t\in T}B(t)<\infty$, then
        \beq \sum_{t\in T}\left(\lim_{n\rightarrow\infty} a_n(t)\right)
            = \lim_{n\rightarrow\infty} \sum_{t\in T} a_n(t).\eeq
\begin{proof}
 Let $a(t)\!=\!\dd\lim_{n\rightarrow\infty}a_n(t)$ and given
         $\vert a(t)\vert\leq B(t)$, then $\sum_{t=1}^\infty a(t)$ 
          converges since $\sum_{t=1}^\infty B(t)$ converges. 
For any integer $M$,
       \beq \Bigl\vert \sum_{t=1}^\infty a_n(t) - \sum_{t=1}^\infty a(t)\Bigr\vert
         \leq \sum_{t=1}^M\vert a_n(t)-a(t)\vert \!+\! \sum_{t=M+1}^\infty
           \Bigl( \vert a_n(t)\vert + \vert a(t)\vert \Bigr). \eeq
 Next, take the limit $n\rightarrow\infty$, and for a finite sum
of positive terms, the summation and limit can be taken in any order:
       \beq \lim_{n\rightarrow\infty}\sum_{t=1}^M\vert a_n(t)\!-\!a(t)\vert
         = \sum_{t=1}^M\Bigl(\lim_{n\rightarrow\infty}\vert 
   a_n(t)\!-\!a(t)\vert\Bigr)=0.\eeq
We also have
 \beq\sum_{t=M+1}^\infty\Bigl( \vert a_n(t)\vert\! +\! \vert a(t)\vert \Bigr) 
           \leq 2\sum_{t=M+1}^\infty B(t), \eeq
  so for any integer $M$,
      \beq \Bigl\vert \lim_{n\rightarrow\infty}\sum_{t=1}^\infty a_n(t)
            - \sum_{t=1}^\infty\lim_{n\rightarrow\infty}a_n(t)\Bigr\vert
              \leq 2\sum_{t=M+1}^\infty B(t).\eeq
 The right-hand side is the remainder of a convergent series, which means
that it must equal zero in the $M\rightarrow\infty$ limit.  The equality
 to be shown easily follows.
\end{proof}
The dominated convergence theorem essentially specifies the conditions
under which the order of taking an asymptotic limit and summing a sequence
does not matter.

And now, without further adieu, we state the very important
\bfalert{fundamental limit theorem for irreducible Markov chains}:
An \alert{irreducible aperiodic} Markov chain with transition matrix $\Pmat$
has a \alert{stationary} distribution $\wvec$ satisfying $w_j>0$, 
$\sum_jw_j=1$,  and $\wvec^T=\wvec^T \Pmat$
     if, and only if, all its states
     are \alert{positive recurrent}, and this stationary distribution is
     \alert{unique} and identical to the limiting distribution
         $w_j=\lim_{n\rightarrow\infty}p_{ij}^{(n)}$ independent of initial
          state $s_i$.
\begin{proof}
For an irreducible aperiodic chain, the following possibilities exist:\\
\hspace*{5mm}(a) \alert{all} states are positive recurrent (an \alert{ergodic} chain),\\
\hspace*{5mm}(b) \alert{all} states are null recurrent,\\
\hspace*{5mm}(c) \alert{all} states are transient.\\
If all states are transient or null recurrent, then
 $\lim_{n\rightarrow \infty}p_{ij}^{(n)}=0$.
If all states are positive recurrent, then since \alert{all} states 
communicate, $f_{ij}=1$ for all $i,j$ and the basic limit theorem of
the renewal equation tells us that
 $\lim_{n\rightarrow\infty} p_{ij}^{(n)}=\mu_j^{-1}$.
Let us define $w_j=\lim_{n\rightarrow\infty} p_{ij}^{(n)}=\mu_j^{-1}$
which is independent of the initial state $s_i$.
For all states positive recurrent, then $0<\mu_j<\infty$
so $w_j>0$ for all $j$.
We have $p_{ij}^{(m+n)}= \sum_{k=1}^\infty p_{ik}^{(n)}p_{kj}^{(m)}$
so using Fatou's lemma:
        \beq \lim_{n\rightarrow\infty} p_{ij}^{(m+n)}= \lim_{n\rightarrow\infty}
          \sum_{k=1}^\infty p_{ik}^{(n)}p_{kj}^{(m)}\geq 
          \sum_{k=1}^\infty \lim_{n\rightarrow\infty} p_{ik}^{(n)}p_{kj}^{(m)}.
          \eeq
Taking the limit $n\rightarrow\infty$ yields
          $ w_j\geq \sum_{k=1}^\infty w_k\ p^{(m)}_{kj}$.
Define $s\equiv \sum_{k=1}^\infty w_k$, then sum the above equation over $j$:
      \beq s=\sum_{j=1}^\infty w_j\geq \sum_{j=1}^\infty
        \sum_{k=1}^\infty w_k\ p^{(m)}_{kj}
        = \sum_{k=1}^\infty w_k \sum_{j=1}^\infty p^{(m)}_{kj}
    =\sum_{k=1}^\infty 
          w_k=s,\eeq
where we used the fact that the rows of the Markov matrix and its powers
sum to unity.
Interchanging the order of the two infinite summations above is possible
since all summands are non-negative (Fubini's theorem).
We have shown that $s\geq s$, which means that the equality must hold:
   \beq \sum_{j=1}^\infty w_j = \sum_{j=1}^\infty
   \sum_{k=1}^\infty w_k\ p_{kj}^{(m)}. \eeq
But for each term, we have already shown that 
 $ w_j\geq \sum_{k=1}^\infty w_k\ p^{(m)}_{kj}$.  Since each one of the
terms in the summation is known to be greater than or equal to zero, we
must conclude that the equality holds term by term for every $j$:
   \beq w_j = \sum_{k=1}^\infty w_k\ p_{kj}^{(m)}. \eeq
For $m=1$, we see that the limiting vector $\wvec$ satisfies the criteria
for a \alert{stationary} vector.
Next, use $\sum_{j=1}^\infty p_{ij}^{(n)}=1$ and Fatou's lemma to show that
    \beq 1 = \lim_{n\rightarrow\infty}\sum_{j=1}^\infty p_{ij}^{(n)}
        \geq \sum_{j=1}^\infty \lim_{n\rightarrow\infty} p_{ij}^{(n)} 
          = \sum_{j=1}^\infty w_j. \eeq
Given $\sum_j w_j\leq 1$, then consider the limit $m\rightarrow\infty$ of
       \beq  w_j = \lim_{m\rightarrow\infty}\sum_{k=1}^\infty 
         w_k\ p_{kj}^{(m)}.\eeq
Since $0\leq p_{kj}^{(m)}\leq 1$, then 
        $\vert w_k p_{kj}^{(m)}\vert \leq w_k$ and 
   $\sum_{k=1}^\infty w_k < \infty$ so
         the dominated convergence theorem can be applied:
      \beq w_j = \lim_{m\rightarrow\infty}\sum_{k=1}^\infty w_k\ p_{kj}^{(m)}=
    \sum_{k=1}^\infty w_k\lim_{m\rightarrow\infty} p_{kj}^{(m)}
      =\Bigl( \sum_{k=1}^\infty w_k \Bigr) w_j. \eeq
We can at last conclude that $\sum_{j=1}^\infty w_j=1$.

Only the uniqueness of the stationary state is left to show.
If another stationary vector $\vvec$ existed, it would have to
satisfy $v_j>0$, $\sum_{j=1}^\infty v_j=1$, and
        $ v_j=\sum_{i=1}^\infty v_ip_{ij}^{(n)}$.
Conditions for the dominated convergence theorem again apply,
so taking the $n\rightarrow \infty$ limit gives
        \beq v_j = \lim_{n\rightarrow\infty} \sum_{i=1}^\infty v_ip_{ij}^{(n)}
          =  \sum_{i=1}^\infty v_i \lim_{n\rightarrow\infty}p_{ij}^{(n)}
          =\Bigl(\sum_{i=1}^\infty v_i\Bigr) w_j = w_j.\eeq
Since $\vvec=\wvec$, then $\wvec$ is unique.
\end{proof}

A simple example may help to understand the above result.
Consider the following transition matrix 
\beq\Pmat=
        \left[\begin{array}{ccc} \frac{3}{4} & \frac{1}{4} & 0\\[4pt]
      0 & \frac{2}{3} & \frac{1}{3}\\[4pt] \frac{1}{4} & \frac{1}{4} 
    & \frac{1}{2}\end{array}\right].\eeq
Since $\Pmat^2$ has all positive entries (greater than zero), this
Markov chain is irreducible.  The eigenvalues of $\Pmat$ are 
$1,\frac{1}{2},\frac{5}{12}$, and the
unnormalized right and left eigenvectors are
 \beq \mbox{right:}\begin{array}{c} 1 \\[6pt]\left[\begin{array}{r} 1\\1\\1\end{array}\right]\end{array}
  \begin{array}{c} \frac{1}{2} \\[6pt]\left[\begin{array}{r} 2\\-2\\1\end{array}\right]\end{array}
  \begin{array}{c} \frac{5}{12} \\[6pt] \left[\begin{array}{r} 3\\-4\\3\end{array}\right]\end{array}
   \qquad\mbox{left:}
 \begin{array}{c} 1 \\[6pt]\left[\begin{array}{r} 2\\3\\2\end{array}\right]\end{array}
  \begin{array}{c} \frac{1}{2} \\[6pt]\left[\begin{array}{r} -1\\ 0\\1\end{array}\right]\end{array}
  \begin{array}{c} \frac{5}{12} \\[6pt] \left[\begin{array}{r} -3\\-1\\4\end{array}\right]\end{array}.
 \eeq
The left fixed-point probability vector and $\lim_{n\rightarrow\infty}\Pmat^n$
are
      \beq \wvec=\frac{1}{7}\left[\begin{array}{r} 2\\3\\2\end{array}\right],
      \quad \lim_{n\rightarrow\infty}\Pmat^n=\Wmat=\frac{1}{7}
        \left[\begin{array}{ccc} 2 & 3 & 2\\
      2 & 3 & 2\\ 2 & 3 & 2\end{array}\right].  
     \eeq

A positive recurrent chain guarantees the existence of at
least one invariant probability vector.  Irreducibility guarantees the
uniqueness of the invariant probability vector.  Aperiodicity guarantees 
that the limit distribution coincides with the invariant distribution.

Suppose a Markov chain is started with a probability vector
given by $\wvec$, the left fixed-point vector of the transition
 matrix $\Pmat$.  This means that the probability of starting in state 
$s_i$ is $w_i$.  Then the probability of being in state $s_j$ after $n$ 
steps is $(\wvec^T\Pmat^n)_j$, but $\wvec^T\Pmat^n=\wvec^T$, so
 this probability is $w_j$.  Thus, the probability vector is always the 
same, that is, it is \alert{stationary} or \alert{invariant}.
When this occurs, the Markov chain is said to be in \alert{equilibrium}.
Recall that an ergodic (aperiodic, irreducible, positive recurrent) Markov 
chain which starts in \alert{any} probability vector $\yvec$ eventually 
tends to equilibrium. The process of bringing the chain into equilibrium 
from a random starting probability vector in known as \alert{thermalization}.

An ergodic Markov chain is \alert{reversible} if the probability
of going from state $s_i$ to $s_j$ is the same as that for
going from state $s_j$ to $s_i$ once the chain is in \alert{equilibrium}.
Since the probability that a transition from $s_i$ to $s_j$ occurs is 
the probability $w_i$ of finding the chain in state $s_i$ in 
equilibrium times the transition probability $p_{ij}$, then
reversibility occurs when $w_i p_{ij}=w_j p_{ji}$.
The above condition is often referred to as \bfalert{detailed balance}.
Note that detailed balance guarantees the fixed-point condition:
since $\sum_j p_{ij}=1$ then 
         \beq \sum_j w_j p_{ji} = \sum_j w_i p_{ij} = w_i. \eeq

Since an irreducible aperiodic Markov chain with positive recurrent
states in equilibrium is a stationary stochastic process, we can
simply adapt the Monte Carlo integration formulas for stationary
stochastic processes.  Hence, the Monte Carlo method of integration 
using a Markov chain in equilibrium is specified by
    \beqs & \displaystyle\int_{\cal V} p(\vec{x})\ f(\vec{x})\ d^Dx
    \approx \langle f\rangle \pm \sqrt{\frac{R_0(f)+2\sum_{h\geq 1}R_h(f)
    }{N}},\\
   &\displaystyle \langle f\rangle \equiv \frac{1}{N}\!\sum_{i=1}^N
     f(\vec{x}_i),\quad
    R_h(f) \equiv \frac{1}{N\!-\!h}\sum_{i=1}^{N-h}
    \Bigl(f(\vec{x}_i)\!-\!\langle f\rangle\!\Bigr)\Bigl( f(\vec{x}_{i+h})
     \!-\!\langle f\rangle\! \Bigr), \eeqs
where the $N$ points $\vec{x}_1,\dots,\vec{x}_N$ in the $D$-dimensional 
volume ${\cal V}$ are elements of an irreducible aperiodic Markov chain with 
positive recurrent states and \alert{stationary} (and limiting) probability 
distribution $p(\vec{x})$ throughout $D$-dimensional volume ${\cal V}$.
Note that the Markov chain must be in equilibrium, and as usual, the
stationary probability distribution must satisfy the
normalization condition $\int_{\cal V} p(\vec{x}) d^Dx=1$.  The autocovariance
must be absolutely summable $\sum_{h=0}^\infty \vert R_h(f)\vert <\infty$.

Once again, let us pause for some reflection.  We have seen that
multi-dimensional integrals can be estimated using the Monte Carlo
method, but importance sampling is often crucial for obtaining
estimates with sufficiently small statistical uncertainty, especially
when the integrand is peaked in one or more regions.  The rejection
method can be used in one or few dimensions, but is difficult or
impossible to apply when the dimensionality of the integration becomes
large.  In such cases, the use of a stationary stochastic process
is often our only option.  A particularly useful type of stationary
stochastic process is an ergodic (positive-recurrent, aperiodic, and 
irreducible) Markov chain in equilibrium.  The amazing fundamental 
limit theorem for ergodic Markov chains tells us that such a Markov
chain has a unique stationary distribution which is also the
limiting distribution.  Hence, we can start the chain with any
initial probability vector and are guaranteed that the probability 
vector will eventually evolve into the required stationary vector.
The uniqueness of the stationary vector and the coincidence of the
stationary vector with the limiting vector make ergodic Markov chains
especially useful for Monte Carlo applications.

Points generated by a Markov process depend on previous elements in the 
chain; as stated earlier, this dependence known as \alert{autocorrelation}.
This autocorrelation depends on the observable (integrand) being
estimated.  For any observable (integrand) $O_i$, the autocorrelation 
$\varrho(\tau)$ is defined by
    \beq \frac{ \langle O_i O_{i+\tau} \rangle - \langle O_i\rangle^2}{
    \langle O_i^2\rangle - \langle O_i\rangle^2}.\eeq
Highly correlated points yield an autocorrelation value near unity;
independent points produce a value near zero.  Decreasing autocorrelations
decreases the Monte Carlo error, as can be seen from the error formula
above.  Usually the dependence decreases as the number of steps between
elements in the chain increases, so a simple way to decrease autocorrelations
is to not use every element in the chain for ``measurements", and instead
skip some number of elements between measurements.

\subsection{The Metropolis-Hastings method}
We generally know the probability density $\pi(\phi)$ that we need to 
sample to evaluate the integral $\int O(\phi)\pi(\phi){\cal D}\phi$,
where $\phi$ represents a vector of integration variables and the
observable $O(\phi)$ is some function of the $\phi$.
For our path integrals, we need to generate paths with 
a probability distribution
    \beq \pi(\phi)= \frac{e^{-S[\phi]/\hbar}}{
     \int {\cal D}\phi^\prime\ e^{-S[\phi^\prime]/\hbar}},\eeq
where $S(\phi)$ is usually a real-valued action.
In the imaginary time formalism, this path integral weight is real and 
positive, allowing a probability interpretation to facilitate importance
sampling in the Monte Carlo method.  In order to sample the
probability density $\pi(\phi)$, we need to construct a Markov
chain whose limiting stationary distribution is $\pi(\phi)$.
But how do we construct the Markov transition matrix 
$P(\widetilde{\phi}\leftarrow\phi)$?  

There are several answers to this question, but we shall focus only
on the simplest answer here:  the \bfalert{Metropolis-Hastings} 
method\cite{metropolis,hastings}.
This method is very simple and very general.  It also has
the advantage that the probability normalization never enters into
the calculation.  Its disadvantage is the presence of strong autocorrelations
since only updates which change the action by a small amount are
allowed.

To describe this method, let us first change to a quantum mechanical
notation of putting earlier states on the right, later states on the left.
The Metropolis-Hastings algorithm uses an auxiliary \alert{proposal} 
density $R(\widetilde{\phi}\leftarrow\phi)$ which\\
\hspace*{4mm}$\bullet$ must be normalized,\\
\hspace*{4mm}$\bullet$ can be evaluated for all $\phi,\widetilde{\phi}$,\\
\hspace*{4mm}$\bullet$ can be easily sampled,\\
\hspace*{4mm}$\bullet$ and needs no relationship to the fixed-point probability 
   density $\pi(\phi)$.\\
Given this proposal density, the Metropolis-Hastings method updates 
the Markov chain $\phi\rightarrow\widetilde{\phi}$ as follows:
    \begin{enumerate}
      \item Use $R(\widetilde{\phi}\leftarrow\phi)$ to propose a new value 
       $\widetilde{\phi}$ from the current value $\phi$.
      \item Accept the new value with probability
         \beq P_{\rm acc}(\widetilde{\phi}\leftarrow\phi)
        =\min\left(1,\frac{R(\phi\leftarrow\widetilde{\phi})\pi(\widetilde{\phi})}{
        R(\widetilde{\phi}\leftarrow\phi)\pi(\phi)}\right).\eeq
      \item If rejected, the original value $\phi$ is retained.
      \end{enumerate}
A rule of thumb is to tweak any parameters in the proposal density
to obtain about a $50\%-60\%$ acceptance rate.  A higher acceptance
rate might indicate that the proposal density is exploring the 
integration volume too slowly, whereas a lower acceptance rate
might indicate that too much computer time is being wasted attempting
updates that get rejected.
If the proposal density satisfies reversibility
$R(\widetilde{\phi}\leftarrow\phi)=R(\phi\leftarrow\widetilde{\phi})$, 
then the acceptance probability reduces to 
     $\min(1,\pi(\widetilde{\phi})/\pi(\phi))$,
which is known as the Metropolis method.

The Metropolis-Hastings method produces a Markov chain which
satisfies detailed balance.  
\begin{proof}
 The (normalized) transition probability density is
    \beqs  W(\widetilde{\phi}\leftarrow\phi)
      &=&P_{\rm acc}(\widetilde{\phi}\leftarrow\phi)
       R(\widetilde{\phi}\leftarrow\phi)\\
       &+& \delta(\widetilde{\phi}-\phi)
       \left(1-\int\!{\cal D}\overline{\phi}
   \ P_{\rm acc}(\overline{\phi}\leftarrow\phi)
       R(\overline{\phi}\leftarrow\phi)\right).\eeqs
Define
     \beqs
       A(\widetilde{\phi}\leftarrow\phi)&\equiv &
       P_{\rm acc}(\widetilde{\phi}\leftarrow\phi)
       R(\widetilde{\phi}\leftarrow\phi)\pi(\phi),\\
       &=& \min\left(1,\frac{R(\phi\leftarrow
   \widetilde{\phi})\pi(\widetilde{\phi})}{
       R(\widetilde{\phi}\leftarrow\phi)\pi(\phi)}\right)
       R(\widetilde{\phi}\leftarrow\phi)\pi(\phi),\\
       &=& \min\left(R(\widetilde{\phi}\leftarrow\phi)\pi(\phi),
       \ R(\phi\leftarrow\widetilde{\phi})\pi(\widetilde{\phi})\right),
        \eeqs
 where the last line follows from 
       $R(\widetilde{\phi}\leftarrow\phi)\pi(\phi)\geq 0$.  Note that
 this quantity is symmetric: $A(\widetilde{\phi}\leftarrow\phi)
 =A(\phi\leftarrow\widetilde{\phi})$. So we have
   \beqs
  W(\widetilde{\phi}\leftarrow\phi)\pi(\phi)
  &=& P_{\rm acc}(\widetilde{\phi}\leftarrow\phi)
R(\widetilde{\phi}\leftarrow\phi)\pi(\phi)\\
&+& \delta(\widetilde{\phi}-\phi)
 \left(1-\int\!{\cal D}\overline{\phi}
\ P_{\rm acc}(\overline{\phi}\leftarrow\phi)
 R(\overline{\phi}\leftarrow\phi)\right)\pi(\phi),\\
&=&A(\widetilde{\phi}\leftarrow\phi)+\delta(\widetilde{\phi}-\phi)
\Bigl(\pi(\phi) -\int\!{\cal D}\overline{\phi}
\ A(\overline{\phi}\leftarrow\phi)
\Bigr),\\
&=&A(\widetilde{\phi}\leftarrow\phi)+\delta(\widetilde{\phi}-\phi)
\ K(\phi),
\eeqs
where 
  \beq K(\phi)=\pi(\phi) -\int\!{\cal D}\overline{\phi}
   A(\overline{\phi}\leftarrow\phi).\eeq
Given the symmetry of $A$ and the Dirac $\delta$-function, then
  detailed balance holds:
  \beq W(\widetilde{\phi}\leftarrow\phi)\pi(\phi)
  =W(\phi\leftarrow\widetilde{\phi})\pi(\widetilde{\phi}),\eeq
as was to be shown.
\end{proof}

\begin{figure}[t]
\begin{center}
  \includegraphics[width=2.5in,bb=3 23 523 532]{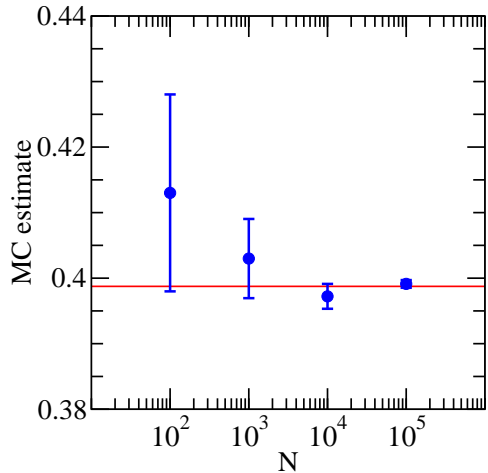}
\end{center}
\caption{Monte Carlo estimates of the ratio of integrals
 $\int_{-\infty}^\infty g(x)h(x)dx\ /  \int_{-\infty}^\infty h(x)dx$, 
 where $g(x)=\cos(\sqrt{1+x^2})$ and $h(x)=e^{-x^2}/(x^2+2)$, against
the number of Markov chain elements $N$ used.  A simple Metropolis method
 was used with $h(x)$ as the sampling probability density.  The
horizontal line indicates the exact answer.
\label{fig:mcint6}}
\end{figure}

Does this really work?  Consider a one dimensional example to 
answer this.  Let $g(x)=\cos(\sqrt{1+x^2})$ and $h(x)=e^{-x^2}/(x^2+2)$.
Notice that $g(x)$ changes sign, but $h(x)\geq 0$ so $h(x)$ is
suitable for importance sampling.  Consider evaluating the ratio of
integrals
  \beq I=\frac{\int_{-\infty}^\infty g(x)h(x)dx}{
   \int_{-\infty}^\infty h(x)dx}=0.3987452\dots,\eeq
using a Markov-chain Monte Carlo method with importance sampling 
density $\dd\pi(x)=Z^{-1}h(x)$, where  $Z=\int_{-\infty}^\infty h(x)dx$.
A simple Metropolis implementation would be as follows:
choose a value $\delta$ with uniform probability in the range
            $-\Delta\leq\delta\leq\Delta$, 
propose $\widetilde{x}=x+\delta$ as the next element in the chain, then
accept with probability $\min(1,\pi(\widetilde{x})
        /\pi(x))\!=\!\min(1,h(\widetilde{x})
        /h(x))$.
Some Metropolis estimates for various values of $N$,
the number of random points used, are shown in Fig.~\ref{fig:mcint6}.
A value of $\Delta=1.5$ was found to yield an acceptance rate near
$50\%$.  Note that we never needed to evaluate $Z$.  The horizontal
line is the exact answer.  One sees that the method really does work.

\section{Monte Carlo study of the simple harmonic oscillator}
\label{sec:sho}

As a first simple example, let us apply the Monte Carlo method
to evaluate path integrals in the one-dimensional simple harmonic
oscillator.  The action in the imaginary time formalism is given by
    \beq S[x(\tau)]=\int_{\tau_a}^{\tau_b}\! d\tau\ \left(
    \textstyle\frac{1}{2}m\dot{x}^2+\frac{1}{2}m\omega^2 x^2\right).\eeq
To carry out a Monte Carlo evaluation, it is necessary to discretize
time $N\varepsilon=\tau_b-\tau_a$:
    \beq \frac{S}{\hbar}=
    \frac{m\varepsilon}{2\hbar}\!\sum_{j=0}^{N-1}\left[
    \left(\frac{x_{j+1}\!-\!x_j}{\varepsilon}\right)^2
    \!+\!\omega^2\left(\frac{
    x_{j+1}\!+\!x_j}{2}\right)^2\right],\eeq
where $\varepsilon$ should be chosen so discretization errors are 
sufficiently small. Introduce the dimensionless parameters
    \beq x_k = d_k \sqrt{\frac{\varepsilon \hbar}{m}},\qquad
    \kappa = \frac{1}{4}\varepsilon^2\omega^2, \eeq
so that the action can be written
    \beq \frac{S}{\hbar}=\frac{1}{2}\!\sum_{j=0}^{N-1}\left[
     (d_{j+1}\!-\!d_j)^2\!+\!\kappa(d_{j+1}\!+\!d_j)^2\right],\eeq
and with a few more manipulations, the action becomes
     \beq \frac{S}{\hbar}=\frac{1}{2}(1+\kappa)(d_0^2+d_N^2)+
      (1\!+\!\kappa)\!\left[\sum_{j=1}^{N-1} d_j^2\right]
     -(1\!-\!\kappa)\!\left[\sum_{j=0}^{N-1}d_jd_{j+1}\right].\eeq
The first constant is irrelevant, so it can be discarded, 
          then one last rescaling
     \beq u_j = d_j\sqrt{1+\kappa},\quad g=\frac{1-\kappa}{1+\kappa},
       \quad d_0=d_N=0,\eeq
yields the final form for the action:
    \beq  \frac{S}{\hbar}=\left[\sum_{j=1}^{N-1} u_j^2\right]
      -g\!\left[\sum_{j=0}^{N-1}u_ju_{j+1}\right].\eeq
In this form, we have set $u_0=u_N=0$, which is tantamount to
requiring $x_a=x_b=0$.  The observables we will compute will be
independent of the choice of the initial $x_a$ and final $x_b$
locations of the particle.  A given path is specified by a 
vector $\uvec$ whose $N-1$ components are $u_j$ for $j=1,2,\dots,N-1$.

We must now devise an auxiliary proposal density in order to
produce a Markov chain using the Metropolis-Hastings method.  There is
considerable freedom in designing such a proposal.  If
we use an auxiliary proposal that simultaneously changes all
$N-1$ components of $\uvec$, one finds that the resulting
changes to the action are rather large, and the acceptance
probability becomes nearly zero.  In order to get a reasonable
acceptance rate, we must make only small changes to the action.
This can be accomplished most easily if we only change \alert{one of
the $u_j$ at a time}.  The most natural way to proceed is to
randomly pick one time slice and perform a local update of that
time slice.  If equal probabilities are assigned to each time 
slice, then detailed balance is maintained and covering the entire
hypervolume of integration is ensured.

A simple procedure for updating the path $\uvec\rightarrow \uvec^{\rm new}$ 
is as follows:
\begin{enumerate}
\item Randomly choose an integer $j$ from 1 to $N_t-1$, where
      $N_t$ is the number of time slices, with equal probability.
\item Propose a random shift $u_j\rightarrow \tilde{u}_j=u_j+\delta$ 
      with $\delta$ chosen with uniform probability
      density in the range $-\Delta\leq\delta\leq\Delta$.
\item Calculate the change $\delta S$ in the action:
       \beq \delta S/\hbar = \delta
     \ \left(\delta+2u_j-g(u_{j-1}+u_{j+1})\right). \eeq
\item Since $R(\tilde{u}_j\leftarrow u_j)=R(u_j\leftarrow \tilde{u}_j)$,
      then accept the proposed value $u_j^{\rm new}=\tilde{u}_j$ with 
      probability
           $\min(1,e^{-\delta S/\hbar})$.  If not accepted, then
       retain the old value: $u_j^{\rm new}=u_j$.
\item Repeat the above procedure
      $N_t$ times to constitute one updating \alert{sweep}.
\end{enumerate}
The rule of thumb for setting the value of $\Delta$ is to achieve an
acceptance rate around 50\%.  A lower rate means that too much time
is being wasted with rejections, whereas a higher rate means that the
Markov chain might be moving through the integration hypervolume too slowly.

To start the Markov chain, one can either choose a random path (hot start)
or choose $u_j=0$ for all $j$ (cold start).  One then updates $N_{\rm therm}$
number of sweeps until the fixed point of the chain is reached (thermalization); 
usually, a few simple observables are monitored.  Once the Markov chain is 
thermalized, the ``measurements" can begin.  The parameters in the 
simulation are chosen according to the following guidelines:
\begin{itemize}
\item  Choose $\varepsilon$ so that discretization errors 
 are sufficiently small.
\item Choose $\Delta$ for an adequate acceptance rate.
\item Choose the number of sweeps $N_{\rm sweeps}$ 
between measurements to achieve sufficiently small autocorrelations.
\item Choose the number of measurements $N_{\rm meas}$ 
to achieve the desired precision in the results.
\end{itemize}

\begin{figure}[t]
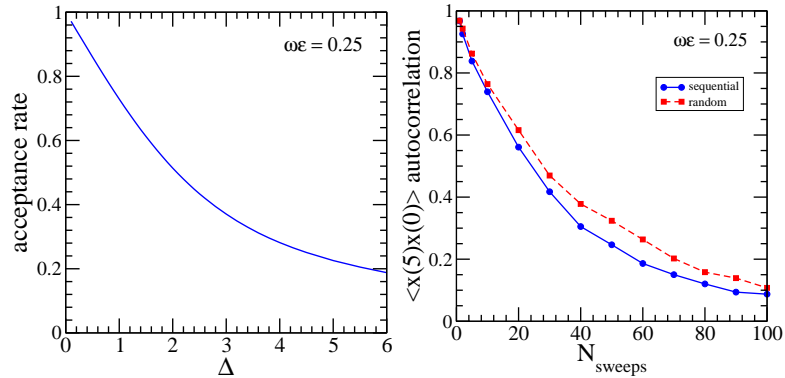

 \begin{center}
 \includegraphics[width=2.0in,bb=18 34 528 531]{acc_rate}
 \includegraphics[width=2.0in,bb=16 16 543 531]{auto_cor}
\end{center}
\caption{Left: Metropolis acceptance rate versus the parameter $\Delta$
for the simple harmonic oscillator with $\omega\varepsilon=0.25$.  
Right: Autocorrelation function
associated with $\langle u(\tau_0+5)x(\tau_0)\rangle$ against the number of 
Metropolis
sweeps for updating the time slices $u_j$ sequentially (solid line) and
randomly (dashed).  When updating the time slices in random order, a sweep
refers to $N_t$ successive local updates, where $N_t$ is the number of 
time slices.
\label{fig:mcsho1}}
\end{figure}

\begin{figure}[t]
 \begin{center}
 \includegraphics[width=3.5in,bb=70 29 523 614]{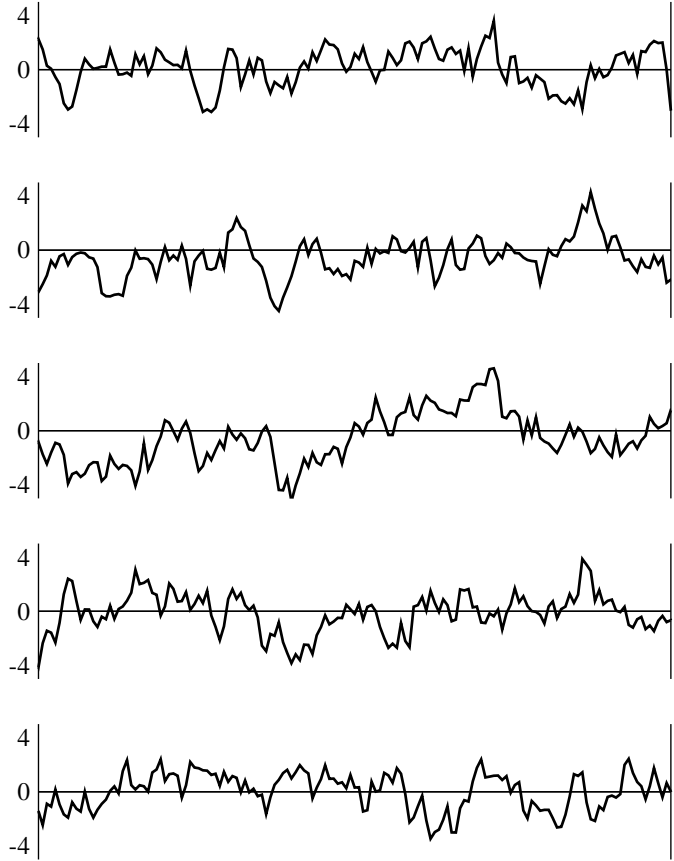}
\end{center}
\caption{A few paths $u_j$ for $j=400-550$ with $N_t=1000$ from an
actual Monte Carlo simulation of the simple harmonic oscillator
with $\omega\varepsilon=0.25$.
\label{fig:mcsho2}}
\end{figure}

\begin{figure}[t]
 \begin{center}
 \includegraphics[width=3.5in,bb=9 34 535 523]{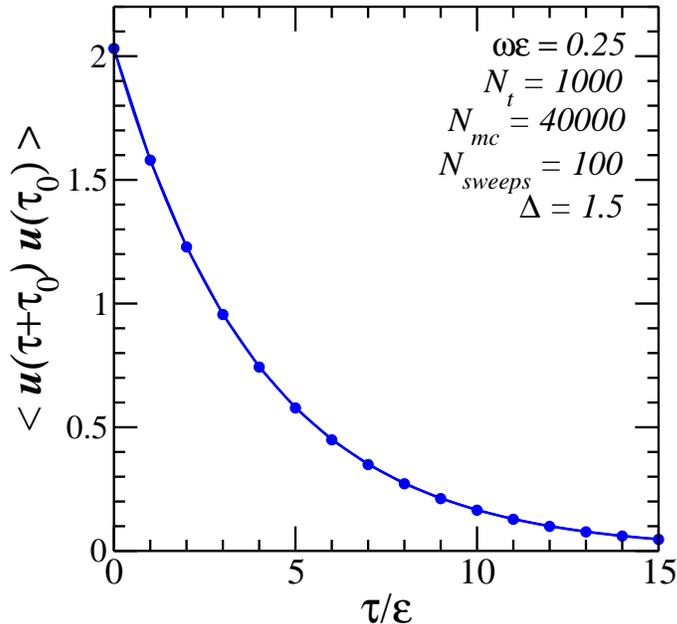}
\end{center}
\caption{Monte Carlo estimates (circles) of the
correlation function $\langle u(\tau_0+\tau)u(\tau_0)\rangle$ are
compared to the exact results (solid line) for the simple harmonic
oscillator with $\omega\varepsilon=0.25$.
\label{fig:mcsho3}}
\end{figure}

The results of some actual Monte Carlo computations using
$\omega\varepsilon=0.25$ with $N_t=1000$ time slices are shown in 
Figs.~\ref{fig:mcsho1}-\ref{fig:mcsho3}.
The Metropolis acceptance rate is shown in the left-hand plot in 
Fig.~\ref{fig:mcsho1}.  To get an acceptance rate around 0.5-0.6, one sees
that $\Delta=1.5$ is a good choice for $\omega\varepsilon=0.25$.  
Autocorrelations were monitored using a typical observable chosen to 
be $\langle u(\tau_0+5)u(\tau_0)\rangle$, where $\tau_0$ is taken
near the midpoint between the path end-points (we actually averaged over
a large number of $\tau_0$ values in the middle region for increased 
statistics).  The autocorrelation function for this observable is shown 
in the right-hand plot of Fig.~\ref{fig:mcsho1} (dashed line). 
One sees that about 100 sweeps are needed to reduce autocorrelations 
down to the level of 0.1.

Updating the path one $u_j$ at a time in random order with each
$j$-value being equally likely to be chosen ensures
detailed balance and coverage of the entire integration volume.
But a simpler method would be to just sequentially sweep through
the $u_j$, updating each $u_j$ one at a time.  With
much less calls to the random number generator, sequential sweeps
would take less computer time.  Coverage of the entire integration
volume is again ensured, but detailed balance is lost.  However,
detailed balance was useful only in that it ensured that the 
ergodic Markov chain had a unique fixed-point.  Even though
detailed balance is lost when sequentially sweeping through the
time slices, the fixed-point condition is maintained.  Thus, sequential
sweeps are totally acceptable.  When updating randomly chosen
$u_j$, the Markov matrix $\Pmat$ is the same for each local 
update. When updating $u_j$ sequentially, the Markov matrix is
different for each local update.  However, the Markov matrix
for an entire sweep $\Pmat_{\rm sweep}$, being the product of the
Markov matrices for each time slice, is the same for each sweep.
So the theoretical foundations described previously still apply,
as long as we apply them to $\Pmat_{\rm sweep}$.  The autocorrelation
function for the observable $\langle u(\tau_0+5)u(\tau_0)\rangle$ when updating
time slices sequentially is shown as a solid line in the right-hand
plot of Fig.~\ref{fig:mcsho1}.  One sees that there is no adverse
affect on the autocorrelations.

Portions of paths produced in an actual Markov chain with
sequential time-slice updating for 
$\omega\varepsilon=0.25, N_t=1000, \Delta=1.5$ are shown in 
Fig.~\ref{fig:mcsho2}.  Monte Carlo estimates for the correlation
function $\langle u(\tau_0+\tau)u(\tau_0)\rangle\rangle$ as a function
of $\tau$ are compared to exact results in Fig.~\ref{fig:mcsho3}.

\section{Monte Carlo calculations in real scalar field theory in 2+1 dimensions}
\label{sec:phi4}

The action for a real scalar field in continuous Euclidean $D$-dimensional 
space-time in the imaginary time formalism is given by
   \beq
    S = \int\! d^Dx\ \left(\frac{1}{2}\partial_\mu\varphi(x)\partial_\mu\varphi(x)
    +\frac{1}{2}m^2\varphi(x)^2 + \frac{g}{4!}\varphi(x)^4\right). 
   \eeq
Note that the action must be dimensionless in natural units $\hbar=c=1$,
and since $m$ has units of a derivative $\partial_\mu$, that is, of a mass,
then the field must have dimension $[\varphi]=[m]^{\frac{1}{2}D-1}$,
requiring that the coupling $g$ has units $[g]=[m]^{4-D}$.  Thus, the
coupling is dimensionless in 4 space-time dimensions, but has units of mass
in 3 space-time dimensions, leaving $g/m$ dimensionless.  We require
$g\geq 0$ or the action will have no minimum.

Quantization of this field theory can be accomplished using path integrals,
but the notion of a ``path" must be generalized: a path here is a field 
configuration in both space and time.  The path integral now consists
of integrations over all field configurations.  For a real scalar field, 
we now have an integral $-\infty \leq \varphi(x)\leq \infty$ at every 
space-time point $x$.  The time-ordered two-point function is given by
    \beq
     \langle T\varphi(x_1)\varphi(x_2)\rangle = \frac{\int {\cal D}\varphi\ \varphi(x_1)
     \varphi(x_2) \exp(-S[\varphi])}{\int {\cal D}\varphi\ \exp(-S[\varphi])},
     \eeq
which generalizes to $n$-point functions, time-ordered product of $n$ fields,
in a straightforward manner.

A Monte Carlo study requires an action on a space-time lattice.
We will use an anisotropic cubic lattice with temporal lattice spacing $a_t$ 
and spatial lattice spacing $a_s$.  Discretization of the action is
achieved by replacing field derivatives by the simplest finite differences,
and integrals over space-time by suitable summations over space-time
lattice sites.  The action is given by
   \beqs
    S &=& a_s^{D-1}a_t\sum_x \!\left(\!\!\sum_\mu\frac{(
       \varphi(x\!+\!a_\mu\hat{\mu})\!-\!\varphi(x))^2}{2a_\mu^2}
   \!+\!\frac{1}{2}m^2\varphi(x)^2 \!+\! \frac{g}{4!}\varphi(x)^4\right), \\
    &=& a_s^{D-1}a_t\!\!\sum_x\! \!\left(\!\!-\!\!
  \sum_\mu\!\frac{\varphi(
      x\!+\!a_\mu\hat{\mu})\varphi(x)}{a_\mu^2}
   \!\!+\!\!\frac{1}{2}\!\left(\!\!m^2\!\!+\!\!\sum_\nu\!\frac{2}{a_\nu^2}
    \!\!\right) \!\!\varphi(x)^2
   \!\! +\!\! \frac{g}{4!}\varphi(x)^4\!\!\right),
    \eeqs
where $a_\mu$ is the lattice spacing in the $\mu$ direction.
Redefine the field $\sqrt{a_s^{D-3}a_t}
       \ \varphi(x) = \sqrt{2\kappa_s}\ \phi(x)$,
where $\kappa_s$ is a dimensionless number, so the new field $\phi(x)$ is
dimensionless.  Then introduce a few more dimensionless parameters:
   \beqs && a_s/a_t=\zeta, \quad \lambda = \frac{g\zeta\kappa_s^2}{6a_s^{D-4}},\\
    && \kappa_s(a_s^2m^2+2\zeta^2+2D-2)=1-2\lambda, \quad
     \kappa = \zeta\kappa_s, \eeqs
to obtain the final form for the lattice action:
   \beqs
      S &=& \sum_x \biggl(-\frac{2\kappa}{\zeta}\sum_{j=1}^{D-1}\phi(x)\phi(x\!+\!a_s\hat{j})
        -2\kappa\zeta\,\phi(x)\phi(x\!+\!a_t\hat{t})\\[-9pt]
     && \qquad\qquad +(1-2\lambda)\phi(x)^2 + \lambda\phi(x)^4\biggr).
    \eeqs
The hopping parameter $\kappa$ essentially sets the mass parameter, 
and $\lambda\geq 0$ is the interaction strength.  In what follows,
we shall focus solely on the above theory in 3 space-time dimensions.

\subsection{Exact results in free field limit $\lambda=0$}
The free field theory $\lambda=0$ is exactly solvable.  In this case, the
path integrals are multivariate gaussians.  The free action can be written 
in the form
     \beq S[\phi] = {\textstyle\frac{1}{2}}\sum_{xy}\phi(x) M(x,y)\phi(y).\eeq
For $N$ lattice sites, $M$ is a real and symmetric $N\times N$ matrix
having positive eigenvalues and given by
   \beqs
     M(x,y) &=&   -\frac{2\kappa}{\zeta}\sum_{j=1}^{D-1}\!\!
     \left(\delta(y,x\!+\!a_s\hat{j})+\delta(x,y\!+\!a_s\hat{j})\right)\\
    && -2\kappa\zeta\, \left(\delta(y,x\!+\!a_t\hat{t})+ \delta(x,y\!+\!a_t\hat{t})\right)
    + 2\delta(x,y).
    \eeqs
The path integrals encountered in the free field theory can be evaluated using
the so-called $J$-trick: use derivatives with respect to an external source $J_k$, followed 
by the limit $J_k\rightarrow 0$, to evaluate all integrals involving any 
number of products of the fields:
     \beqs
      &&\prod_{i=1}^N\left(\int_{-\infty}^\infty\!\! d\phi_i\right)
       \phi_{m_1}\phi_{m_2}\dots\phi_{m_r}
     \  \exp( -{\textstyle\frac{1}{2}}\phi_j M_{jk}\phi_k)\\ &=& 
    \lim_{J_n\rightarrow 0} \frac{\delta}{\delta J_{m_1}} 
   \cdots \frac{\delta}{\delta J_{m_r}} 
       \prod_{i=1}^N\left(\int_{-\infty}^\infty\!\! d\phi_i\right)
     \  \exp( -{\textstyle\frac{1}{2}}\phi_j M_{jk}\phi_k +J_n\phi_n),\\
    &=&\lim_{J_n\rightarrow 0}
    \frac{\delta}{\delta J_{m_1}} \cdots \frac{\delta}{\delta J_{m_r}} 
      \left(\det\left(\frac{M}{2\pi}\right)\right)^{-1/2}\!\! \exp\left(\textstyle\frac{1}{2}
      J_j M^{-1}_{jk} J_k\right).
   \eeqs 
This trick does the Wick contractions automagically!

The two-point function is given by $\langle T\phi(x_1)\phi(x_2)\rangle 
= M^{-1}(x_1,x_2)$.  The inverse of $M$ can be obtained
by the method of Green functions and using Fourier transforms.
For an $L_x\times L_y\times L_t$ lattice, the result is
   \beq
    M^{-1}(x,y)=\frac{\zeta}{2\kappa L_x L_y L_t}\sum_{k_\mu}  
    \frac{\cos(k\!\cdot\! (x\!-\!y))}{  (a_s^2m^2
    +4\sum_{j=1}^2\sin^2({\textstyle\frac{1}{2}}k_j)
    +4\zeta^2\, \sin^2({\textstyle\frac{1}{2}}k_t))},
   \eeq
where $k_\mu = 2\pi n_\mu/L_\mu$ for $n_\mu=0,1,2,\dots,L_\mu-1$.
The pole in the two-point function gives the energy $a_t E_p$ of a 
single particle of momentum $a_sp$:
   \beq   a_tE_p
        =2\sinh^{-1}\left(\frac{1}{2\zeta}\sqrt{a_s^2m^2
          +4\sin^2({\textstyle\frac{1}{2}}a_sp_x)
          +4\sin^2({\textstyle\frac{1}{2}}a_sp_y)}\right).\eeq
For small $a_t,a_s$, this becomes $E_p=\sqrt{m^2+p_x^2+p_y^2}$.
The spectrum is the sum of free particle energies.

\subsection{Metropolis updating}

The Metropolis-Hastings method is useful only when the auxiliary
proposal density leads to a reasonable acceptance rate, typically
around 0.5.  If we simultaneously change field values at all lattice
sites, the value of the action most likely changes by a large amount,
and the Metropolis-Hastings acceptance probability plummets to zero.
However, if we propose a change only to the field value on one site,
a reasonable acceptance rate can be achieved.  To maintain detailed
balance and ensure coverage of the entire integration region, the
site to be updated should be chosen randomly, with each site being equally
likely to be selected.  However, as in the case of the simple harmonic
oscillator, one finds that updating the fields at sites selected 
sequentially,
sweeping through the lattice, works just as well.  Although detailed
balance is lost, the crucial fixed-point condition is maintained,
and by sequentially visiting every site, coverage of the entire
integration region is ensured.

Thus, we shall use an auxiliary proposal probability density
that \alert{sweeps} through the lattice, visiting each lattice site
sequentially, updating each and every site one at a time.  In the
battle against autocorrelations, we expect that such a local
updating scheme should be effective in treating the small wavelength 
modes of the theory, but the long wavelength modes may not be dealt 
with so well.

Recall that the action is
   \beqs
      S &=& \sum_x \biggl(-\frac{2\kappa}{\zeta}\sum_{j=1}^{D-1}
      \phi(x)\phi(x\!+\!a_s\hat{j})
        -2\kappa\zeta\,\phi(x)\phi(x\!+\!a_t\hat{t})\\
     && \qquad\qquad +(1-2\lambda)\phi(x)^2 + \lambda\phi(x)^4\biggr).
    \eeqs
Define the neighborhood $N(x)$ of the site $x$ by
  \beq
     N(x)=-\frac{2\kappa}{\zeta}\sum_{j=1}^{D-1}\Bigl(\phi(x+a_s\hat{j})
   +\phi(x-a_s\hat{j})\Bigr)
     -2\kappa\zeta\,\Bigl(\phi(x+a_t\hat{t})+\phi(x-a_t\hat{t})\Bigr).
    \eeq
If the field at the one site $x$ is changed $\phi(x)\rightarrow\phi(x)+\Delta$,
      then the change in the action is
     \beq
     \delta S = \Delta\biggl( N(x)+(\Delta+2\phi(x)) 
      \biggl(1+\lambda\Bigl( (\Delta+2\phi(x)) \Delta +2(\phi(x)^2-1) \Bigr)
        \biggr)\biggr).
     \eeq
This change in the action can also be written
\beqs
\delta S &=& \Delta\,( a_0 + a_1\Delta + a_2\Delta^2+ a_3\Delta^3),\\
     a_0 &=& N(x)+2\phi(x) (1+2\lambda(\phi(x)^2-1)),\\
     a_1 &=& 1 + 2\lambda (3\phi(x)^2-1),\\
     a_2 &=& 4\lambda\phi(x),\\
     a_3 &=& \lambda.
\eeqs
Single-site updates involve a single continuous real variable $\phi$.
A simple proposal density is then
    \beq
       R(\widetilde{\phi}\leftarrow\phi)=\left\lbrace
        \begin{array}{l@{\hspace{1cm}}l} \displaystyle\frac{1}{2\Delta_0},   
        & -\Delta_0\leq(\widetilde{\phi}-\phi)
         \leq\Delta_0,\\[10pt]
        0, & \vert\widetilde{\phi}-\phi\vert > \Delta_0.\end{array}
       \right.
     \eeq
In other words, a value for $\Delta$ is chosen randomly with uniform
probability density in the range $-\Delta_0\leq \Delta\leq\Delta_0$,
and the proposed value for the field is $\phi(x)+\Delta$.
The width $\Delta_0$ is chosen to obtain an acceptance rate around 50\%.
The proposed new value is accepted with probability $\min(1,\exp(-\delta S))$.
If rejected, the current field value is retained.  This single-site
procedure is repeated at every site on the lattice, sequentially sweeping
through the lattice.

\subsection{Microcanonical updating}

When the single particle mass $a_t m_{\rm gap}$ is small, the
\alert{coherence length} $\xi=1/(a_t m_{\rm gap})$ becomes large.
The so-called continuum limit of the theory is reached when
the coherence length is large compared to the lattice spacing,
that is, when $\xi\rightarrow\infty$.  However, $\xi\rightarrow\infty$
only occurs near a second order phase transition (a critical point).
One finds that autocorrelations with the above Metropolis updating scheme
become long ranged as $\xi$ becomes large; this is known as \alert{critical
slowing down}.  Autocorrelations are problematic even for $\xi\approx 5$ 
with the above Metropolis updating.  In such cases, we will need to use
some other procedure to better update the long wavelength modes.

Long wavelength modes are associated with lower frequencies and lower 
energies. In other words, long-wavelength modes are associated with very 
small changes to the action.  A possible way to improve autocorrelations 
is to make \alert{large} but \alert{action preserving} $\delta S=0$ changes
to the field at one site.  We shall refer to this as a \alert{microcanonical}
update, but such schemes are often referred to as \alert{overrelaxation}
in the literature.  Local updating is so easy, we do not want to give up on
it yet!  In devising our microcanonical updating scheme, we must ensure
that the fixed-point of our Markov chain is unaffected.  Note that
microcanonical updating cannot be used just by itself since it does not
explore the entire integration region.  Microcanonical updating must
be used in combination with some scheme that does cover the entire
integration volume, such as the above-described Metropolis sweeps.
 
To facilitate the discussion of such microcanonical updating, let us
first revisit the Metropolis-Hastings method, examining the case of
a sharply-peaked proposal probability density.  Suppose $f(\phi)$ is
a well-behaved, single-valued, invertible function, then consider a 
proposal density given by a Breit-Wigner peaked about $f(\phi)$:
\[ R_f(\widetilde{\phi}\leftarrow\phi)
 = \frac{1}{\pi}\frac{\varepsilon}{\Bigl(\widetilde{\phi}-f(\phi)\Bigr)^2
+\varepsilon^2},
\]
where $\varepsilon$ is a constant.  Notice that this probability density
is properly normalized: 
\[\int_{-\infty}^\infty\! d\widetilde{\phi}\ R_f(
\widetilde{\phi}\leftarrow\phi)=1.\]
With such a proposal density, the standard choice for the acceptance 
probability which satisfies detailed balance is given, as usual, by
     \beq P_{\rm acc}(\widetilde{\phi}\!\leftarrow\!\phi)
       \!=\!\min\!\!\left(\!\!1,\!\frac{R_f(\phi\!\leftarrow\!
        \widetilde{\phi})\pi(\widetilde{\phi})}{
       R_f(\widetilde{\phi}\!\leftarrow\!\phi)\pi(\phi)}\!\!\right)
      \!\!=\!\min\!\!\left(\!\!1,\!\frac{\Bigl(\!(\widetilde{\phi}\!-\!f(\phi))^2
       \!+\!\varepsilon^2\!\Bigr)\pi(\widetilde{\phi})}{
      \Bigl(\!(\phi\!-\!f(\widetilde{\phi}))^2\!+\!\varepsilon^2\!\Bigr)\pi(\phi)}
       \!\!\right).\eeq

As $\varepsilon$ becomes very small, the proposal density becomes very sharply
peaked about $\widetilde{\phi}=f(\phi)$.  In fact, we are interested in
taking the limit $\varepsilon\rightarrow 0$ to obtain a Dirac 
$\delta$-function:
\[ \delta(x)=\frac{1}{\pi}\lim_{\varepsilon\rightarrow 0}\frac{\varepsilon}{x^2+
\varepsilon^2}.\]
The probability of proposing a $\widetilde{\phi}$ value between 
$f(\phi)-\varepsilon\leq \widetilde{\phi}\leq f(\phi)+\varepsilon$ is
given by
\[  \int_{f(\phi)-\varepsilon}^{f(\phi)+\varepsilon}\!\! d\widetilde{\phi}
\ R_f(\widetilde{\phi}\leftarrow\phi)=\frac{1}{2}.
\]
However, we need to consider a range of values which tends to a single
value but for which the probability tends
to unity as $\varepsilon\rightarrow 0$.  Clearly, a larger range is
needed.  The probability of proposing a value between
$f(\phi)-\sqrt{\varepsilon}\leq \widetilde{\phi}\leq f(\phi)
+\sqrt{\varepsilon}$ is
\[  \int_{f(\phi)-\sqrt{\varepsilon}}^{f(\phi)+\sqrt{\varepsilon}}
\!\! d\widetilde{\phi}
\ R_f(\widetilde{\phi}\leftarrow\phi)=\frac{2}{\pi}\tan^{-1}\left(\frac{1}{
\sqrt{\varepsilon}}\right),
\]
which does tends to unity as $\varepsilon\rightarrow 0$.  If $f(\phi)$
is more than $\sqrt{\varepsilon}$ away from $\phi$, then the probability
that the transition is actually made is
\[
 \int_{f(\phi)-\sqrt{\varepsilon}}^{f(\phi)+\sqrt{\varepsilon}}
\!\! d\widetilde{\phi}
\ W_f(\widetilde{\phi}\leftarrow\phi)
=  \int_{f(\phi)-\sqrt{\varepsilon}}^{f(\phi)+\sqrt{\varepsilon}}
\!\! d\widetilde{\phi}
\ P_{\rm acc}(\widetilde{\phi}\leftarrow\phi)
R_f(\widetilde{\phi}\leftarrow\phi).
\]
Given that $R_f(\widetilde{\phi}\leftarrow\phi)$ is always positive,
the above integral is given by
\[
\min\left(\frac{2}{\pi}\tan^{-1}\left(\frac{1}{
\sqrt{\varepsilon}}\right),
\ \frac{1}{\pi}\int_{f(\phi)-\sqrt{\varepsilon}}^{f(\phi)+\sqrt{\varepsilon}}
\!\! d\widetilde{\phi}
\frac{\varepsilon\ \pi(\widetilde{\phi})}{
  \Bigl((\phi-f(\widetilde{\phi}))^2+\varepsilon^2\Bigr)\pi(\phi)}\right).
\]
If we write $\widetilde{\phi}=f(\phi)+y$, then the remaining integral
above becomes
\[
\frac{1}{\pi}\int_{-\sqrt{\varepsilon}}^{\sqrt{\varepsilon}}
\!\! dy
\frac{\varepsilon\ \pi(f(\phi)+y)}{
  \Bigl((\phi-f(f(\phi)+y))^2+\varepsilon^2\Bigr)\pi(\phi)}.
\]
Now consider two cases: (a) $f(f(\phi))\neq\phi$, and (b) $f(f(\phi))=\phi$.

For the first case when $f(f(\phi))\neq\phi$, since we are integrating over
such a small range, the series expansion of the integrand about the central
$y=0$ point should approximate the true integrand well, assuming no 
singularities.  Performing this expansion about $y=0$, then integrating, 
one finds a zero probability as $\varepsilon\rightarrow 0$, as long as 
$\pi(f(\phi))/\pi(\phi)$ is finite.  To see this, begin by noting that
the integral has the form
\[
\frac{\varepsilon}{\pi}\int_{-\sqrt{\varepsilon}}^{\sqrt{\varepsilon}}
\!\! dy\frac{(a_0+a_1y+a_2y^2+\dots)}{(\varepsilon^2+b_0+b_1y+b_2y^2
+b_3y^3+\dots)},
\]
where the $a_j,b_j$ are constants.
If the denominator does not become zero anywhere in the integration
range, then the integral can be well approximated by
\begin{eqnarray*}
&&\frac{\varepsilon a_0}{\pi(b_0+\varepsilon^2)}
\int_{-\sqrt{\varepsilon}}^{\sqrt{\varepsilon}}
\!\! dy \Bigl(1+c_1(\varepsilon)y+c_2(\varepsilon)y^2
 +c_3(\varepsilon)y^3+c_4(\varepsilon)y^4+\dots\Bigr)\\
&&=\frac{2\varepsilon^{3/2}a_0}{\pi(b_0+\varepsilon^2)}\Bigl(1
+\textstyle\frac{1}{3}\varepsilon c_2(\varepsilon)
+\frac{1}{5}\varepsilon^2 c_4(\varepsilon)+\cdots\Bigr)
\rightarrow 0\ \mbox{as}\ \varepsilon\rightarrow 0 \ \mbox{if}
\ b_0\neq 0,
\end{eqnarray*}
where the $c_j(\varepsilon)$ tend to finite constants as 
$\varepsilon\rightarrow 0$.

For the second case when $f(f(\phi))=\phi$, more care is needed when
expanding about $y=0$ since the integral has the form
\[
\frac{\varepsilon}{\pi}\int_{-\sqrt{\varepsilon}}^{\sqrt{\varepsilon}}
\!\! dy\frac{(a_0+a_1y+a_2y^2+\dots)}{(\varepsilon^2+b_2y^2
+b_3y^3+b_4y^4\dots)}.
\]
To reproduce the integrand to a good approximation over the
entire integration range, the $b_2y^2$ term must be retained
in the denominator, and the rest of the function can be
expanded about $y=0$:
\[
\frac{\varepsilon}{\pi}\int_{-\sqrt{\varepsilon}}^{\sqrt{\varepsilon}}
\!\! dy\frac{a_0}{(\varepsilon^2+b_2y^2)}\left\{1
+\frac{a_1}{a_0}y+\frac{a_2}{a_0}y^2+ \left(\frac{a_3}{a_0}-
\frac{b_3}{\varepsilon^2}\right)y^3
+\dots\right\}.
\]
For $b_2>0$, then the result of the integration is
\[
\frac{2a_0}{\pi\sqrt{b_2}}\tan^{-1}\left(\sqrt{\frac{b_2}{\varepsilon}}
\right)\left\{ 1 + d_1\sqrt{\varepsilon}+d_2\varepsilon+d_3\varepsilon^{3/2}
+\cdots\right\}.
\]
Hence, the acceptance probability is given in the
limit $\varepsilon\rightarrow 0$ by
\[ P_{\rm acc}=\min\left( 1, \frac{a_0}{\sqrt{b_2}}\right).\]
In this case, $a_0=\pi(f(\phi))/\pi(\phi)$ and $b_2=(f^\prime(f(\phi)))^2$,
where the $^\prime$ indicates the derivative function.
If we differentiate both sides of $f(f(\phi))=\phi$ with respect to $\phi$,
we see that for a self-inverse function,
\begin{equation}
 1=\frac{d}{d\phi}\biggl( f(f(\phi))\biggr)=f^\prime(f(\phi))\ f^\prime(\phi),
\end{equation}
so that for a self-inverse function,
\begin{equation}
 \frac{1}{(f^\prime(f(\phi)))^2}= \left\vert
 \frac{f^\prime(\phi)}{f^\prime(f(\phi))}\right\vert
\qquad\mbox{(self-inverse function)}.
\end{equation}
Taking the limit $\varepsilon\rightarrow 0$, we have a proposal density 
and an acceptance probability given by
\begin{eqnarray}
  R_f(\widetilde{\phi}\leftarrow\phi)
 &=& \delta\Bigl(\widetilde{\phi}-f(\phi)\Bigr),\qquad f(f(\phi))=\phi,\\
P_{\rm acc}(\widetilde{\phi}\leftarrow\phi)
&=&\min\left(1,\frac{\sqrt{\vert f^\prime(\phi)\vert}
 \ \pi(\widetilde{\phi})}{
  \sqrt{\vert f^\prime(\widetilde{\phi})\vert}\ \pi(\phi)}\right).
\end{eqnarray}

Now specialize to the case of microcanonical updating in which the
self-inverse function $f(\phi)$ reflects the field in some way so
as to preserve the action.  Let $S(\phi)$ denote that part of the 
action which involves the field at the site $x$ being updated.  
If $\phi$ is the current value of the field at site $x$, then let 
$f(\phi)$ denote another value of the field for which 
$S(\phi)=S(f(\phi))$ so that $\pi(f(\phi))=\pi(\phi)$. For an 
infinitesimal change $\phi\rightarrow\phi+\delta\phi$, we have
\[
 S(\phi+\delta\phi)=S(f(\phi+\delta\phi)).\]
Expanding both sides,
\begin{eqnarray*}
 S(\phi)+S^\prime(\phi)\delta\phi + O(\delta\phi^2)
 &=& S(f(\phi)+f^\prime(\phi)\delta\phi+O(\delta\phi^2))\\
 &=&S(f(\phi))+S^\prime(f(\phi))\ f^\prime(\phi)\delta\phi
 +O(\delta\phi^2)\\
 &=& S(\phi) + S^\prime(f(\phi))\ f^\prime(\phi)\delta\phi
 +O(\delta\phi^2).
\end{eqnarray*}
Solving this equation order by order in $\delta\phi$ leads to
\[  
 S^\prime(\phi)
 = S^\prime(f(\phi))\ f^\prime(\phi)\quad\rightarrow\quad
 f^\prime(\phi) = \frac{S^\prime(\phi)}{S^\prime(f(\phi))}.
\]
Hence,
\[
  f^\prime(f(\phi)) = \frac{S^\prime(f(\phi))}{
  S^\prime(f(f(\phi)))} = \frac{S^\prime(f(\phi))}{
  S^\prime(\phi)}.
\]
So the proposal and acceptance probability densities are
\beqs
  R_f(\widetilde{\phi}\leftarrow\phi)
 &=& \delta\Bigl(\widetilde{\phi}-f(\phi)\Bigr),
 \qquad f(f(\phi))=\phi,\quad S(f(\phi))=S(\phi),
\label{eq:microA1}\\
P_{\rm acc}(\widetilde{\phi}\leftarrow\phi)
&=&\min\left(1, \left\vert\frac{S^\prime(\phi)}{
 S^\prime(\widetilde{\phi})}\right\vert\right),\qquad
\label{eq:microA2}
 \pi(\phi)= \frac{\exp(-S[\phi])}{\displaystyle
 \int{\cal D}\widetilde{\phi}\ \exp(-S[\widetilde{\phi}])}.
\eeqs

So far, we have examined only the case of applying a single
self-inverse function which leaves the action invariant.
In the $\phi^4$ field theory, the action preserving equation $\delta S=0$
will often have {\em four} solutions.  In other words, more than one
self-inverse action-preserving function are possible.  The
simplest way to proceed is to randomly pick one of the
self-inverse action-preserving functions $f_j(\phi)$ with equal 
probabilities, then apply the accept-reject condition on the 
resulting field value $f_j(\phi)$.

The above procedure {\em always} proposes a change to the field,
then an accept-reject step is applied.  A straightforward modification
of the above procedure is to {\em not always} propose a change.  
The above method can be generalized to include a probability 
$\mu$ of proposing a change; we will find that sometimes we will 
need $\mu<1$ to prevent (damped) oscillations in the autocorrelation 
function.

The summary of our microcanonical updating process is as follows:
\begin{enumerate}
\item
Decide to propose a new field value with probability $\mu$.
If the random decision is to retain the current field value,
the steps below can be skipped.
\item
Given initial value $\phi$ of the field at site $x$, solve 
$\delta S(\phi)=0$. Let $\phi_j$ denote the real solutions which
differ from $\phi$.  These will be the roots of a cubic
polynomial.  Sometimes there will be three such real distinct
solutions, other times there will be only one. (The case
of degenerate solutions is highly unlikely.)
\item
With equal probability, randomly choose one of the $\phi_j$
as the proposed new field value. Let $\widetilde{\phi}$
denote the chosen value.
\item
Accept this value with probability
\[
P_{\rm acc}(\widetilde{\phi}\leftarrow\phi)
=\min\left(1, \left\vert\frac{S^\prime(\phi)}{
 S^\prime(\widetilde{\phi})}\right\vert\right).
\]
If rejected, the original value $\phi$ is retained.
\end{enumerate}
The above procedure is repeated for each site, sequentially sweeping
through the lattice.
All of the above formulas have assumed that $S^\prime(\phi)\neq 0$
and $S^\prime(\phi_j)\neq 0$.  In the extremely unlikely case that any 
such minima, maxima, or inflection points are encountered, simply
retain the original field value and move on to the next site.

\subsection{Autocorrelations in the free field $\lambda=0$ theory}

\begin{figure}[t]
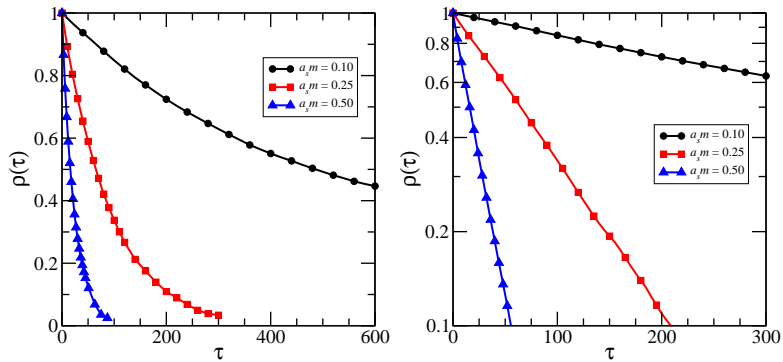

 \begin{center}
 \includegraphics[width=2.0in,bb=18 44 541 531]{autocorr1}
 \includegraphics[width=2.0in,bb=18 44 541 531]{autocorr2}
 \end{center}
\caption{The autocorrelation function $\rho(\tau)$ for
 $\langle\Phi(t)\Phi(0)\rangle$ with $t=1/(2a_sm)$ and 
$\Phi(t)=\sum_{xy}\phi(x,y,t)$ using $a_s m = 0.10, 0.25, 0.50$ 
 and $\lambda=0$ on $24^3$ isotropic lattices.  The parameter
 $\tau$ refers to the number of Metropolis sweeps.  The right
 plot has a logarithmic vertical scale.
\label{fig:phi4autocorr1}}
\end{figure}

\begin{figure}[tb]
 \begin{center}
 \includegraphics[width=2.0in,bb=18 44 541 531]{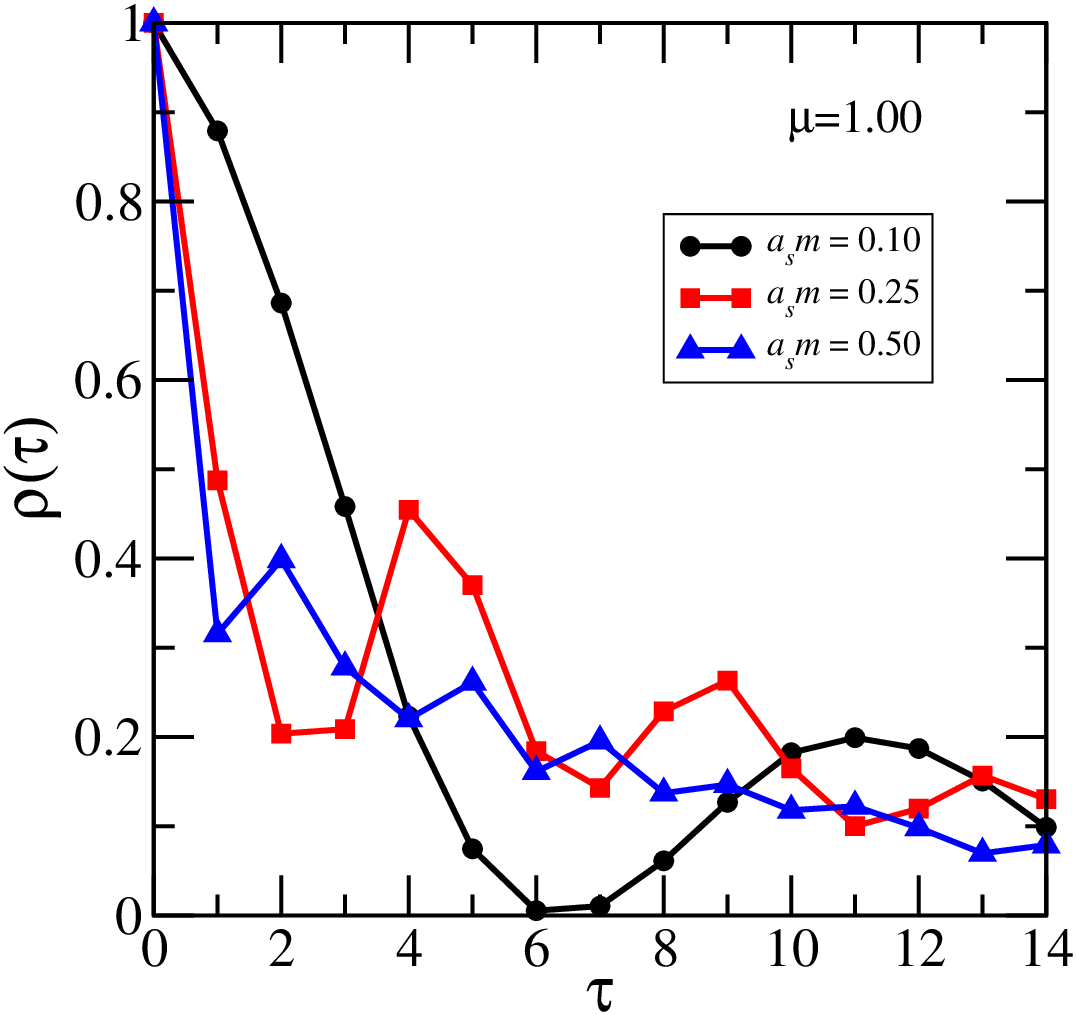}
 \includegraphics[width=2.0in,bb=18 44 541 531]{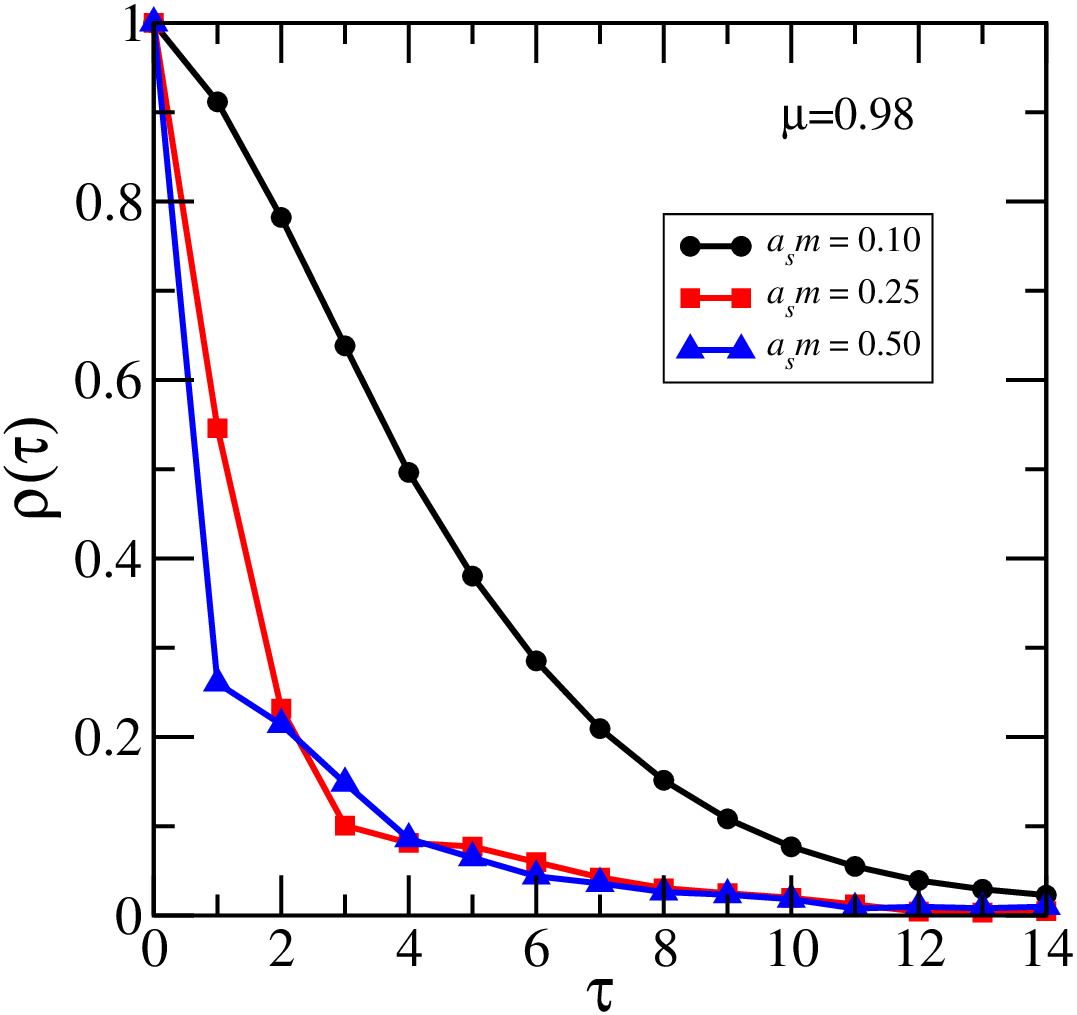}
 \end{center}
\caption{The autocorrelation function $\rho(\tau)$ for
 $\langle\Phi(t)\Phi(0)\rangle$ with $t=1/(2a_sm)$ and 
$\Phi(t)=\sum_{xy}\phi(x,y,t)$ using $a_s m = 0.10, 0.25, 0.50$ 
 and $\lambda=0$ on $24^3$ isotropic lattices.  The parameter
 $\tau$ refers to the number of compound sweeps.  Each compound
sweep is one Metropolis sweep, followed by one microcanonical
 sweep with probability $\mu$ of proposing a change.  In the
left plot, $\mu=1$, whereas $\mu=0.98$ in the right plot.
\label{fig:phi4autocorr2}}
\end{figure}

\begin{figure}[tb]
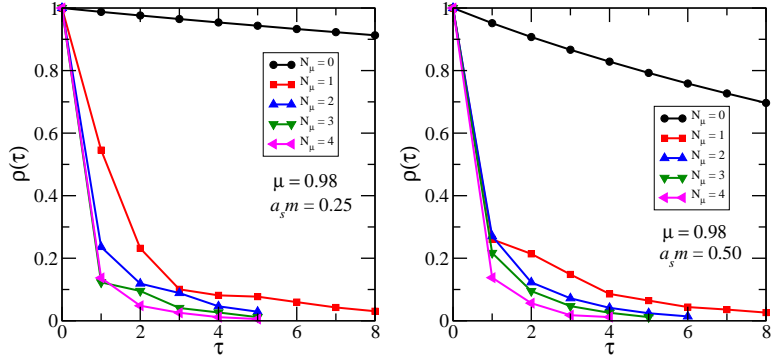

 \begin{center}
 \includegraphics[width=2.0in,bb=18 44 541 531]{autocorr7}
 \includegraphics[width=2.0in,bb=18 44 541 531]{autocorr8}
 \end{center}
\caption{The autocorrelation function $\rho(\tau)$ for
 $\langle\Phi(t)\Phi(0)\rangle$ with $t=1/(2a_sm)$ and 
$\Phi(t)=\sum_{xy}\phi(x,y,t)$ using $a_s m = 0.25$ (left plot)
 and $a_s m= 0.50$ (right plot) and $\lambda=0$ on $24^3$ isotropic 
 lattices.  The parameter
 $\tau$ refers to the number of compound sweeps.  Each compound
 sweep is one Metropolis sweep, followed by $N_\mu$ microcanonical
 sweeps with probability $\mu=0.98$ of proposing a change.
\label{fig:phi4autocorr4}}
\end{figure}

Let us first study autocorrelations in the free field $\lambda=0$
theory. The autocorrelation function $\rho(\tau)$ for the observable
 $\langle\Phi(t)\Phi(0)\rangle$ with $t=1/(2a_sm)$ and 
$\Phi(t)=\sum_{xy}\phi(x,y,t)$ is shown in 
Figs.~\ref{fig:phi4autocorr1}-\ref{fig:phi4autocorr4}
for various different updating schemes.
The parameter $\tau$ is the number of compound sweeps, and
$a_s m = 0.10, 0.25, 0.50$ are used for $\lambda=0$ on $24^3$ isotropic 
lattices.  

\begin{figure}[t]
 \begin{center}
  \includegraphics[width=2.0in,bb=18 44 541 531]{autocorr5}
  \includegraphics[width=2.0in,bb=18 44 541 531]{autocorr6}
 \end{center}
\caption{The autocorrelation function $\rho(\tau)$ for
 $\langle\Phi(t)\Phi(0)\rangle$ with $t=1/(2a_sm)$ and 
$\Phi(t)=\sum_{xy}\phi(x,y,t)$ using $a_s m = 0.10$ only
 and $\lambda=0$ on $24^3$ isotropic lattices.  The parameter
 $\tau$ refers to the number of compound sweeps.  Each compound
 sweep is one Metropolis sweep, followed by $N_\mu$ microcanonical
 sweeps with probability $\mu$ of proposing a change.  In the
 left plot, $N_\mu=1$ and $\mu$ is varied; in the right plot,
 $\mu=0.98$ and $N_\mu$ is varied.
\label{fig:phi4autocorr3}}
\end{figure}

In Fig.~\ref{fig:phi4autocorr1}, each compound sweep is just
one Metropolis sweep.  One sees that nearly 2200 sweeps are needed to reduce 
autocorrelations down to 0.1 for $a_sm=0.10$.  Fig.~\ref{fig:phi4autocorr2}
shows the dramatic reduction in autocorrelations by including microcanonical 
sweeps in the updating.  Each compound sweep in this figure is one
Metropolis sweep, followed by one microcanonical sweep with probability
$\mu$ of proposing a change.  The autocorrelation function has undesirable
oscillations when $\mu=1$ in the free field theory, as shown in the
left hand plot in this figure.  These oscillations can be removed by
using $\mu=0.98$, as shown in the right-hand plot of this figure.

In Figs.~\ref{fig:phi4autocorr3} and \ref{fig:phi4autocorr4}, each 
compound sweep is one Metropolis, followed by $N_\mu$ microcanonical
sweeps with probability $\mu$ of proposing a change.  In the left-hand
plot of Fig.~\ref{fig:phi4autocorr3}, $N_\mu=1$ is used and $\mu$ is varied.
One sees that in the free scalar field theory, autocorrelations improve
as $\mu$ increases towards unity, but $\mu=1$ introduces undesirable
oscillations.  Setting $\mu=0.98$ seems to be ideal.  This value is used
in the right-hand plot in this figure, and $N_\mu$ is varied.  This plot
shows autocorrelations improving as $N_\mu$ increases, but there are
diminishing returns.  As $N_\mu$ increases, each compound sweep takes
significantly more time, so this has to be weighed against the improvement
in the autocorrelations.  One finds that $N_\mu=3$ seems optimal
in the case of $a_s m=0.10$.   The autocorrelations for different $N_\mu$
with $\mu=0.98$ fixed are also shown in Fig.~\ref{fig:phi4autocorr4}
for $a_s m=0.25$ (left plot) and $a_s m=0.50$ (right plot).  There is
a dramatic improvement in going from $N_\mu=0$ to $N_\mu=1$, but there
is no further gain in increasing $N_\mu$ any further for these mass
parameters.

\subsection{Extracting observables}

The stationary-state energies can be extracted from
the asymptotic decay rates of temporal correlations of the fields.
The temporal evolution of the field as a Heisenberg-picture quantum operator is
  \beq \phi(t)=e^{Ht}\phi(0)e^{-Ht}, \eeq
and under certain general assumptions (an action satisfying
both link and site reflection positivity) and ignoring temporal boundary 
conditions, then for $t\geq 0$,
  \beqs \langle 0\vert\phi(t)\phi(0)\vert 0\rangle
 &=& \sum_n \langle 0\vert e^{Ht}\phi(0)e^{-Ht}\vert n\rangle\langle n\vert
\phi(0)\vert 0\rangle,\\
&=& \sum_n \Bigl\vert\langle n\vert\phi(0)\vert 0\rangle\Bigr\vert^2
 e^{-(E_n-E_0)t}=\sum_n A_n e^{-(E_n-E_0)t},
\eeqs
where a complete set of (discrete) eigenstates of $H$ satisfying
$H\vert n\rangle = E_n\vert n\rangle$ has been inserted.  Note that on a lattice
with periodic boundary conditions, momentum is discrete, so the energy
eigenstates are also discrete.
If
  $\langle 1\vert\phi(0)\vert 0\rangle\neq 0$, then $A_1$ and
  $E_1-E_0$ can be extracted as $t$ becomes large, assuming $\langle 0\vert
\phi(0)\vert 0\rangle=0$.
One can use any operator $O(t)$ which is a function
    of the field $\phi(t)$ only on a time slice $t$.
The extraction of $A_1$ and $E_1-E_0$ is done using a correlated$-\chi^2$ fit:
   \beq
    \chi^2 = \sum_{tt^\prime}\Bigl(C(t)-M(t,\alpha)\Bigr)\sigma_{tt^\prime}^{-1}
   \Bigl( C(t^\prime)-M(t^\prime,\alpha)\Bigr),
   \eeq
where $C(t)$ represents the Monte Carlo estimates of the correlation
    function with covariance matrix $\sigma_{tt^\prime}$ and the model
   function is $M(t,\alpha)=\alpha_1 e^{-\alpha_0 t}$.  The covariance
matrix $\sigma_{tt^\prime}$ is determined using the standard Monte
Carlo variance formula.
To determine estimates of the model parameters $\alpha_0$ and $\alpha_1$,
one minimizes the above $\chi^2$ with respect to the model parameters.
Uncertainties in the best-fit parameters
      $\alpha_0 = E_1-E_0$ and $\alpha_1=A_1$ are usually obtained by a 
       \alert{jackknife} or \alert{bootstrap} procedure (more on this
in a moment).  The fit must be done for a time range
      $t_{\rm min}\leq t\leq t_{\rm max}$ such that an acceptable fit
      quality is obtained, that is, $\chi^2/\mbox{dof}\approx 1$.
A sum of two-exponentials as a model function can be used to
     minimize sensitivity to $t_{\rm min}$,
   but the fit parameters associated
   with faster-decaying exponential are generally {\em not} good
  estimates of the gap to the next energy level and should be discarded.

The Monte Carlo method exploits the central limit theorem to
determine the statistical uncertainty in the covariance matrix
$\sigma_{tt^\prime}$.  But how can one determine the uncertainty
in the fit parameters $\alpha_0$ and $\alpha_1$?  These are not
simple quantities that can be defined properly on a single path;
the Monte Carlo variance formula cannot be easily applied.
The use of resampling schemes solves this problem.
Let $\langle f\rangle$ denote the Monte Carlo estimate of
      some quantity $f$ using all $X_k$ elements in a Markov chain,
for $k=1,2,\dots,N$, and let $\langle f\rangle_J$ denote the Monte 
Carlo estimate of $f$ \alert{omitting} $X_J$  (so only the other 
$N-1$ $X_k$ values are used).
The so-called \bfalert{jackknife} error estimate in $\langle f\rangle$
is given by
       \beqs \sigma^{(J)} &=& \left( \frac{N-1}{N}
           \sum_{J=1}^N ( \langle f\rangle_J - \langle f\rangle)^2
           \right)^{1/2},\\
   \langle f\rangle &=& \frac{1}{N}\sum_{k=1}^N f(X_k),\qquad
  \langle f\rangle_J = \frac{1}{N-1}\sum_{k\neq J} f(X_k).
           \eeqs
Again, the Monte Carlo variance formula can be used to determine the
covariance matrix $\sigma_{tt^\prime}$ for the correlation function itself
in $\chi^2$, and the jackknife method gives an estimate of the errors
in the model fit parameters.

Another resampling scheme is the \bfalert{bootstrap}.
Again, let $\langle f\rangle$ denote the Monte Carlo estimate of
      some quantity $f$ using all $X_k$ for $k=1,2,\dots,N$,
and let $\langle f\rangle_b$ denote the Monte Carlo estimate of
        $f$ using a new set $\widehat{X}_k$, for $k=1,2,\dots,N$, where
        each $\widehat{X}_k$ is one of the original $X_j$ chosen
        randomly with equal probability (a bootstrap sample).
A given $X_j$ can occur multiple times in the bootstrap sample.
After one obtains a large number $B$ of such estimates,
then the following quantity
 $\widehat{\langle f\rangle}=(1/B)\sum_{b=1}\langle f\rangle_b$
is evaluated.  The bootstrap error is given by
         \beq \sigma^{(B)} = \left( \frac{1}{B-1} \sum_{b=1}^B
            (\langle f\rangle_b -  \widehat{\langle f\rangle})^2
         \right)^{1/2}.\eeq
For a given quantity, a plot of its probability distribution can
be obtained from its $B$ bootstrap estimates.

A particularly good visual tool to see how well an energy can be
extracted is the so-called \bfalert{effective mass}.
For a correlator $C(t)$, the effective mass is defined by
    \beq m_{\rm eff}(t)=\ln\left(\frac{C(t)}{C(t+a_t)}\right).\eeq
This is a function which tends to $E_1-E_0$ as $t$ becomes large:
\beqs
 m_{\rm eff}(t)&=&
 \ln\left(\frac{
 A_1e^{-(E_1-E_0)t}\Bigl(1+(A_2/A_1)e^{-(E_2-E_1)t}+\dots \Bigr)}{
 A_1e^{-(E_1-E_0)(t+a_t)}\Bigl(1\!+\!(A_2/A_1)e^{-(E_2-E_1)(t+a_t)}\!+\!\dots 
\!\!\!\Bigr)}\right),\\ &=&
 \ln\left(\frac{
 e^{(E_1-E_0)a_t}\Bigl(1+(A_2/A_1)e^{-(E_2-E_1)t}+\dots \Bigr)}{
 \Bigl(1\!+\!(A_2/A_1)e^{-(E_2-E_1)(t+a_t)}\!+\!\dots 
\!\!\!\Bigr)}\right),\\
 &\stackrel{t\rightarrow\infty}{\longrightarrow}&
  \ln\left(e^{(E_1-E_0)a_t}\right)=a_t(E_1-E_0).
\eeqs
The value $E_1-E_0$ is seen as a large-time \alert{plateau} in the effective
 mass. Contributions from faster-decaying exponentials are seen
   as deviations of the effective mass from its asymptotic plateau value.
A ``good'' operator with little coupling to higher-lying states results
in a  rapid onset of the plateau. Statistical noise generally grows with $t$.

Extracting more than just the lowest energy in a symmetry
channel requires a hermitian \alert{matrix} of correlation functions 
 $C_{ij}(t)$.  Let $\lambda_n(t,t_0)$ denote the eigenvalues of 
 $C(t_0)^{-1/2}\,C(t)\,C(t_0)^{-1/2}$, where $t_0$ is some fixed reference 
time. These eigenvalues can be viewed as \alert{principal} 
       correlators.  Assume that they are ordered such that 
 $\lambda_0\geq\lambda_1\geq\cdots$  as $t$ becomes large,
 then one can show that
   \beqs
   \lim_{t\rightarrow\infty}\lambda_n(t,t_0) &=& e^{-E_n (t-t_0)}\Bigl(
    1 + O(e^{-\Delta_n (t-t_0)})\Bigr),\\
    \Delta_n &=& \min_{k\neq n}\vert E_k-E_n\vert.
   \eeqs
The effective masses associated with these principal correlators
are known as \alert{principal effective masses}:
    \beq
     m^{(n)}_{\rm eff}(t) = \ln\left(\frac{\lambda_n(t,t_0)}{
     \lambda_n(t+a_t,t_0)}\right).\eeq
For an $N\times N$ correlation matrix, these functions tend, as $t$ 
becomes large, to the energies of the $N$ lowest-lying states that 
couple to the operators whose correlations are being evaluated.  Examples
of such effective masses will be shown in the next section.

\subsection{Spectrum of the free field $\lambda=0$ theory}

To determine the spectrum for the free-field case $(\lambda=0)$ on an 
$N_x\times N_y\times N_t$ lattice, define
    \beq \Phi(t,n_x,n_y)=\sum_{x,y} \phi(x,y,t) 
      \ e^{2\pi i x n_x/N_x+2\pi i n_y/N_y}.\eeq
The lowest six levels having total zero momentum can be extracted 
     using the following set of six operators:
\beqs
 O_0(t) &=& \Phi(t,0,0),\\
 O_1(t) &=& \Phi(t,0,0)\ \Phi(t,0,0),\\
 O_2(t) &=& \Phi(t,1,0)\ \Phi(t,-1,0),\\
 O_3(t) &=& \Phi(t,0,1)\ \Phi(t,0,-1),\\
 O_4(t) &=& \Phi(t,1,1)\ \Phi(t,-1,-1),\\
 O_5(t) &=& \Phi(t,1,-1)\ \Phi(t,-1,1).
\eeqs

\begin{figure}
  \begin{center}
  \includegraphics[width=2.2in,bb=18 144 592 718]{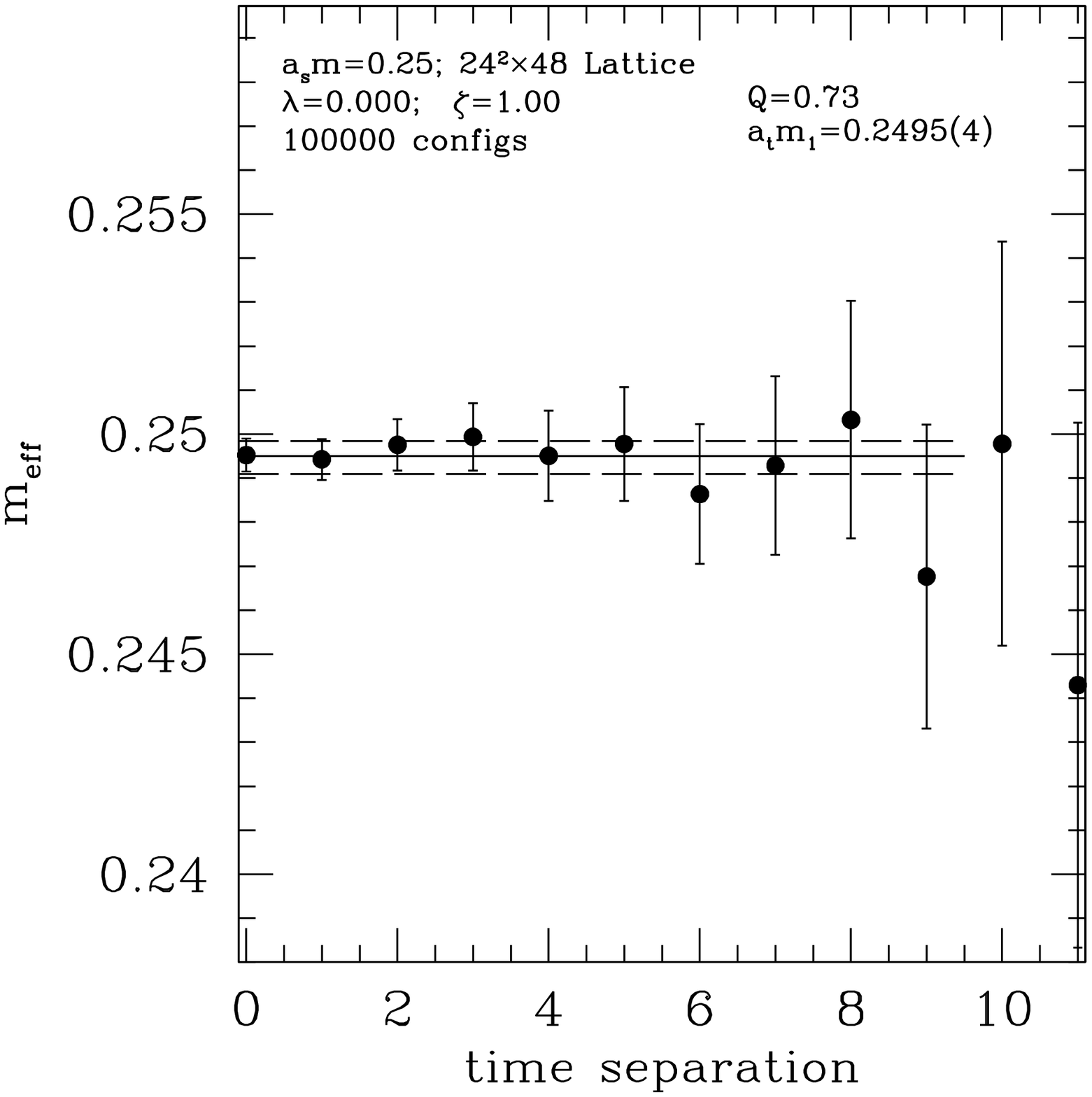}
  \includegraphics[width=2.2in,bb=18 144 592 718]{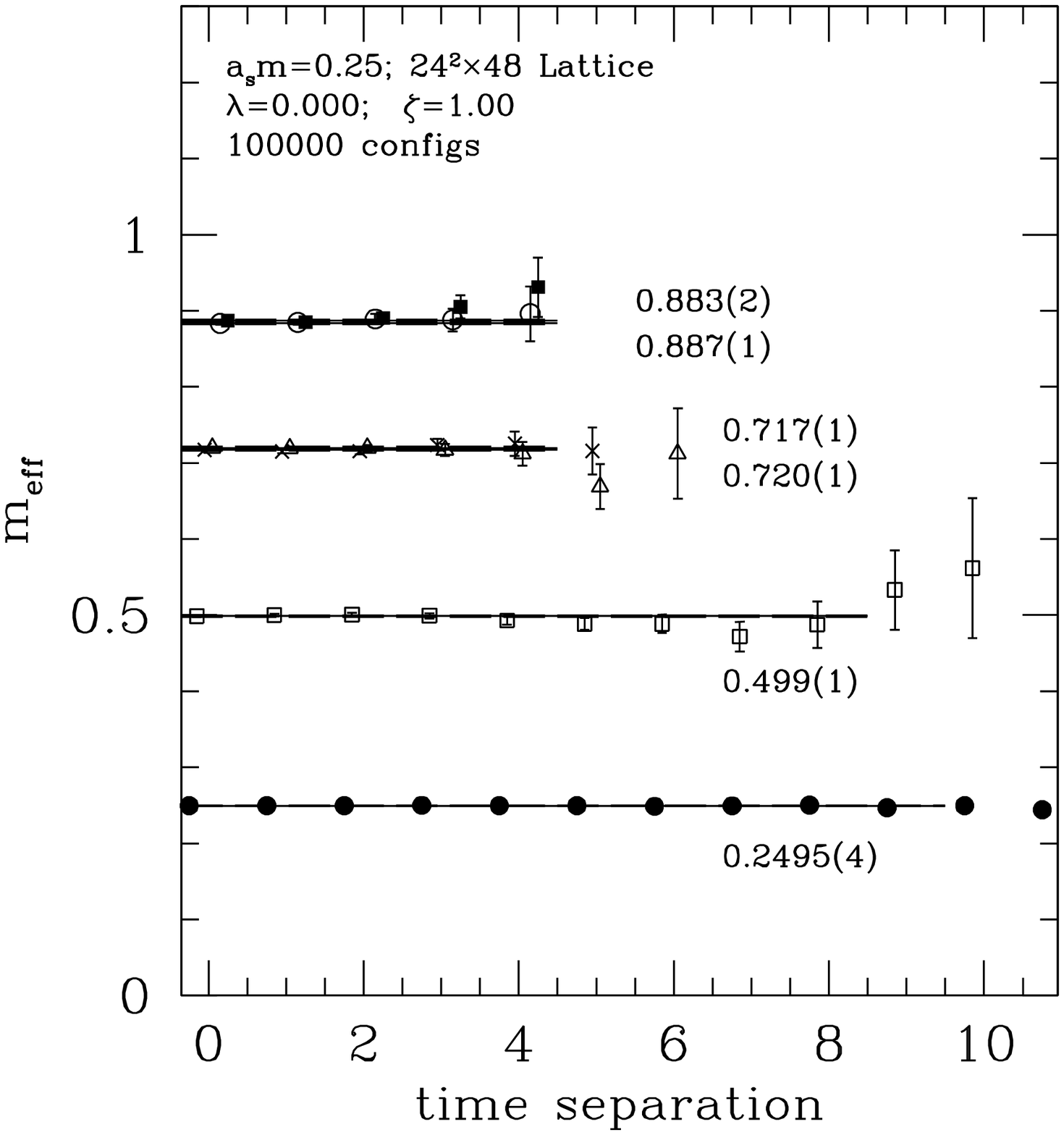}
  \end{center}
\caption{Principal effective masses from the $6\times 6$ matrix of
temporal correlations of the operators discussed in the text for the 
free field theory on a $24^2\times 48$ isotropic lattice with $a_sm=0.25$.
The left-hand plot shows the lowest-lying level, a single particle
at rest.  The right-hand plot shows the six lowest-lying levels.
All energies agree with exactly determined values: 0.24935 for the
single particle mass, 0.49871 for two particles at rest, 0.71903 for
the two states consisting of two particles having equal and opposite
minimal momenta, and 0.88451 for the next two energies.
\label{fig:phi4spectrum}}
\end{figure}

Principal effective masses from an actual Monte Carlo calculation
using these operators in the $\lambda=0$ scalar field theory are shown 
in Fig.~\ref{fig:phi4spectrum}.  The six lowest-lying levels are
extracted, and a $24^2\times 48$ isotropic lattice with $a_sm=0.25$
is used.  In this figure, the Monte Carlo results are compared with 
the known exact results: $0.24935$ for the mass, $0.49871$ for twice 
the mass, $0.71903$ for the two states having minimal relative momenta, 
and $0.88451$ for the next two states.  Note that the Monte Carlo
calculation only gives results for energies in terms of $a_t^{-1}$.
To obtain energies in terms of MeV, it is necessary to set the
value of $a_t^{-1}$; this is known as setting the scale.  Experimental
input is needed for this.

\subsection{The interacting theory}

We now introduce particle interactions by allowing the coupling $\lambda$
to be greater than zero.  Negative values of $\lambda$ are not allowed
as the energy would not be bounded from below.

\begin{figure}
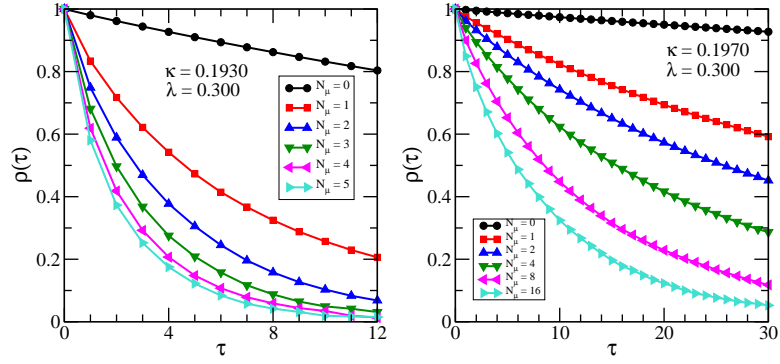

  \begin{center}
  \includegraphics[width=2.0in,bb=18 44 541 531]{autocorr9}
  \includegraphics[width=2.0in,bb=18 44 541 531]{autocorr10}
  \end{center}
\caption{Autocorrelation function $\rho(\tau)$ of the quantity
$\langle\Phi(t)\Phi(0)\rangle$ for $t\sim 1/(2a_sm)$ and $\Phi(t)=\sum_{xy}\phi(x,y,t)$
against the number of compound sweeps $\tau$, each consisting of one Metropolis
sweep followed by $N_\mu$ microcanonical sweeps with probability $\mu$
of proposing a change at each site.  In both plots, $\mu=1.00$ and $N_\mu$ is varied.
(a) In the left-hand plot, $t=2a_t$ is used with $\kappa=0.1930$ and $\lambda=0.300$ on 
a $24^2\times 48$ isotropic lattice such that the mass gap is about $0.25$; the
microcanonical acceptance rate is $0.807$. (b) In the 
right-hand plot, $t=5a_t$ is used with $\kappa=0.1970$ and $\lambda=0.300$ on a 
$32^2\times 96$ isotropic lattice so that the mass gap is about $0.10$; the
microcanonical acceptance rate is $0.804$.
\label{fig:phi4intautocorr1}}
\end{figure}

The autocorrelation function $\rho(\tau)$ of $\langle\Phi(t)\Phi(0)\rangle$ 
for $t\sim 1/(2a_sm_{\rm gap})$ in the interacting theory is shown
in Fig.~\ref{fig:phi4intautocorr1} for two sets of $\kappa,\lambda$
parameters and various values of $N_\mu$.  In these figures, $\tau$ refers
to the number of compound sweeps, and each compound sweep is one Metropolis
sweep, followed by $N_\mu$ microcanonical sweeps with probability $\mu=1$
of proposing a change.  In the left plot, $t=2a_t$ is used with 
$\kappa=0.1930$ and $\lambda=0.300$ on $24^2\times 48$ isotropic lattices 
and $a_sm_{\rm gap}\sim 0.25$.  In the right plot, $t=5a_t$ is used with 
$\kappa=0.1970$ and $\lambda=0.300$ on $32^2\times 96$ isotropic lattices 
and $a_sm_{\rm gap}\sim 0.10$.  The microcanonical acceptance rate is
about $80 \% $ in both cases.  These plots again show how important
the microcanonical sweeps are in reducing autocorrelations in the Markov
chains.

\begin{figure}
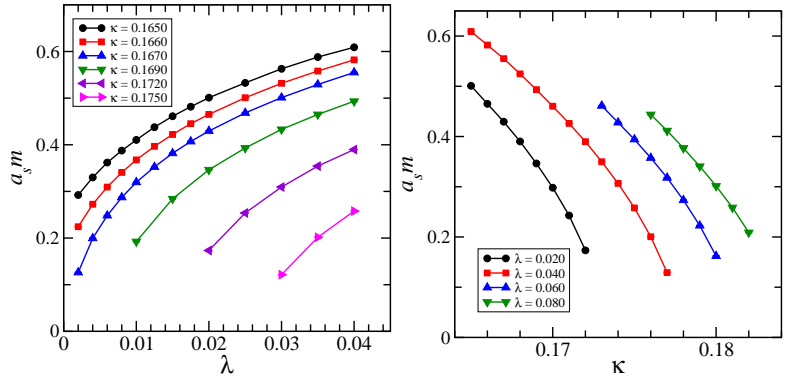

  \begin{center}
  \includegraphics[width=2.0in,bb=22 35 523 523]{massgaps1}
  \includegraphics[width=2.0in,bb=22 43 523 523]{massgaps2}
  \end{center}
\caption{The mass gap $a_s m_{\rm gap}$ in $2+1$-dimensional $\phi^4$
theory on $24^3$ isotropic lattices for various values of the hopping
parameter $\kappa$ and the coupling $\lambda$.
\label{fig:massgaps}}
\end{figure}

\begin{figure}[t]
  \begin{center}
  \includegraphics[width=2.5in,bb=2 30 529 523]{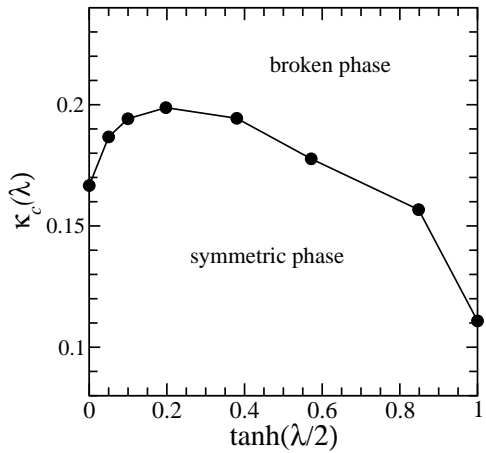}
  \end{center}
\caption{The phase boundary for 2+1-dimensional $\phi^4$ theory,
shown as a critical value $\kappa_c(\lambda)$ of the hopping
parameter as a function of the coupling $\lambda$.  The symmetry
$\phi\rightarrow -\phi$ is spontaneously broken above the
phase boundary.  The theory becomes the Ising model as 
$\lambda\rightarrow\infty$.
\label{fig:phases}}
\end{figure}

Single particle masses on $24^3$ isotropic lattices are shown for various
values of $\kappa,\lambda$ in Fig.~\ref{fig:massgaps}.  One sees that the
mass parameter $m$ in the Lagrangian is no longer the mass of the particle.
This $2+1$-dimensional $\phi^4$ theory has two phases separated by a line of 
critical points.  For each value of $\lambda$, there exists a critical 
value  $\kappa_c(\lambda)$ at which the mass gap goes to zero.  The 
so-called symmetric phase occurs for $\kappa < \kappa_c(\lambda)$; in this
phase, the $\phi\rightarrow -\phi$ symmetry holds, and $\langle\phi\rangle =0$.
The so-called broken phase occurs for $ \kappa > \kappa_c(\lambda)$;
in this phase, the symmetry $\phi\rightarrow -\phi$ of the Lagrangian
is spontaneously broken, such that there is a nonzero vacuum expectation
value for the field: $\langle\phi\rangle\neq 0$.  A somewhat qualitative
sketch of the phase boundary is shown in Fig.~\ref{fig:phases}.

Although the physics of this field theory are very interesting, the focus
of these lectures is the Monte Carlo method, which we have already
discussed in great detail.  In these six lectures, there is inadequate
time to study the renormalization group flows of this model, nor to
describe the calculational techniques needed to probe the phase
transition.  Phase transitions only occur in systems having an 
infinite number of degrees of freedom, requiring an infinite number
of lattice sites.  Monte Carlo studied must be performed with a finite
number of lattice sites out of necessity, so the study of the phase
transition requires a few clever tricks.  The interested reader is 
invited to further explore this field theory after these lectures.

\section{Monte Carlo calculations in lattice quantum chromodynamics}
\label{sec:qcd}

The field of lattice QCD began with the famous paper of Ken Wilson
in 1974\cite{wilson}.  Wilson found a way of formulating QCD on
a hypercubic space-time lattice which preserved local gauge
invariance, an important property linked to the renormalizability
of the theory.  In lattice QCD, the \alert{quarks} reside on the
sites, while the \alert{gluon} field resides on the links between sites.
Wilson advocated the use of link variables $U_\mu(x)$ which are 
path-ordered exponentials of the gauge field from one site to its
neighbor.  In other words, the $U_\mu(x)$ are parallel transport matrices.
In QCD, the link variables are elements of the group $SU(3)$.
For gluons, the path integration involves an eight-dimensional integral
on each link, so the path integral has dimension $32N_xN_yN_zN_t$.  
For a $24^4$ lattice, the path integral has a dimension near 10.6 million.
Of course, the fermion quark fields have to be dealt with, too.
The quark fields are Grassmann valued and obey Fermi-Dirac statistics.
This introduces major complications into the Monte Carlo updating
procedure.

Current Monte Carlo updating methods in pure gauge theories work
very well.  The best methods are similar to that already described
in the $\phi^4$ theory.  One sweeps through the lattice using
either a Metropolis or a heat bath local updating scheme, then
sweeps some number of times with an action-preserving local updating
method.  In some field theories, it is possible to sample the probability
density associated with a local update, either using a transformation
method or the rejection method.  Such a local updating method is
called a heat bath\cite{heatbath}.  A heat bath method for $SU(2)$
lattice gauge theory was proposed in Ref.~\refcite{creutzheatbath}
and was later improved in Ref.~\refcite{heatbathsu2}.  A heat bath
method for $SU(3)$ exists\cite{heatbathsu3}, but a more efficient
pseudo-heatbath method\cite{cabmar} of updating an $SU(3)$ matrix
by successive $SU(2)$ subgroups is usually used.  Microcanonical
(overrelaxation) updating of $SU(2)$ subgroups was proposed in
Ref.~\refcite{brown}, and a microcanonical procedure for $SU(3)$
is described in Ref.~\refcite{patel}.  Local microcanonical
updating algorithms for general $SU(N)$ gauge theories have been
devised in Refs.~\refcite{creutzOR,forcrand}.

Monte Carlo updating methods including quarks in lattice QCD 
are steadily improving.  The methods currently used are based on
the so-called Hybrid Monte Carlo\cite{HMC} method and a variant
known as RHMC\cite{RHMC}.  Fermions present special challenges
in terms of formulating them on a lattice and carrying out
Monte Carlo calculations.  Unfortunately, since my time is nearly
up, I will not be able to discuss these issues any further
here.

Before concluding, I would like to present the results of two Monte
Carlo studies in lattice gauge theory and lattice QCD that I have
been personally involved in.  An amazing feature of nonabelian
gauge theories is that the massless gluons can bind to form 
rather massive objects known as \alert{glueballs}.
The analog of such particles in electromagnetism would be massive
globules of pure light!  The mass spectrum of such objects from
Ref.~\refcite{glueballs} is shown in Fig.~\ref{fig:glueballs}. 
States are labeled by $J^{PC}$, where $J$ is the spin, $P$ is the
parity, and $C$ is the charge conjugation quantum number. The scale
has been set using $r_0^{-1}=410(20)$~MeV, where $r_0$ is a convenient
hadron scale defined in terms of the static quark potential.  These
results were computed using Cabibbo-Marinari pseudo-heatbath and 
Creutz microcanonical overrelaxation.  A $24\times 24$ correlation
matrix was used in each symmetry channel.  The mass of the lowest-lying
scalar glueball is the only particle mass that I know of that has a 
bounty on its head: the Clay Mathematics Institute will pay \$1 million
for a mathematical proof that this mass is nonzero.

\begin{figure}[t]
\begin{center}
 \includegraphics[width=3.0in,bb=11 98 535 631]{glueballs}
 \end{center}
\caption{ The mass spectrum of glueballs in the pure $SU(3)$ gauge theory
 from Ref~\protect\refcite{glueballs}.
 The masses are given in terms of the hadronic scale $r_0$ along the
 left vertical axis and in terms of GeV along the right vertical axis
 (assuming $r_0^{-1}=410$ MeV).  The mass uncertainties indicated by the
 vertical extents of the boxes do {\em not} include the uncertainty in
 setting $r_0$.  The locations of states whose interpretation requires further 
 study are indicated by the dashed hollow boxes.
\label{fig:glueballs}}
\end{figure}

\begin{figure}[t]
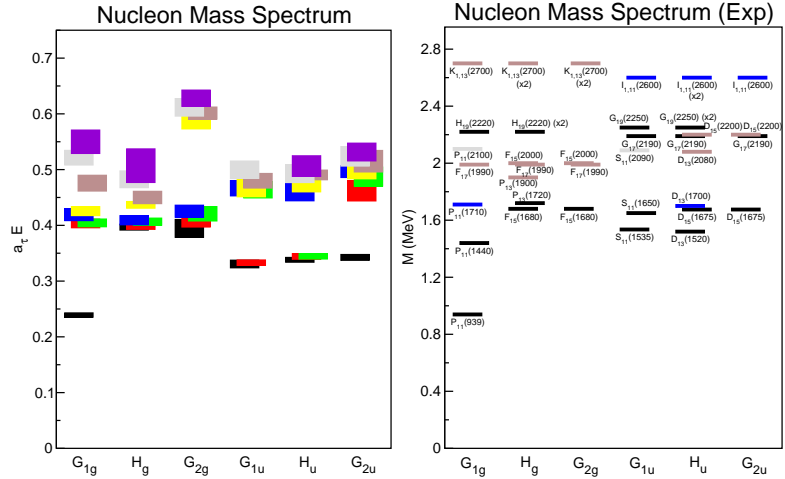

\begin{center}
\includegraphics[width=2.0in, bb=6 84 532 695]{nucleon_spectrum}
\includegraphics[width=2.0in, bb=10 84 532 695]{nucleon_spectrum_expt}
\end{center}
\caption[figwf]{(Left) An exploratory first computation of the
nucleon spectrum from 200 quenched configurations
on a $12^3\times 48$ anisotropic lattice using the Wilson gauge and quark
actions with $a_s \sim 0.1$ fm,
$a_s/a_t \sim 3.0$ and $m_\pi\sim 700$~MeV. The results
are from Ref.~\protect\refcite{lichtl2006}. (Right) The currently
known spectrum from experiment. Black denotes four-star states, blue
denotes three-star states, tan denotes two-star states, and gray
denotes a one-star state. 
\label{fig:nucleonspectra}}
\end{figure}

I am currently a member of a collaboration of lattice QCD theorists
known as the Lattice Hadron Physics Collaboration (LHPC). One of
our current goals is to use the Monte Carlo method to predict the
baryon and meson spectrum of QCD.  Our plans on how we intend to
accomplish this are outlined in Ref.~\refcite{baryons}.
We have recently completed exploratory studies\cite{lichtl2006} on small 
anisotropic lattices in the quenched approximation to QCD (ignoring
quark loops) at relatively large quark masses. These
studies demonstrated our ability to isolate up to nine energy 
eigenvalues from our
correlation functions.  The spectrum of isospin $I = 1/2$ states,
albeit at an unphysically large quark mass corresponding to $m_\pi
\simeq 700~{\rm MeV}$, is shown as the left-hand panel in
Fig.~\ref{fig:nucleonspectra}. The right-hand panel shows
the experimental spectrum, with states assigned according to the
irreducible representations of the cubic group; even in this quenched
calculation, at unphysical pion masses, there are tantalizing
suggestions of the existence of a band of negative-parity states well
separated from the higher excitations, as observed experimentally.
These calculations are ongoing.

\section{Conclusion}
\label{sec:conclude}

This series of six lectures was an introduction to using the Monte Carlo
method to carry out nonperturbative studies in quantum field theories.  
First, the path integral method in nonrelativistic quantum mechanics was
reviewed and several simple examples were studied: a free particle in one
dimension, the one-dimensional infinite square well, a free particle
in one dimension with periodic boundary conditions, and the
one-dimensional simple harmonic oscillator.  The extraction
of observables from correlation functions or vacuum expectation
values was discussed, and the evaluation of these correlation
functions using ratios of path integrals was described, introducing
Wick rotations to imaginary time.

A major portion of this series of lectures was then devoted to
the evaluation of path integrals in the imaginary time formalism
using the Monte Carlo method with Markov chains.  After a brief
review of probability theory, the law of large numbers and the 
central limit theorem were used to justify simple Monte Carlo 
integration.  However, the estimation of path integrals with
sufficient accuracy required the need for clever importance sampling,
which led to the use of stationary stochastic processes, and in
particular, ergodic Markov chains.  Several properties of Markov
chains were discussed and proven, particularly the fundamental limit
theorem for ergodic Markov chains.  The Metropolis-Hastings method 
of constructing a Markov chain was described.

Next, the one-dimensional simple harmonic oscillator was studied using
the Metropolis method.  One of the simplest quantum field theories,
a real scalar field theory with a $\phi^4$ interaction in three
space-time dimensions, was then studied.  The Metropolis method was 
seen to be plagued by strong autocorrelations.  An action-preserving 
(microcanonical) updating method was described,
and its effectiveness in reducing autocorrelations was demonstrated.
The extraction of stationary-state energies was detailed, introducing
correlated-$\chi^2$ fitting, as well as jackknife and bootstrap error
estimates.  The lectures concluded with some comments and results
from lattice QCD.

In addition to the references already cited, see
Ref.~\refcite{snell,ross,billingsley} for an introduction to probability,
Refs.~\refcite{parzen,prabhu} for more details concerning
stochastic processes, and Ref.~\refcite{montvay} for a very
good textbook on quantum fields on a lattice.

This work was supported by the U.S.\ National Science Foundation
through grant PHY-0354982.  I would especially
like to thank Robert Edwards, Jimmy Juge, Julius Kuti, Adam Lichtl,
Mike Peardon, and David Richards for their helpful comments
and suggestions.

\bibliographystyle{ws-procs9x6}
\bibliography{HUGS_montecarlo}

\end{document}